\newif\ifprd
\prdtrue

\ifprd
\documentclass[prd,twocolumn,showpcs,amsmath,amssymb,letterpaper,superscriptaddress]{revtex4}
\else
\RequirePackage{lineno}
\documentclass[letterpaper,superscriptaddress]{revtex4}
\usepackage{doublespace}
\fi

\usepackage{graphicx}
\usepackage{dcolumn}
\usepackage{bm}

\usepackage{amssymb}
\usepackage{amsmath}
\usepackage{slashed}
\usepackage{multirow}

\newcommand{\ifb}[0]{\ensuremath{\rm fb^{-1}}}
\newcommand{\ipb}[0]{\ensuremath{\rm pb^{-1}}}
\newcommand{\gev}[0]{\ensuremath{\rm GeV}}
\newcommand{\met}[0]{\ensuremath{\slashed{E}_T}}
\newcommand{\Z}[0]{\ensuremath{Z}}
\newcommand{\zee}[0]{\ensuremath{Z \rightarrow ee}}
\newcommand{\zmm}[0]{\ensuremath{Z \rightarrow \mu\mu}}
\newcommand{\zll}[0]{\ensuremath{Z \rightarrow \ell\ell}}
\newcommand{\mll}[0]{\ensuremath{M_{\ell\ell}}}
\newcommand{\ttbar}[0]{\ensuremath{t\bar{t}}}
\newcommand{\njet}[0]{\ensuremath{N_{\rm jet}}}

\newcommand{\njett}[0]{\ensuremath{N_{\rm jet}^{30}}}
\newcommand{\jt}[0]{\ensuremath{J_T}}
\newcommand{\jten}[0]{\ensuremath{J_T^{10}}}
\newcommand{\jtt}[0]{\ensuremath{J_T^{30}}}
\newcommand{\srbfom}[0]{\ensuremath{S/(1.5+\sqrt{B})}}

\begin{document}

\ifprd
\else
\pagewiselinenumbers
\fi

\title{
Search for New Particles Leading to Z+jets Final States in $p\bar{p}$ Collisions at $\sqrt{s}=1.96$ TeV
}

\affiliation{Institute of Physics, Academia Sinica, Taipei, Taiwan 11529, Republic of China} 
\affiliation{Argonne National Laboratory, Argonne, Illinois 60439} 
\affiliation{Institut de Fisica d'Altes Energies, Universitat Autonoma de Barcelona, E-08193, Bellaterra (Barcelona), Spain} 
\affiliation{Baylor University, Waco, Texas  76798} 
\affiliation{Istituto Nazionale di Fisica Nucleare, University of Bologna, I-40127 Bologna, Italy} 
\affiliation{Brandeis University, Waltham, Massachusetts 02254} 
\affiliation{University of California, Davis, Davis, California  95616} 
\affiliation{University of California, Los Angeles, Los Angeles, California  90024} 
\affiliation{University of California, San Diego, La Jolla, California  92093} 
\affiliation{University of California, Santa Barbara, Santa Barbara, California 93106} 
\affiliation{Instituto de Fisica de Cantabria, CSIC-University of Cantabria, 39005 Santander, Spain} 
\affiliation{Carnegie Mellon University, Pittsburgh, PA  15213} 
\affiliation{Enrico Fermi Institute, University of Chicago, Chicago, Illinois 60637} 
\affiliation{Comenius University, 842 48 Bratislava, Slovakia; Institute of Experimental Physics, 040 01 Kosice, Slovakia} 
\affiliation{Joint Institute for Nuclear Research, RU-141980 Dubna, Russia} 
\affiliation{Duke University, Durham, North Carolina  27708} 
\affiliation{Fermi National Accelerator Laboratory, Batavia, Illinois 60510} 
\affiliation{University of Florida, Gainesville, Florida  32611} 
\affiliation{Laboratori Nazionali di Frascati, Istituto Nazionale di Fisica Nucleare, I-00044 Frascati, Italy} 
\affiliation{University of Geneva, CH-1211 Geneva 4, Switzerland} 
\affiliation{Glasgow University, Glasgow G12 8QQ, United Kingdom} 
\affiliation{Harvard University, Cambridge, Massachusetts 02138} 
\affiliation{Division of High Energy Physics, Department of Physics, University of Helsinki and Helsinki Institute of Physics, FIN-00014, Helsinki, Finland} 
\affiliation{University of Illinois, Urbana, Illinois 61801} 
\affiliation{The Johns Hopkins University, Baltimore, Maryland 21218} 
\affiliation{Institut f\"{u}r Experimentelle Kernphysik, Universit\"{a}t Karlsruhe, 76128 Karlsruhe, Germany} 
\affiliation{High Energy Accelerator Research Organization (KEK), Tsukuba, Ibaraki 305, Japan} 
\affiliation{Center for High Energy Physics: Kyungpook National University, Taegu 702-701, Korea; Seoul National University, Seoul 151-742, Korea; SungKyunKwan University, Suwon 440-746, Korea} 
\affiliation{Ernest Orlando Lawrence Berkeley National Laboratory, Berkeley, California 94720} 
\affiliation{University of Liverpool, Liverpool L69 7ZE, United Kingdom} 
\affiliation{University College London, London WC1E 6BT, United Kingdom} 
\affiliation{Centro de Investigaciones Energeticas Medioambientales y Tecnologicas, E-28040 Madrid, Spain} 
\affiliation{Massachusetts Institute of Technology, Cambridge, Massachusetts  02139} 
\affiliation{Institute of Particle Physics: McGill University, Montr\'{e}al, Canada H3A~2T8; and University of Toronto, Toronto, Canada M5S~1A7} 
\affiliation{University of Michigan, Ann Arbor, Michigan 48109} 
\affiliation{Michigan State University, East Lansing, Michigan  48824} 
\affiliation{University of New Mexico, Albuquerque, New Mexico 87131} 
\affiliation{Northwestern University, Evanston, Illinois  60208} 
\affiliation{The Ohio State University, Columbus, Ohio  43210} 
\affiliation{Okayama University, Okayama 700-8530, Japan} 
\affiliation{Osaka City University, Osaka 588, Japan} 
\affiliation{University of Oxford, Oxford OX1 3RH, United Kingdom} 
\affiliation{University of Padova, Istituto Nazionale di Fisica Nucleare, Sezione di Padova-Trento, I-35131 Padova, Italy} 
\affiliation{LPNHE, Universite Pierre et Marie Curie/IN2P3-CNRS, UMR7585, Paris, F-75252 France} 
\affiliation{University of Pennsylvania, Philadelphia, Pennsylvania 19104} 
\affiliation{Istituto Nazionale di Fisica Nucleare Pisa, Universities of Pisa, Siena and Scuola Normale Superiore, I-56127 Pisa, Italy} 
\affiliation{University of Pittsburgh, Pittsburgh, Pennsylvania 15260} 
\affiliation{Purdue University, West Lafayette, Indiana 47907} 
\affiliation{University of Rochester, Rochester, New York 14627} 
\affiliation{The Rockefeller University, New York, New York 10021} 
\affiliation{Istituto Nazionale di Fisica Nucleare, Sezione di Roma 1, University of Rome ``La Sapienza," I-00185 Roma, Italy} 
\affiliation{Rutgers University, Piscataway, New Jersey 08855} 
\affiliation{Texas A\&M University, College Station, Texas 77843} 
\affiliation{Istituto Nazionale di Fisica Nucleare, University of Trieste/\ Udine, Italy} 
\affiliation{University of Tsukuba, Tsukuba, Ibaraki 305, Japan} 
\affiliation{Tufts University, Medford, Massachusetts 02155} 
\affiliation{Waseda University, Tokyo 169, Japan} 
\affiliation{Wayne State University, Detroit, Michigan  48201} 
\affiliation{University of Wisconsin, Madison, Wisconsin 53706} 
\affiliation{Yale University, New Haven, Connecticut 06520} 
\author{T.~Aaltonen}
\affiliation{Division of High Energy Physics, Department of Physics, University of Helsinki and Helsinki Institute of Physics, FIN-00014, Helsinki, Finland}
\author{A.~Abulencia}
\affiliation{University of Illinois, Urbana, Illinois 61801}
\author{J.~Adelman}
\affiliation{Enrico Fermi Institute, University of Chicago, Chicago, Illinois 60637}
\author{T.~Affolder}
\affiliation{University of California, Santa Barbara, Santa Barbara, California 93106}
\author{T.~Akimoto}
\affiliation{University of Tsukuba, Tsukuba, Ibaraki 305, Japan}
\author{M.G.~Albrow}
\affiliation{Fermi National Accelerator Laboratory, Batavia, Illinois 60510}
\author{S.~Amerio}
\affiliation{University of Padova, Istituto Nazionale di Fisica Nucleare, Sezione di Padova-Trento, I-35131 Padova, Italy}
\author{D.~Amidei}
\affiliation{University of Michigan, Ann Arbor, Michigan 48109}
\author{A.~Anastassov}
\affiliation{Rutgers University, Piscataway, New Jersey 08855}
\author{K.~Anikeev}
\affiliation{Fermi National Accelerator Laboratory, Batavia, Illinois 60510}
% ...
\collaboration{CDF Collaboration\footnote{With visitors from $^a$University of Athens, 15784 Athens, Greece, 
$^b$University of Bristol, Bristol BS8 1TL, United Kingdom, 
$^c$University Libre de Bruxelles, B-1050 Brussels, Belgium, 
$^d$Cornell University, Ithaca, NY  14853, 
$^e$University of Cyprus, Nicosia CY-1678, Cyprus, 
$^f$University College Dublin, Dublin 4, Ireland, 
$^g$University of Edinburgh, Edinburgh EH9 3JZ, United Kingdom, 
$^h$University of Heidelberg, D-69120 Heidelberg, Germany, 
$^i$Universidad Iberoamericana, Mexico D.F., Mexico, 
$^j$University of Manchester, Manchester M13 9PL, England, 
$^k$Nagasaki Institute of Applied Science, Nagasaki, Japan, 
$^l$University de Oviedo, E-33007 Oviedo, Spain, 
$^m$University of London, Queen Mary College, London, E1 4NS, England, 
$^n$University of California Santa Cruz, Santa Cruz, CA  95064, 
$^o$Texas Tech University, Lubbock, TX  79409, 
$^p$University of California, Irvine, Irvine, CA  92697, 
$^q$IFIC(CSIC-Universitat de Valencia), 46071 Valencia, Spain, 
}}
\noaffiliation

\date{\today}

\begin{abstract}

We present the results of a search for new particles that lead to a \Z\ boson plus jets in $p\bar{p}$ collisions at $\sqrt{s}=1.96$ TeV using the Collider Detector at Fermilab (CDF II).  A data sample with a luminosity of 1.06 \ifb\ collected using \Z\ boson decays to $ee$ and $\mu\mu$ is used.  We describe a completely data-based method to predict the dominant background from standard-model \Z+jet events.  This method can be similarly applied to other analyses requiring background predictions in multi-jet environments, as shown when validating the method by predicting the background from $W$+jets in \ttbar\ production.  No significant excess above the background prediction is observed, and a limit is set using a fourth generation quark model to quantify the acceptance.  Assuming $BR(b' \rightarrow b\Z) = 100\%$ and using a leading-order calculation of the $b'$ cross section, $b'$ quark masses below 268 $\gev/c^2$ are excluded at 95\% confidence level.


\end{abstract}

\maketitle


\section{Introduction}
\label{sec:intro}

This paper presents a search for new particles decaying to \Z\ gauge bosons created in $p\bar{p}$ collisions at $\sqrt{s}=1.96$~TeV
with the CDF II detector at the Fermilab Tevatron, extending and complementing other work with such final states
\cite{bib:cdfxsec,bib:runibp,bib:zresd0,bib:lxy}.  A variety of extensions to the standard model predict new particles with couplings
to \Z\ bosons \cite{bib:bprime,bib:susygmsbz,bib:susysugraz,bib:bpt,bib:strassler}.  We wish to discover or rule out these types of
models, while maintaining model independence in the search.  That is, while these theories offer guidance about the possible
characteristics of physics beyond the standard model, they do not necessarily correspond to what actually exists in nature, and so
the analysis is not tailored to specific models.

Of course, some assumptions are necessary in choosing how to discriminate between the standard model background and new signals.  We
examine final states with \Z\ bosons and additional jets.  In particular, we focus on final states in which there are at least 3
jets, each with at least 30 GeV of transverse energy $E_T$.  This assumption was motivated by studying the optimal kinematic
selection of a specific model, the fourth generation model \cite{bib:bprime}.  In the fourth generation model, an additional pair of
heavy quarks is added to the standard model's three.  The production mechanisms of the new down-type quark (called the $b'$) would be
identical to that of the top quark, with pair-production having the largest cross section.  Depending on its mass, the direct
tree-level decays of the $b'$ could be either kinematically forbidden or heavily Cabibbo-suppressed.  These situations could give
rise to a large branching ratio of $b' \rightarrow b Z$ via a loop diagram.  While the selection was chosen as the optimal set of
kinematic cuts using this model as a signal, this analysis constrains all models with \Z+3 jet final states.


The dominant background for this final state is from standard model \Z\ production with jets from higher order QCD processes. A
leading order calculation of this background is insufficient. Use of higher order calculations is complicated because it involves
hard-scattering matrix elements in combination with soft non-perturbative QCD processes. Recent NLO predictions \cite{bib:mrennaqcd}
have been used \cite{bib:alpgen_analyses} with the aid of Monte Carlo simulations to account for the
non-perturbative overlap.  Any such method requires validation with data. In this paper, we develop a different approach that uses
the data as more than a validation tool, and uses it alone for the background estimation.  In this approach, we extrapolate the jet
transverse energy distributions from a low energy control region of the data into the high energy signal region.

This paper is organized as follows.  Section~\ref{sec:cdfdet} contains a brief overview of the portions of the CDF II detector
relevant to this measurement.  Section~\ref{sec:data} lists the trigger requirements, and describes and motivates the signal sample
selections.  Section~\ref{sec:bkga} lists the backgrounds.  Section~\ref{sec:bkgmethod} describes, validates, and applies the method
of predicting the dominant background.  In Sec.~\ref{sec:bkgb} the predictions for the remaining backgrounds are described.  In
Sec.~\ref{sec:results} we present the results of the search, and conclude in Sec.~\ref{sec:conc}.

\section{The CDF II Detector}
\label{sec:cdfdet}

The CDF II detector is described in detail elsewhere \cite{bib:cdfdet}; here, only the portions required for this analysis are
described.  We first describe the coordinate system conventions.  In the CDF coordinate system, the origin is the center of the
detector, and the $z$ axis is along the beam axis, with positive $z$ defined as the proton beam direction.  The $x$ axis points radially
outward from the Tevatron ring, leaving the $y$ axis direction perpendicular to the earth's surface with positive direction upward.
Spherical coordinates are used where appropriate, in which $\theta$ is the polar angle (zero in the positive $z$ direction), $\phi$
is the azimuthal angle (zero in the positive $x$ direction), and the pseudorapidity $\eta$ is defined by $\eta \equiv - \ln [
\tan(\theta/2) ]$.  At hadron colliders, transverse energies and momenta are usually the appropriate physical quantities, defined by
$E_T \equiv E \sin \theta$ and $p_T \equiv p \sin \theta$ (where $E$ is a particle's energy and $p$ is the magnitude of a particle's
momentum).

A tracking system is situated directly outside the beam pipe and measures the trajectories and momenta of charged particles.  The
innermost part of the tracking system is the silicon detector, providing position measurements on up to 8 layers of sensors in the
radial region $1.3 < r < 28$~cm and the polar region $|\eta| \lesssim 2.5$.  Outside of this detector lies the central outer tracker
(COT), an open-cell drift chamber providing measurements on up to 96 layers in the radial region $40 < r < 137$~cm and the polar
region $|\eta| \lesssim 1$.  Directly outside of the COT a solenoid provides a $1.4$~T magnetic field, allowing particle momenta to
be obtained from the trajectory measurements in this known field.

Surrounding the tracking system, segmented electromagnetic (EM) and hadronic calorimeters measure particle energies.  In the central
region, the calorimeters are arranged in a projective barrel geometry and cover the polar region $|\eta| < 1.2$.  In the forward
region, the calorimeters are arranged in a projective ``end-plug'' geometry and cover the polar region $1.2 < |\eta| < 3.5$.  Two
sets of drift chambers, one directly outside the hadronic calorimeter and another outside additional steel shielding, measure muon
trajectories in the region $|\eta| < 0.6$; another set of drift chambers similarly detects muons in the region $0.6 < |\eta| < 1$.
Muon scintillators surround these drift chambers in the region $|\eta| < 1$ for trigger purposes.  A luminosity measurement is
provided by Cherenkov detectors in the region $3.7 < |\eta| < 4.7$ via a measurement of the average number of $p\bar{p}$ collisions
per crossing \cite{bib:clc}.

Collision events of interest are selected for analysis offline using a three level trigger system, with each level accepting events
for processing at the next level.  At level 1, custom hardware enables fast decisions using rudimentary tracking information and a
simple counting of reconstructed objects.  At level 2, trigger processors enable decisions based on partial event reconstruction.  At
level 3, a computer farm running fast event reconstruction software makes the final decision on event storage.

\section{Data Sample and Event Selection}
\label{sec:data}

We first describe the baseline \Z\ selection, and then describe the kinematic selection used to discriminate the potential signal
from the standard model background.  The kinematic selection is chosen and backgrounds are predicted \textit{a priori}, before
looking in the signal region.  While remaining as data-driven as possible throughout the analysis, Monte Carlo simulation is used in
some studies, consistency checks, and for illustration purposes.  In all cases, the Monte Carlo events are generated with
\textsc{pythia} \cite{bib:pythia} and the detector responses are modeled with a \textsc{geant} simulation \cite{bib:geant} as
described in \cite{bib:cdfxsecprd}.

\subsection{Baseline $Z$ Selection}
\label{subsec:basesel}

The data sample consists of $\zee$ and $\zmm$ candidate events collected using single electron and muon triggers.  The electron
trigger requires at least one central electromagnetic energy cluster with $E_T > 18$~GeV and a matching track with $p_T > 9\ \gev/c$.
The muon trigger requires at least one central track with $p_T > 18\ \gev/c$ with matching hits in the muon drift chambers.  The
average integrated luminosity of these data samples is 1.06 $\ifb$ \cite{bib:lumiave}.

\Z\ candidate events are selected offline by requiring at least one pair of electron or muon candidates both with $p_T > 20\ \gev/c$
and invariant mass in the range $81 < M_{\ell\ell} < 101\ \gev/c^2$.  The electron and muon identification variables are described in
detail in Refs.~\cite{bib:cdfxsecprd,bib:mythesis}.  The selection is described briefly here.  To increase efficiency, only one of
the lepton pair has stringent identification requirements (the ``tight'' candidate), while on the other lepton the identification
requirements are relaxed (the ``loose'' candidate).  

``Loose'' electron candidates consist of well-isolated EM calorimeter clusters with low energy in the hadronic calorimeter; in the
central part of the detector ($|\eta|<1.2$) well-measured tracks from the COT are required; in the forward parts of the detector
($|\eta|>1.2$) no track is required, but the shower shape in the EM calorimeter is required to be consistent with that expected from
electrons.  ``Tight'' electron candidates have all the requirements of ``loose'' candidates, and are additionally required to be
central ($|\eta|<1.2$), to have a shower shape consistent with that expected from electrons, to have calorimeter position and energy
measurements consistent with its matching track, and to have no nearby tracks consistent with that expected in electrons from photon
conversions.

``Loose'' muon candidates consist of well-measured tracks in the COT and well-isolated EM and hadronic calorimeter clusters with
minimal energy deposits.  ``Tight'' muon candidates have all the requirements of ``loose'' candidates, and are additionally required
to have matching hits in the muon drift chambers.

\begin{figure}
\begin{center}
\includegraphics[width=2.5in]{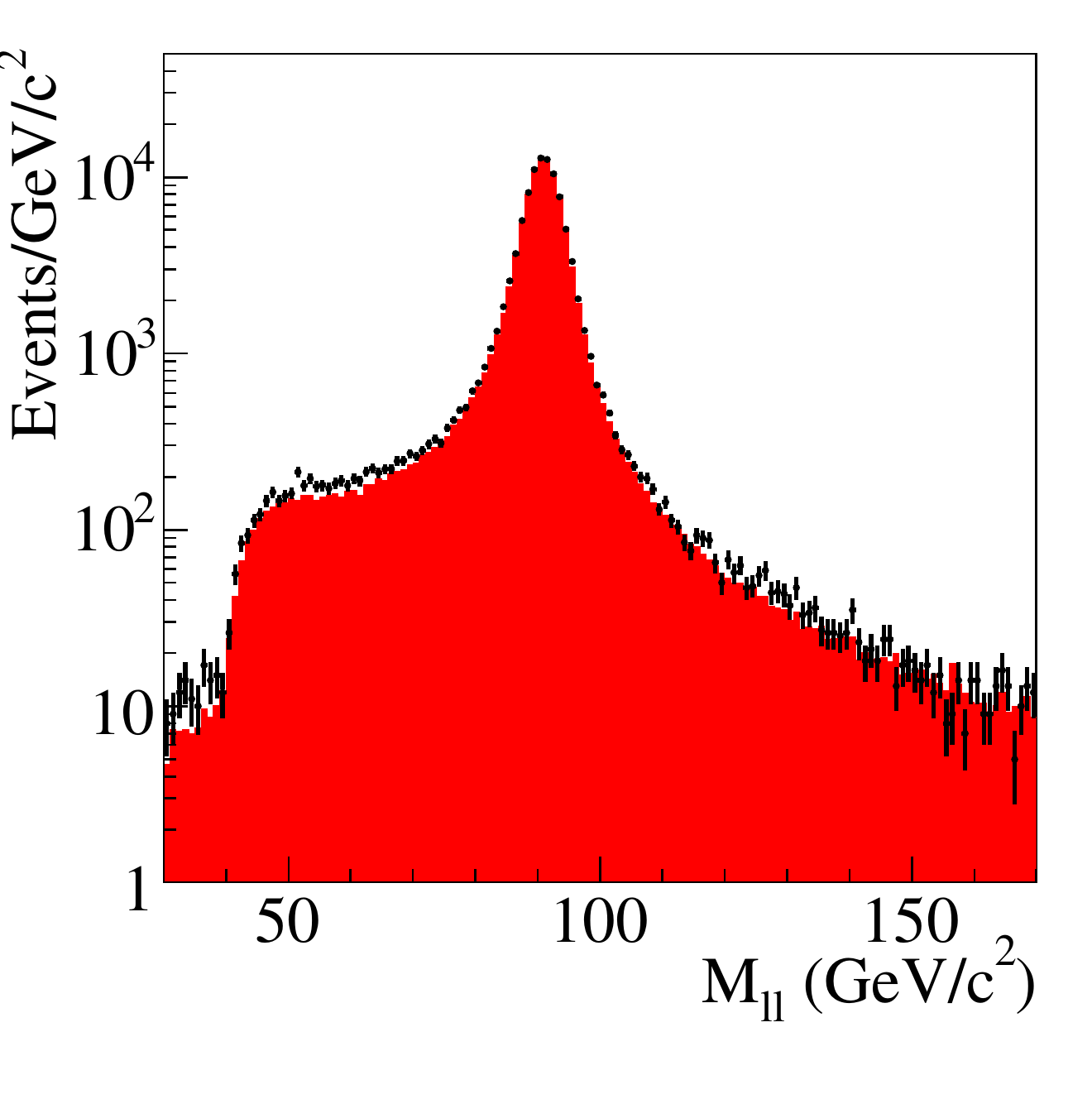}
\end{center}
\caption[]{
Distribution of $M_{\ell\ell}$ of \zee\ and \zmm\ data (black points and errors) using the baseline \Z\ selection described in the text.
Overlaid are standard model \zee\ and \zmm\ Monte Carlo events, normalized to the number of events expected with the given
luminosities using the expected cross section of 250 pb.
}
\label{fig:prd_datacomb_mz}
\end{figure}

Finally, all electron and muon pairs are required to be consistent with originating from the same $z$ vertex and to have a
time-of-flight difference (as measured by the COT) inconsistent with that expected for cosmic rays.  They are also required to be
separated in $\phi$ by an angle greater than $5^{\circ}$ to remove two lepton candidates mis-reconstructed from a single lepton.

Using this selection, the distribution of $M_{\ell\ell}$ is plotted and compared to standard model \Z\ Monte Carlo simulation in
Fig.~\ref{fig:prd_datacomb_mz}.

\subsection{Kinematic Selection}
\label{subsec:kinsel}


The analysis focuses on topologies with large numbers of highly energetic jets in the final state, for which the signal (from the
decay of heavy objects) can be better separated from standard model \Z+jet production.  Jets are clustered using the
``\textsc{midpoint}'' clustering algorithm \cite{bib:midpoint} with a cone size of 0.4 radians.  Corrections are applied to
extrapolate the jet energies back to the parton level using a generic jet response \cite{bib:jetcorr}.  Jets are required to have
$|\eta| < 2$.

The following discriminators are used:
\begin{center}
\begin{tabular}{rcp{2in}}
$N_{\rm jet}^{X}$ & $=$ & Number of jets in the event with $E_T > X$~GeV \\
$J_{T}^{X}$       & $=$ & Scalar sum of $E_T$ of jets in the event with $E_T > X$~GeV
\end{tabular}
\end{center}
The thresholds $X$ as well as the cut values on these variables are determined by optimization \cite{bib:thresh}.  In the
optimization we use the figure of merit \srbfom\ (where $S$ is the expected number of signal events and $B$ is the expected number of
background events) to quantify the sensitivity as a compromise between best discovery and best limit potential
\cite{bib:sigbkg1p5,bib:1p5explan}.  In the low background region ($B \ll 1$), maximizing this figure of merit is equivalent to
maximizing the signal efficiency.  In the high background region ($B \gg 1$), this figure of merit has the same behavior as
$S/\sqrt{B}$.  For the optimization study, $p\bar{p}\rightarrow b' \bar{b'}$ Monte Carlo simulations with a range of masses are used
as the signal $S$.  Standard model \Z\ Monte Carlo simulations are used for the background $B$.

In order to be sensitive to a range of masses, we must take into account the generic behavior of new signals: as mass increases the
cross section decreases while the transverse energy spectra become harder.  Therefore, to be optimally sensitive to higher mass
signals, we cut at larger values of \njet\ and \jt\, thus removing more of the background to give sensitivity to the lower cross
sections.

For the sake of simplicity, we desire that our selection only changes gradually with mass and uses the same $E_T$ threshold on all
jets.  With a simple selection, the data-based background prediction method becomes easier.  To confirm that this desire for
simplicity does not considerably reduce the search sensitivity, and to understand what cut values and thresholds to use, we first
establish a ``target'' selection.  The ``target'' selection is defined as the selection with the highest sensitivity when placing
cuts on the individual jet $E_T$'s and \jt.  This is found by scanning through all possible cuts on \jten\ (that is, \jt\ is
calculated with a 10 GeV threshold on the jets) and all possible $E_T$ thresholds for up to 4 jets (ordered by $E_T$), and finding
the point with the optimal sensitivity.  In this scan, step sizes of 10 GeV are used for the jet $E_T$ thresholds, and a step size of
50 GeV is used for \jten.  This scan is done independently for $b'$ masses in the range $100 \leq m_{b'} \leq 350\ \gev/c^2$ with a
step size of 50 $\gev/c^2$.

The optimal points found by this scan for a $b'$ mass of 150 $\gev/c^2$ are shown in column 2 of Table~\ref{tab:optimtable2}.  These
cut values give the best possible sensitivity at this mass point when placing cuts on the individual jet $E_T$'s and \jten.  Again,
we wish to choose a simple selection that gradually changes as a function of mass, and use the target sensitivities at all mass
points for comparison.  Based on the optimal target points for $b'$ masses in the range $100 \leq m_{b'} \leq 350\ \gev/c^2$, we
choose the simpler requirements of $\njett \geq 3$ and $\jten > m_{b'} c^2$.  The sensitivity of the simple requirements is compared
to the target sensitivity in column 3 of Table~\ref{tab:optimtable2} for the 150 $\gev/c^2$ mass point.

\begin{table}
\begin{center}
\begin{tabular}{ ccc }
\hline \hline
                         & Values    & Values of simple \\
Variable                 & from scan & selection \\
\hline
$E_T^{\rm jet\ 1}$ thresh.:  & $50$ &  $30$ \\
$E_T^{\rm jet\ 2}$ thresh.:  & $30$ &  $30$ \\
$E_T^{\rm jet\ 3}$ thresh.:  & $30$ &  $30$ \\
$E_T^{\rm jet\ 4}$ thresh.:  & $20$ &   $0$ \\
$J_T^{10} $     cut:     & $0$ & $150$ \\ \hline
$N_{\rm sig}$:  & $48.5$ & $75.5$ \\
$N_{\rm bkg}$:  & $2.60$ & $13.8$ \\
$\srbfom$: & $15.6$ & $14.5$ \\
\hline \hline
\end{tabular}
\end{center}
\caption{Optimal point compared with the simple selection of $N_{jet}^{30} \geq 3$ and $J_T^{10} > 150$, for the $m_{b'} = 150\
\gev/c^2$ mass point.  Here, $N_{\rm sig}$ is the number of signal events expected in 1\ifb\ after the given selection using $b'$ Monte
Carlo simulations.  $N_{\rm bkg}$ is the number of background events expected in 1\ifb\ after the given selection using standard model
\Z\ Monte Carlo simulations.  In this optimization study, $2.7 \times 10^{5}$ standard model \Z\ events were used; 1500 signal events
were used (both counted before jet selection).}
\label{tab:optimtable2}
\end{table}

From the table it is apparent that, for $m_{b'} = 150\ \gev/c^2$, the sensitivity of the simple cuts is only negligibly less than the
target sensitivity.  We find the same to be true for all mass points studied, except for the $m_{b'} = 100\ \gev/c^2$ mass point.  In
that case, however, the sensitivity of the simple cuts is still adequate because of the larger cross sections for lower mass
particles \cite{bib:senslowmass}.  In addition, low masses near 100 $\gev/c^2$ are less interesting as they are already more tightly
excluded~\cite{bib:pdgbprime}.  Thus, we conclude that the simpler selection of $\njett \geq 3$ and $\jten > m_{b'} c^2$ is nearly
optimal for the mass range of interest.

In the above, \jt\ was calculated using a 10 GeV $E_T$ threshold on the jets.  For the purposes of the background estimation, it is
simpler to use the same $E_T$ threshold on \jt\ as one uses on the \njet\ variable.  Therefore, a 30 GeV threshold is used when
calculating \jt.  This was found to give a small decrease in sensitivity in the $b'$ model with the benefit of a gain in simplicity.

The kinematic jet selection was found to be optimal when using the fourth generation model as the signal.  When optimizing using the
figure of merit \srbfom, the optimal point is independent of the normalization of the signal.  That is, any model with a different
cross section but the same kinematic distributions will give the same optimal point.  In addition, the shape of the kinematic
distributions are mostly determined by the $b'$ mass.  We therefore expect that this selection is nearly optimal for all models with
heavy particles produced in pairs and decaying to \Z+jet.  In general, this selection is sensitive to any model with high $E_T$ jets
in the final state.  It may not be optimal for an arbitrary model, but designing a simple selection that is optimal for the entire
class of \Z+high $E_T$ jet models is not possible.

In this optimization, we assumed new signals would lead to final states consisting of a \Z\ boson and many high $E_T$ jets.  Of
course, some assumption about signal characteristics must be made in order to understand how to separate signal from background.
These assumptions will naturally reduce the model independence of the search.  There is a trade-off between the specificity of these
assumptions and the sensitivity to a particular model.  For example, in nearly all new physics models with \Z\ boson final states,
the transverse momentum spectrum of the \Z\ is harder than for standard model \Z\ production.  This is because, in these models, the
\Z\ is usually a decay product of a massive particle.  One would conclude that the \Z\ transverse momentum is a very
model-independent variable, and therefore well-motivated.  However, we find, in the $b'$ model sensitivity study, that the jet
kinematic requirements have much higher sensitivity than the \Z\ transverse momentum.  The cost of this sensitivity is a loss of
generality: with this assumption we are no longer sensitive to \Z\ final states without high $E_T$ jets.  The sensitivity of the $b'$
model can be further enhanced by requiring $b$ jets using displaced vertices (because of the $b' \rightarrow bZ$ decay), again with a
cost to generality.  In our analysis, as a compromise between model independence and sensitivity, we choose to require additional
jets in the event.

To summarize, after selecting \zee\ and \zmm\ events, the kinematic selection is:
\begin{itemize}
\item{$\njett \geq 3$}, and
\item{$\jtt > m_{b'} c^2$}.
\end{itemize}
That is, \Z\ events with $\njett \geq 3$ are selected, and the $\jtt$ distribution is scanned for an excess.  Step sizes of 50 GeV
are used.

\section{Backgrounds}
\label{sec:bkga}

In the signal region described above, there are potential backgrounds from the following sources:
\begin{itemize}
\item single-\Z\ production in conjunction with jets,
\item multi-jet events, where two jets fake leptons,
\item cosmic rays coincident with multi-jet events,
\item $WZ$+jets, where the $W$ decays to jets,
\item $ZZ$+jets, where one of the $Z$'s decays to jets,
\item $WW$+jets, where both $W$'s decay to leptons, and
\item $t\bar{t}$+jets, where both $W$'s decay to leptons.
\end{itemize}

The dominant background is from standard model single-\Z\ production in conjunction with jets.  Since beyond leading-log order
diagrams make potentially large contributions to events with $\njett \geq 3$, calculation of this background from theoretical first
principles is extremely difficult, and therefore would require careful validation with data.  Rather than using data as merely a
validation tool we take a different approach, and instead measure the background directly from data, and with data alone.  The
following section is devoted to describing this prediction technique for the dominant background from \Z+jet.  As this technique has
not been applied previously, it is explained thoroughly, with careful validation studies described.  The remaining backgrounds are
estimated in Sec.~\ref{sec:bkgb}.

\section{Data-Based \Z+jet Background Prediction Technique}
\label{sec:bkgmethod}

Given the above selection, there are two tasks: the total number of background events with $\njett \geq 3$ must be predicted, and the
shape of the \jtt\ distribution after this cut must be predicted.  When combined, these two components give the full normalized \jtt\
distribution prediction.  The background for events with $\njett \geq 3$ and any \jtt\ cut can be obtained from this distribution.
The method for predicting each of the two components is described separately in the following two sections.

In each of the prediction methods, fits to various jet $E_T$ distributions are used.  A parameterization that describes the shapes of
these jet $E_T$ distributions well is therefore required.  The parameterization used is:

\begin{equation}
f(E_T) = p_0 \frac{e^{-E_T/p_1}}{(E_T)^{p_2}} ,
\label{eqn:jetetparamfinal}
\end{equation}
where the $p_i$ are fitted parameters.  This parameterization was motivated by observations in Monte Carlo simulations, control
regions of data, and phenomenological studies that: at low $E_T$, the jet $E_T$ shape follows a power law function; at high $E_T$, it
follows an exponential decay function.  The above parameterization satisfies these limiting behaviors.  With the above convention,
the parameter $p_1$ has dimensions of energy, the parameter $p_2$ is dimensionless, and both parameters are positive.  Further
discussion and motivation for this parameterization is provided in \cite{bib:mythesis}.

\subsection{Number of Events with $\njett \geq 3$}
\label{subsec:njetnorm}

In order to predict the total number of events with $\njett \geq 3$, we use the jet $E_T$ distributions in the $\njett \leq 2$
control regions.  Since jets are counted above an $E_T$ threshold (in this case 30 GeV), the $\njet$ distribution is completely
determined from the jet $E_T$ distributions.  To illustrate this, and to describe the method, standard model \zmm\ Monte Carlo
simulations are used.  After validation with control samples, the method is applied to the \Z\ data.

In Fig.~\ref{fig:smzmm_blindsep_et2}, the $E_T$ distribution of the third highest jet is shown.  By construction, a cut on $\njett
\leq 2$ separates this distribution into two regions. This distribution can be fit in the $E_T < 30$~GeV region and extrapolated to
the $E_T > 30$~GeV region to get the expected number of background events with $\njett \geq 3$.

\begin{figure}
\begin{center}
\includegraphics[width=2.5in]{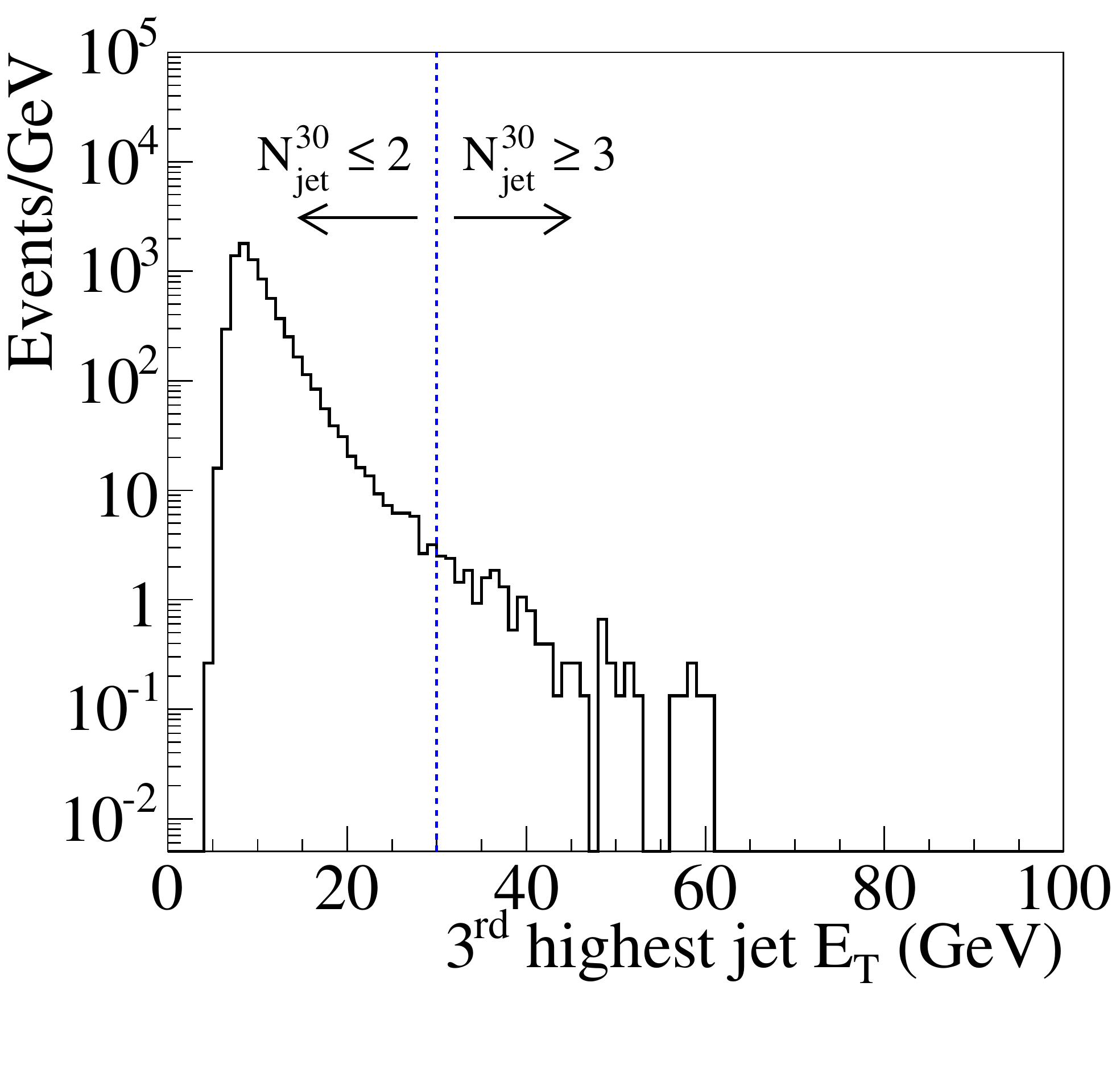}
\end{center}
\caption[]{
$E_T$ distribution of the third highest $E_T$ jet in standard model \zmm\ Monte Carlo simulations.  Events with $\njett \leq 2$ have $E_T
< 30\ \gev$; events with $\njett \geq 3$ have $E_T > 30\ \gev$.
}
\label{fig:smzmm_blindsep_et2}
\end{figure}

We fit the parameterization from Eq.~(\ref{eqn:jetetparamfinal}) to the jet $E_T$ distribution of Fig.~\ref{fig:smzmm_blindsep_et2},
and show the results in Fig.~\ref{fig:smzmm_bkgfit_et2} \cite{bib:fnlikemaxfit}.  The fit matches well the broad features of the
distribution above 30 GeV.  The number of events with $\njett \geq 3$ is then predicted by integrating the fitted distribution from
30 GeV to infinity.  The fit prediction obtained with this method (with its uncertainty from fit parameter error propagation
described in Sec.~\ref{subsec:fituncert}) is $116^{+10}_{-13}$ events (with the number of generated Monte Carlo events having an
equivalent luminosity of 7~\ifb).  The number of events observed in the simulated data with $\njett \geq 3$ is 152.  In this case,
the extrapolation predicts the background to within $31 \pm 16$\%.  The level of consistency will be evaluated further in the
validation studies with data in Sec.~\ref{subsec:fitval}.

\begin{figure}
\begin{center}
\includegraphics[width=2.5in]{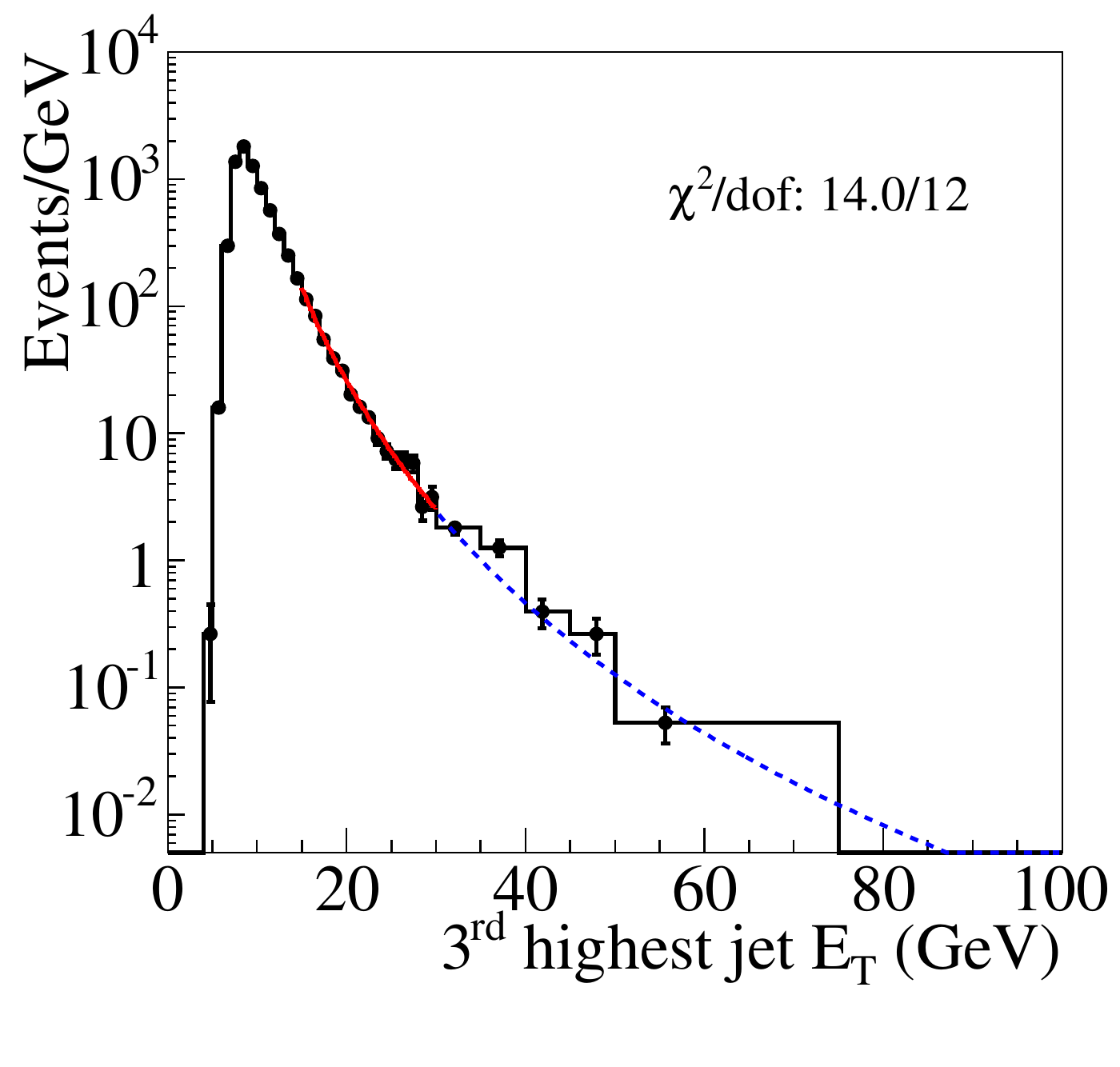}
\end{center}
\caption[]{
$E_T$ distribution of the third highest $E_T$ jet in standard model \zmm\ Monte Carlo events.  The distribution is fit to
Eq.~(\ref{eqn:jetetparamfinal}) in the range $15 < E_T < 30$~GeV, and extrapolated to the $E_T > 30$~GeV region.
}
\label{fig:smzmm_bkgfit_et2}
\end{figure}

This method, using the jet $E_T$ distributions to predict integrals of the \njet\ distribution, can clearly be extended to other
analyses as well.  For illustration purposes only we describe other examples here, still using standard model \zmm\ Monte Carlo
simulation.  Consider predicting the total number of events with $\njet^{80} \geq 1$ (that is, we require at least one jet with an
$E_T$ threshold of $80$~GeV).  In this case, a fit to the highest $E_T$ jet distribution below $80$~GeV can be extrapolated to above
that threshold, as in Fig.~\ref{fig:bkgfit_ordered_jetet0fit_comb_njet40_80_smz}.  (Note that the highest $E_T$ distribution in this
figure is harder than the third highest $E_T$ jet distribution, as one expects when ordering the jets by $E_T$).  It is clear that
the extrapolation describes the distribution reasonably well.  

If we instead wish to predict the number of events with $\njet^{40} \geq 1$, we must fit the same $E_T$ distribution below $40$~GeV
and extrapolate it to above that threshold, also shown in Fig.~\ref{fig:bkgfit_ordered_jetet0fit_comb_njet40_80_smz}.  It is clear
that the extrapolation does not describe the high $E_T$ portion of the distribution well.  There is a large systematic uncertainty
present in extrapolations that use such a small portion of the distribution that the shape can not be reliably  obtained.  This can
be mitigated by raising the $E_T$ threshold, unless the shape of the jet $E_T$ distribution at high $E_T$ can be otherwise
constrained.  In the case examined in this analysis, we fit the third highest $E_T$ jet (which has a softer $E_T$ distribution than
the highest $E_T$ jet) in the region $E_T < 30$ GeV.  We have checked that the data in this region constrains the shape sufficiently
with validation studies using control samples of data and Monte Carlo simulations, described later in Sec.~\ref{subsec:fitval}.

From the above, it is apparent that one can estimate the background for events with $\njet^{X} \geq n$ by fitting the $E_T$
distribution of the $n^{\rm th}$ highest $E_T$ jet in the region $E_T < X$ and extrapolating the fit to the region $E_T > X$, as long
as the fit region $E_T < X$ constrains the shape sufficiently.

\begin{figure}
\begin{center}
\includegraphics[width=2.5in]{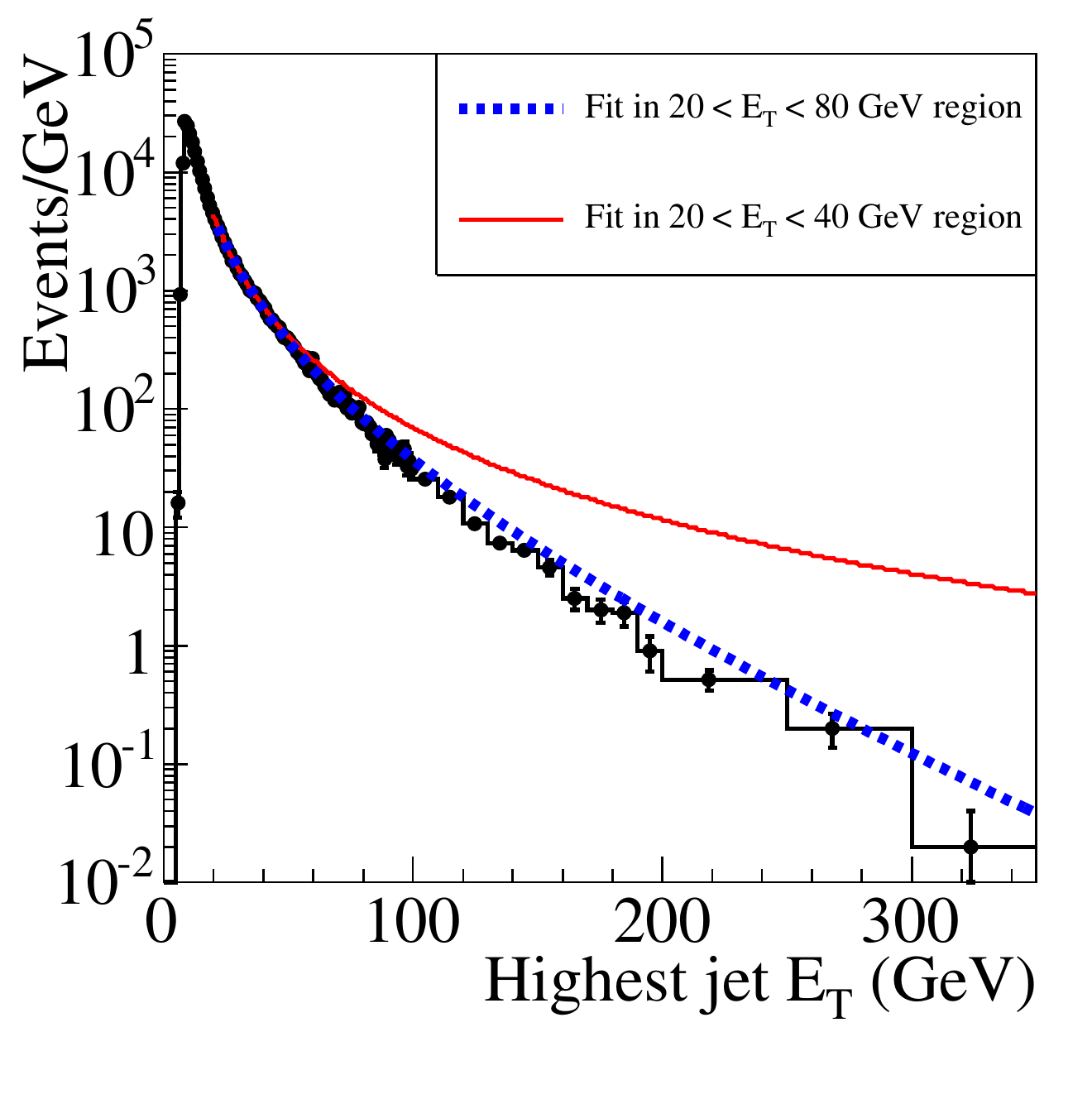}
\end{center}
\caption[]{
$E_T$ of the highest $E_T$ jet in standard model $Z\rightarrow\mu\mu$ Monte Carlo events.  The distribution is fit to
Eq.~(\ref{eqn:jetetparamfinal}) in the region $20<E_T<80$~GeV (dotted line), and again in the region $20<E_T<40$~GeV (solid line).
}
\label{fig:bkgfit_ordered_jetet0fit_comb_njet40_80_smz}
\end{figure}

\subsection{$J_T$ Shape Determination}
\label{subsec:jtshape}

We now describe the method used to determine the shape of the \jtt\ distribution of events with $\njett \geq 3$.  After finding the
shape, it is then normalized to the number of events with $\njett \geq 3$ found by the above method.  We again use standard model
\zmm\ Monte Carlo events to explain the method, and later will apply it to data.

Since \jtt\ is simply the sum of the individual jet transverse energies above 30 GeV, if the $E_T$ distributions of jets for events
with $\njett \geq 3$ are known, the \jtt\ distribution can be predicted for these events.  We extrapolate the shape of these jet
$E_T$ distributions from the jet $E_T$ distributions of $\njett \leq 2$ events.  In order to do such an extrapolation, we must
understand the variation of the jet $E_T$ distribution as a function of $\njett$.

The $E_T$ distributions of all jets in events with $\njett = $ 1 and 2, normalized to have equal area, is shown in
Fig.~\ref{fig:prd_datacomb_jetetcomp} using \zll\ data.  The general shape is similar, though jets in $\njett = 2$ events have a
slightly harder tail at high $E_T$.  We model this by fitting to each jet $E_T$ distribution (using Eq.~(\ref{eqn:jetetparamfinal}))
and extrapolating the fit parameters to $\njett \geq 3$ events.  To avoid simultaneously extrapolating two fit parameters we only
extrapolate the exponential parameter ($p_1$), as this parameter governs the high $E_T$ behavior in our parameterization.  In order
to extrapolate only this parameter, we fit the $\njett = 1$ $E_T$ spectrum allowing both parameters to float freely, then fix the
power law parameter ($p_2$) in the fit to the $\njett = 2$ $E_T$ spectrum.  We then extrapolate the $p_1$ parameter of
Eq.~(\ref{eqn:jetetparamfinal}) linearly as a function of \njett, from their fitted values at $\njett = 1$ and $\njett = 2$ into the
region $\njett \geq 3$.

Figures~\ref{fig:prd_smzmm_bkgfit_njet30_1} and \ref{fig:prd_smzmm_bkgfit_njet30_2} show the fits of the spectra for events with 1
and 2 jets.  Figure~\ref{fig:prd_smzmm_bkgfit_exp_vs_njet} shows the linear extrapolation of the exponential parameters.  For
illustration, the exponential parameter obtained from a fit to the $E_T$ distribution in $\njett = 3$ events (again fixing the power
law parameter to that found in the $\njett = 1$ events) is shown on the same figure.  The extrapolation reasonably predicts the
parameter for events with $\njett = 3$ \cite{bib:fnnofitnjetfour}.

\begin{figure}
\begin{center}
\includegraphics[width=2.5in]{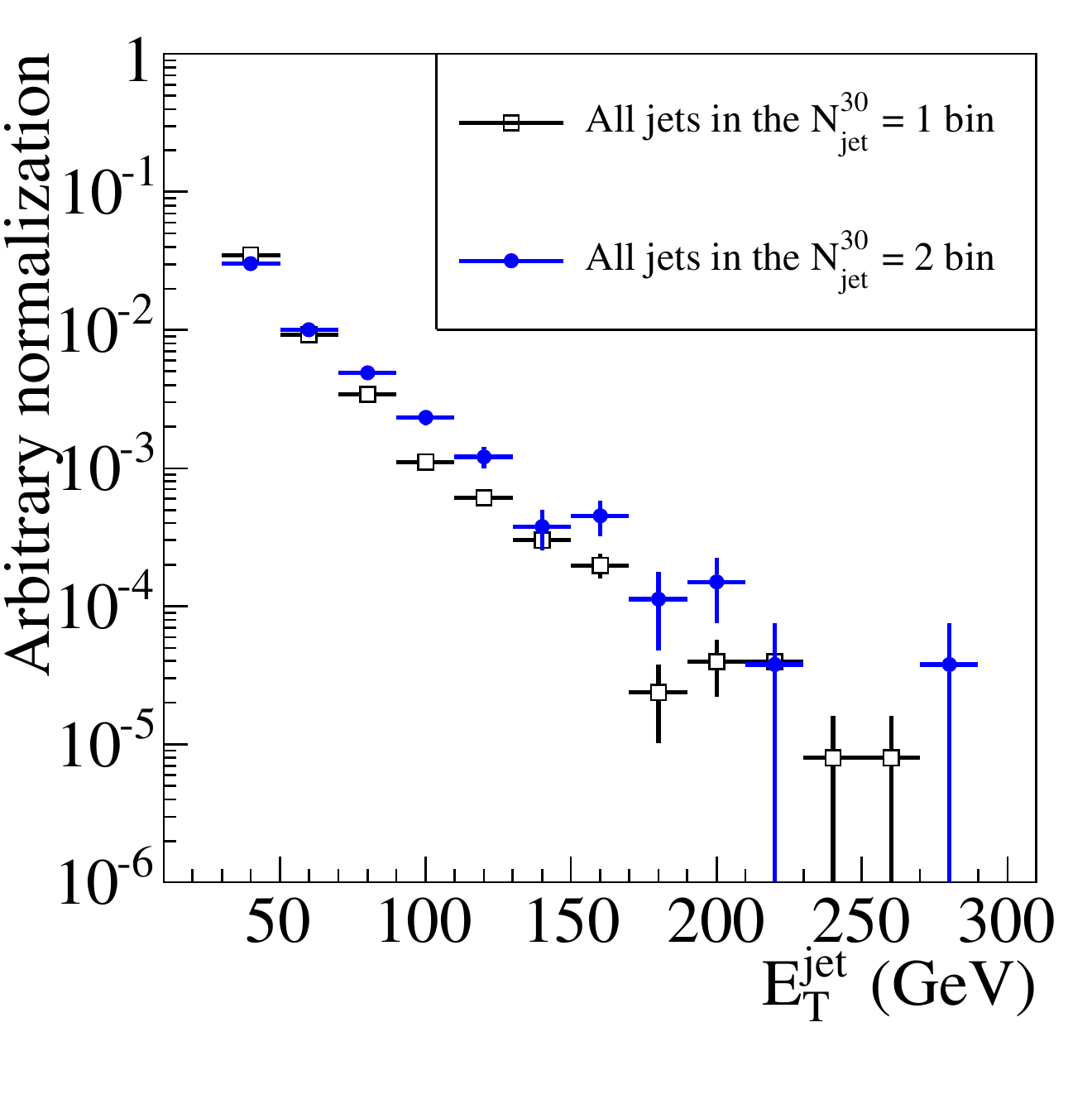}
\end{center}
\caption[]{
$E_T$ distribution of jets in $\njett = 1$ events (open squares) and in $\njett = 2$ events (solid circles) in \zll\
data.  Events with higher $\njett$ have harder $E_T$ spectra.
}
\label{fig:prd_datacomb_jetetcomp}
\end{figure}

\begin{figure}
\begin{center}
\includegraphics[width=2.5in]{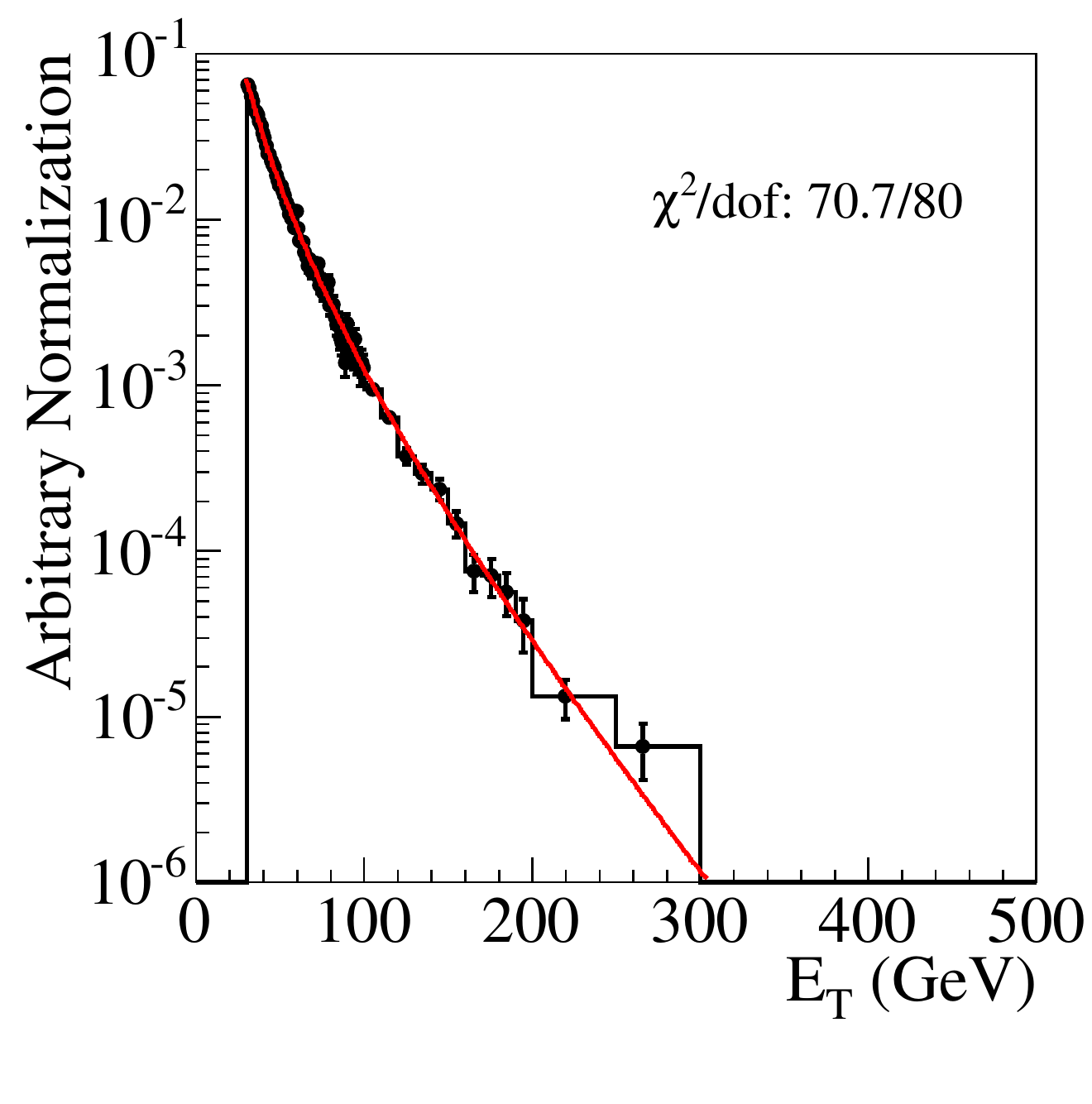}
\end{center}
\caption[]{
$E_T$ distribution of jets in $\njett = 1$ events in standard model \zmm\ Monte Carlo events.  The distribution is fit to
Eq.~(\ref{eqn:jetetparamfinal}) in the range $E_T > 30$.
}
\label{fig:prd_smzmm_bkgfit_njet30_1}
\end{figure}

\begin{figure}
\begin{center}
\includegraphics[width=2.5in]{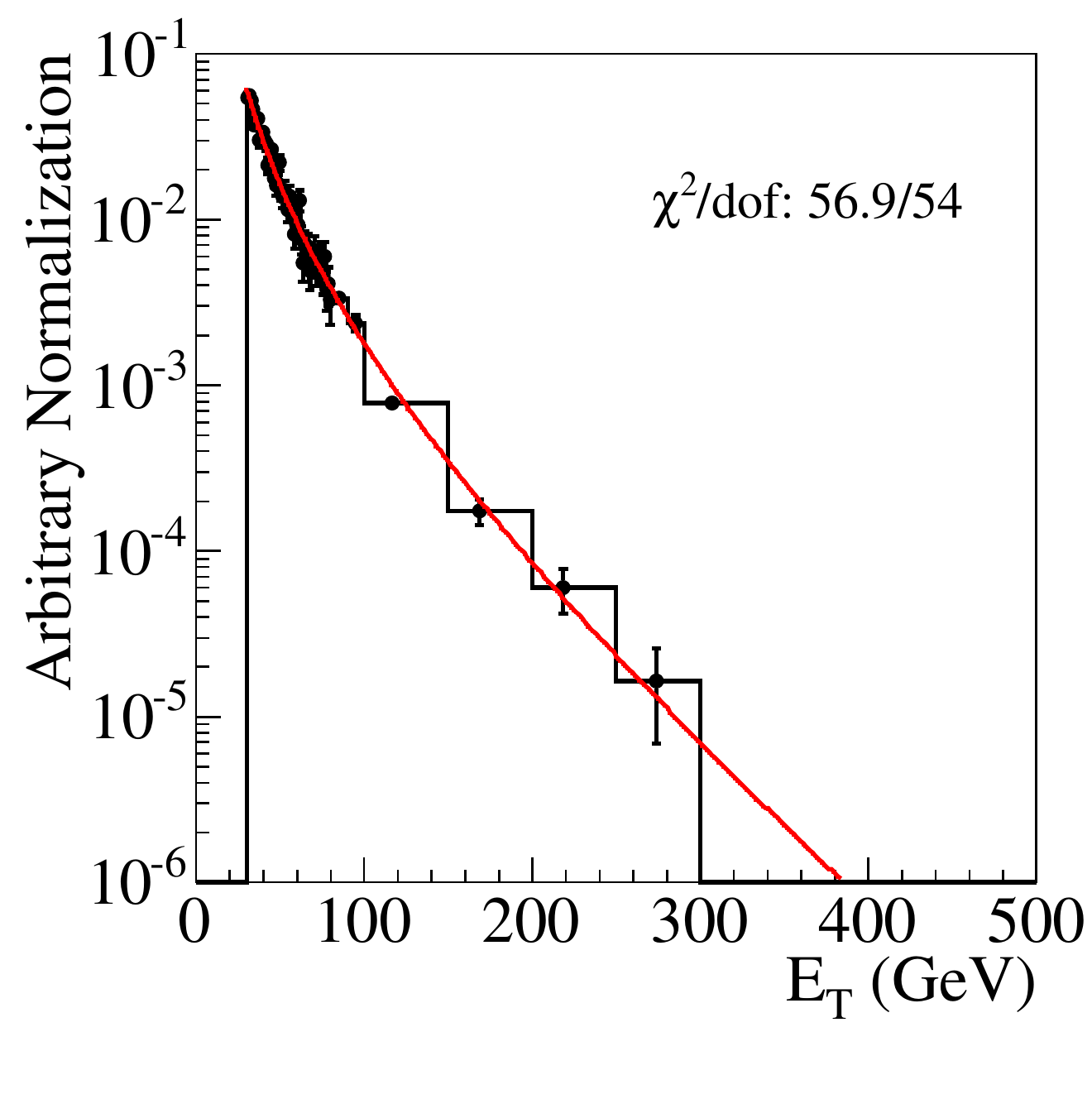}
\end{center}
\caption[]{
$E_T$ distribution of jets in $\njett = 2$ events in standard model \zmm\ Monte Carlo events.  The distribution is fit to
Eq.~(\ref{eqn:jetetparamfinal}) in the range $E_T > 30$, with the parameter $p_2$ fixed to that obtained from Fig.
\ref{fig:prd_smzmm_bkgfit_njet30_1}.
}
\label{fig:prd_smzmm_bkgfit_njet30_2}
\end{figure}

\begin{figure}
\begin{center}
\includegraphics[width=2.5in]{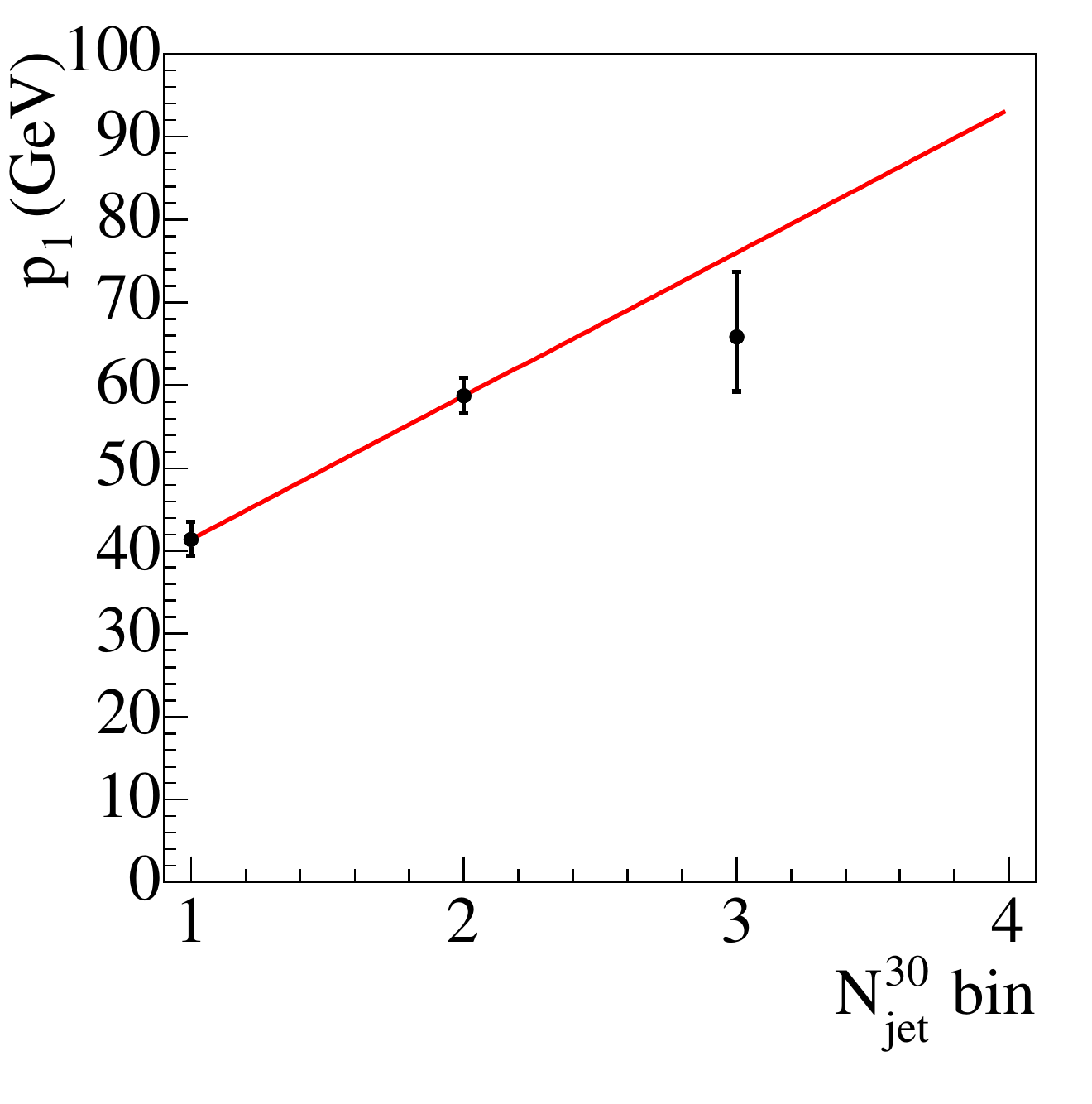}
\end{center}
\caption[]{
The extrapolation of the exponential parameter $p_1$ vs. $\njett$ in standard model \zmm\ Monte Carlo events.
}
\label{fig:prd_smzmm_bkgfit_exp_vs_njet}
\end{figure}

This dependence of the jet $E_T$ spectra on $\njett$ is modeled as described by our parameter extrapolation, allowing us to predict
the shapes of the jet $E_T$ spectra for events with $\njett \geq 3$.  The \jtt\ distribution is now almost completely determined.
Only an estimate for the relative fractions of events with 3, 4, 5, ...  jets is needed.  For this, we use an exponential fit
parameterization, fit to the $\njett$ distribution in the region $\njett \leq 2$, and use this shape in the $\njett \geq 3$ region.
This fit is shown in Fig.~\ref{fig:prd_smzmm_bkgfit_njet30_shape}.  There is no theoretical motivation for an exponential shape; we
merely use it as an estimate, and verify that the \jtt\ prediction does not strongly depend on the chosen parameterization.  As the
total number of events with $\njett \geq 3$ is already constrained using the method from Sec.~\ref{subsec:njetnorm}, the dependence
of the $\jtt$ distribution on the exponential parameterization of the $\njett$ distribution is small.

\begin{figure}
\begin{center}
\includegraphics[width=2.5in]{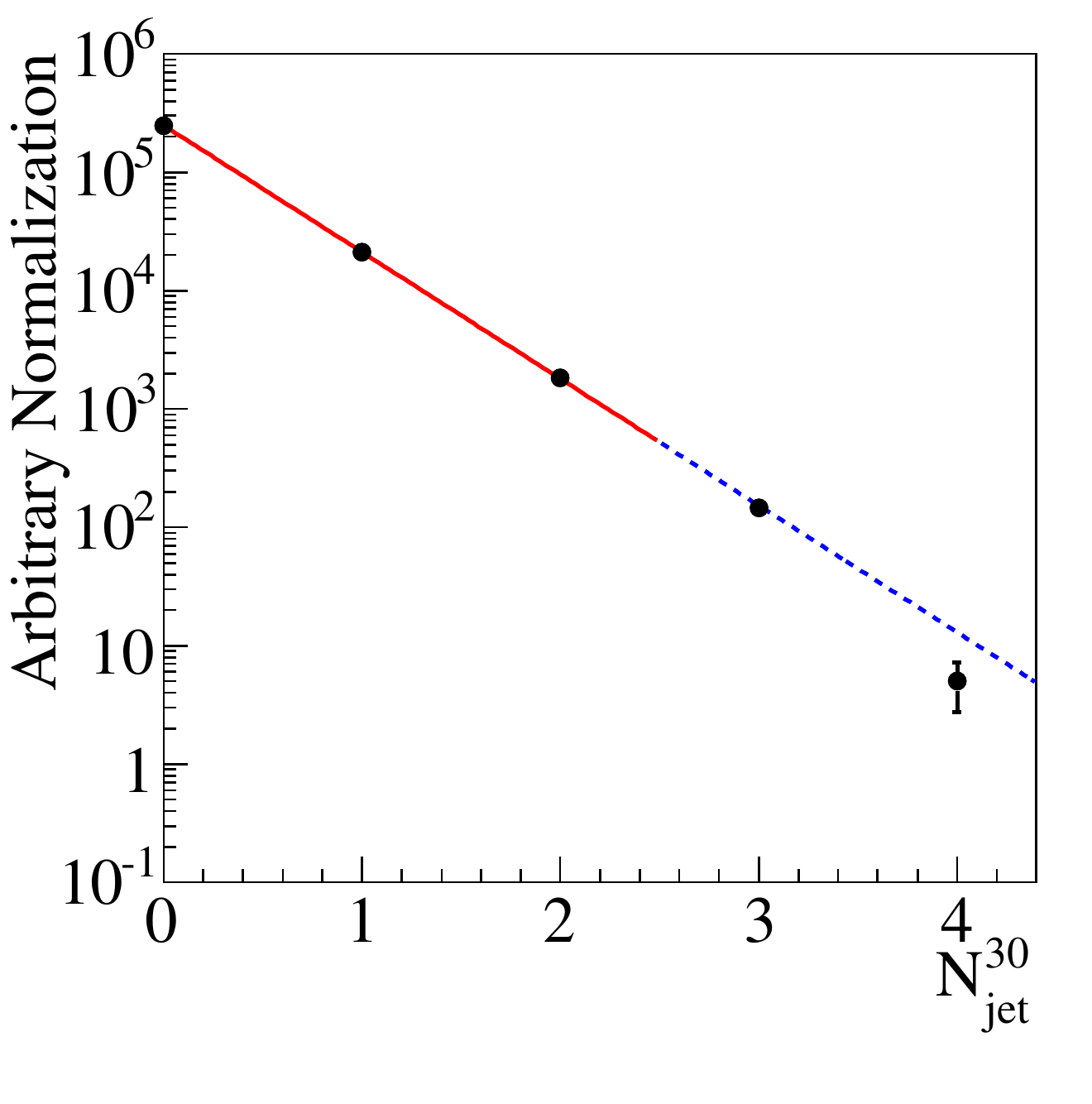}
\end{center}
\caption[]{
$\njett$ distribution in standard model \zmm\ Monte Carlo events, fit to an exponential in the range $\njett \leq 2$.  This shape is
used to estimate the relative fractions of events with 3, 4, 5, ... jets.
}
\label{fig:prd_smzmm_bkgfit_njet30_shape}
\end{figure}

Finally, given the above shapes, it is straightforward to make a simple Monte Carlo program that samples these shapes to get the
\jtt\ distribution.  The steps required to make this \jtt\ prediction are:
\begin{enumerate}
\item{\label{lista1} For each event, generate the number of jets by randomly sampling the predicted $\njett$ distribution in the
range $\{3,4,5,...\}$.}
\item{\label{lista2} Take the appropriate jet $E_T$ distribution for this number of jets after extrapolating the exponential fit
parameter.  Independently sample this jet $E_T$ distribution for each jet.}
\item{\label{lista3} Sum these jets to obtain the $\jtt$.}
\end{enumerate}
The process is repeated as necessary until the \jtt\ shape is obtained to the desired level of statistical precision.  

On step \ref{lista2}, the jet $E_T$ shapes are independently sampled; however, there is potentially some correlation between the
individual jet energies.  Including this correlation in the $\jtt$ shape prediction would have the effect of making the tail at large
values of $\jtt$ slightly harder.  In the validation studies in Sec.~\ref{subsec:fitval} we verify that the correlation is below the
level necessary to affect the fit prediction.  To understand this further, in Fig.~\ref{fig:prd_zdata_njet2_et_vs_et}, we plot the
$E_T$ of one the jets versus the other in events with $\njett = 2$ in the \zll\ data.  There is no correlation evident in the plot;
in the $663$ events with $\njett = 2$, only a small correlation of 25\% is found, indicating that independently sampling the $E_T$
distribution is a reasonable approximation.

\begin{figure}
\begin{center}
\includegraphics[width=2.5in]{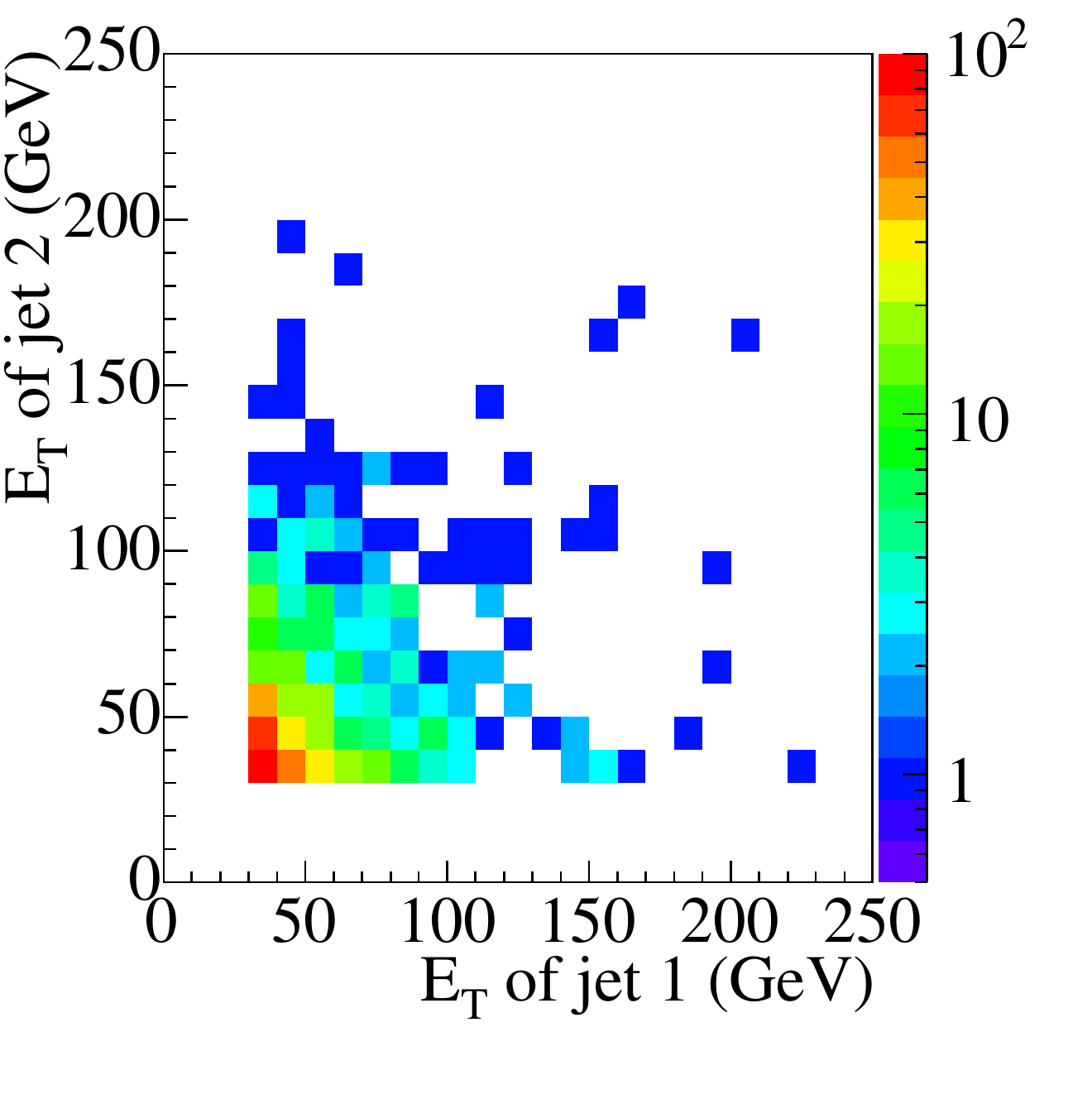}
\end{center}
\caption[]{
The $E_T$ of a random jet vs. the $E_T$ of the other, using jets with $\njett = 2$ in \zll\ data.
}
\label{fig:prd_zdata_njet2_et_vs_et}
\end{figure}

\subsection{Uncertainties on Fit Prediction}
\label{subsec:fituncert}

There are two sources of uncertainty on the mean background prediction: the statistical uncertainty from the finite amount of data in
the fits, and the systematic uncertainty from imperfect modeling of the various shapes in the fits.

\subsubsection{Statistical Uncertainty on Fit Prediction}
\label{subsubsec:statuncert}

The third highest $E_T$ jet normalization fit predicts the total number of events with $\njett \geq 3$, using the parameter values at
the minimum $-\log L$, where $L$ is the likelihood (or equivalently, the maximum likelihood).  The $1 \sigma$ uncertainty on the
number of events is simply obtained from its values at the minimum $-\log L + \frac{1}{2}$.  Since the total number of events with
$\njett \geq 3$ is given by a single fit, its uncertainty is easily determined with this method.

The $\jtt$ prediction is obtained by extrapolating the behavior of multiple distributions, and to estimate its shape uncertainty we
vary each fit parameter independently within its uncertainty (output by the fit) and re-do the extrapolation procedure.  The
individual uncertainties are combined in quadrature to obtain the total uncertainty.  The normalization error is then added in
quadrature as well to obtain the uncertainty on the fully-normalized $\jtt$ distribution.

\subsubsection{Systematic Uncertainty on Fit Prediction}
\label{subsubsec:systuncert}

As the background from \Z+jet events is determined from a fit to the data, the only source of systematic uncertainties is
mis-parameterization of those data.  If the data were poorly parameterized, fitting a subset of the data would give a large change in
the background prediction.  We therefore estimate the size of the mis-parameterization uncertainties by changing the range of each
fit and re-doing the fit procedure to obtain the $\jtt$ normalization and shape prediction.  Both uncertainties, that on the total
number of events with $\njett \geq 3$ (from the third highest $E_T$ jet fit), and that on the $\jtt$ shape, are estimated in this
way.  The variations from each fit range change are then added in quadrature to obtain the full uncertainty.  The fit range changes
are summarized in Table~\ref{tab:fitranges}.  The ``$\pm 1 \sigma$'' range changes are chosen to give sufficient coverage when
observed in control samples of data.

\begin{table}[htb]
\begin{center}
\begin{tabular}{cccc}
\hline \hline
Distribution            & nominal range     & ``$-1\sigma$'' range & ``$+1\sigma$'' range \\ 
\hline
Third highest $E_T$ jet   & $(15, 30)\ \gev$        & $(15, 26)\ \gev$           & $(17, 30)\ \gev$       \\
$\njett = 1$ jet $E_T$  & $(30, \infty)\ \gev$    & $(30, 150)\ \gev$          & $(70, \infty)\ \gev$       \\
$\njett = 2$ jet $E_T$  & $(30, \infty)\ \gev$    & $(30, 80)\ \gev$           & $(50, \infty)\ \gev$       \\
$\njett$ shape          & $[0, 2]\ \mathrm{jets}$          & $[0, 1]\ \mathrm{jets}$             & $[1, 2]\ \mathrm{jets}$             \\
\hline \hline
\end{tabular}
\end{center}
\caption[]{ 
Nominal fit ranges and the fit range changes used to estimate systematic uncertainties.  The nominal fit range of each distribution
is shown in the second column.  The third and fourth columns show the ranges used to estimate the uncertainty from a
mis-parameterization of that distribution.
}
\label{tab:fitranges}
\end{table}

Finally, using the technique and the uncertainties developed above in the Monte Carlo simulation, we can demonstrate that the method
is self-consistent by checking the normalized $\jtt$ prediction for events with $\njett \geq 3$ matches that observed in Monte Carlo
events.  This comparison is shown in Fig. \ref{fig:prd_smzmm_bkgfit_jt_unblind}.  The observed distribution agrees well with the
prediction.

\begin{figure}
\begin{center}
\includegraphics[width=2.5in]{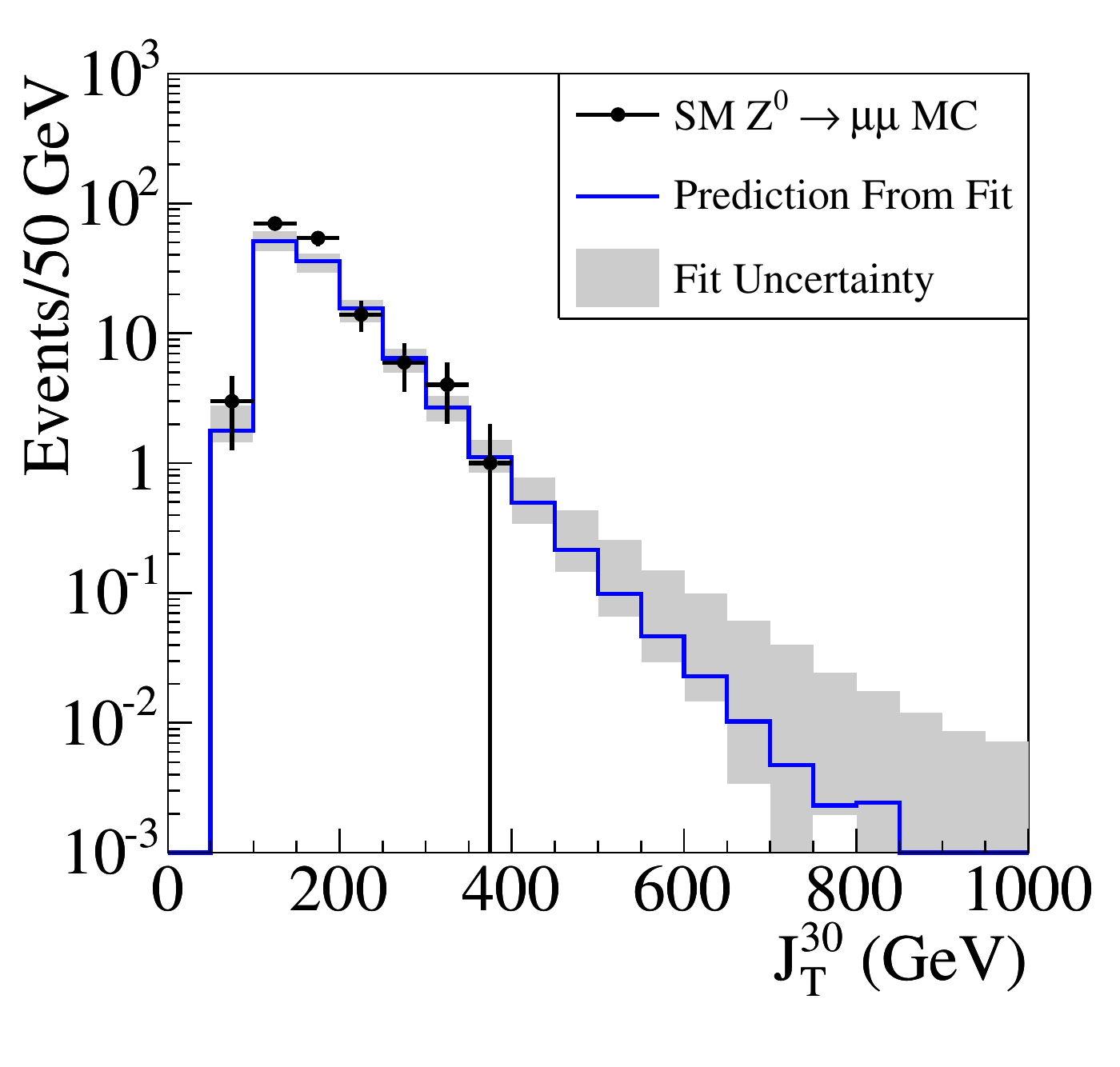}
\end{center}
\caption[]{
The prediction for the \jtt\ distribution (blue line) of standard model \Z\ Monte Carlo and its uncertainty (gray band), compared to
the actual distribution (black points with errors).
}
\label{fig:prd_smzmm_bkgfit_jt_unblind}
\end{figure}

\subsection{Validation of Technique}
\label{subsec:fitval}

Having demonstrated and described the procedure for obtaining the \Z+jet background using Monte Carlo simulation, its validation,
done predominantly in data, is now described.  The \Z+jet data cannot be used as a validation sample because of potential signal
bias, so we must test on other data samples.  We use two sets of multi-jet data as background-only validation samples, and $W$+jet
data as a background sample containing a real heavy quark signal from \ttbar\ production.  Finally, we do signal-injection studies
with Monte Carlo simulations to understand the effect of signal bias on the fit procedure.

\subsubsection{Multi-Jet Data}
\label{subsubsec:fitqcdval}

The \Z+jet background extrapolation only requires information about the jet $E_T$ distributions, and not the \Z.  It should therefore
perform similarly well not only for \Z+jet events, but ``$X$''+jet events, provided that the ``$X$'' has a similar transverse
momentum spectrum to the \Z.  For example, if the ``$X$'' has a minimum $p_T$ threshold, the $E_T$ distributions of the jets will be
sculpted such that they no longer follow the power law $\times$ exponential parameterization of Eq.~(\ref{eqn:jetetparamfinal}).

We first obtain ``$X$''+jet events from multi-jet data dominated by QCD interactions using prescaled jet triggers that require at
least one jet with $E_T > 20$~GeV \cite{bib:fnprescale}.  An ``$X$'' is then constructed by picking two random jets in the event,
requiring they both have $E_T > 20$~GeV (to match the electron and muon $p_T$ cuts), and requiring $M_{X} > 70\ \gev/c^2$ to remove
the invariant mass turn-on.  The invariant mass is not further restricted to the region $81 < M_X < 101\ \gev/c^2$ to maximize
statistics; in any case the \jtt\ distribution is observed to not depend on $M_X$ in this sample.

\newcommand{\jettrigbkgfitpred}{$97 \pm 27$}
\newcommand{\jettrigobserve}{$80$}
\newcommand{\jettrigprob}{$0.73$}
\newcommand{\jettrigsigma}{$0.6 \sigma$}
\newcommand{\jettrigjtcutbkgfitpred}{$19.7^{+9.2}_{-9.0}$}
\newcommand{\jettrigjtcutobserve}{$20$}
\newcommand{\jettrigjtcutnsigma}{$0.004 \sigma$}

Given this ``$X$'' selection, the remaining jets in the event are used to validate the procedure.
Figure~\ref{fig:prd_jet20data_bkgfit_et2} shows the third highest $E_T$ jet distribution.  We extrapolate this distribution above 30
GeV using Eq.~(\ref{eqn:jetetparamfinal}).  A prediction of \jettrigbkgfitpred\ (statistical uncertainty only) events with $\njett
\geq 3$ is obtained.  \jettrigobserve\ events are observed.  This is consistent within the uncertainties.  To quantitatively evaluate
the level of consistency we calculate the probability to measure the observed number of events or higher given the background
prediction, as well as convert this probability to units of standard deviations \cite{bib:fnprobcalc}.  This calculation gives a
corresponding probability of \jettrigprob; this is a \jettrigsigma\ level of consistency.

\begin{figure}
\begin{center}
\includegraphics[width=2.5in]{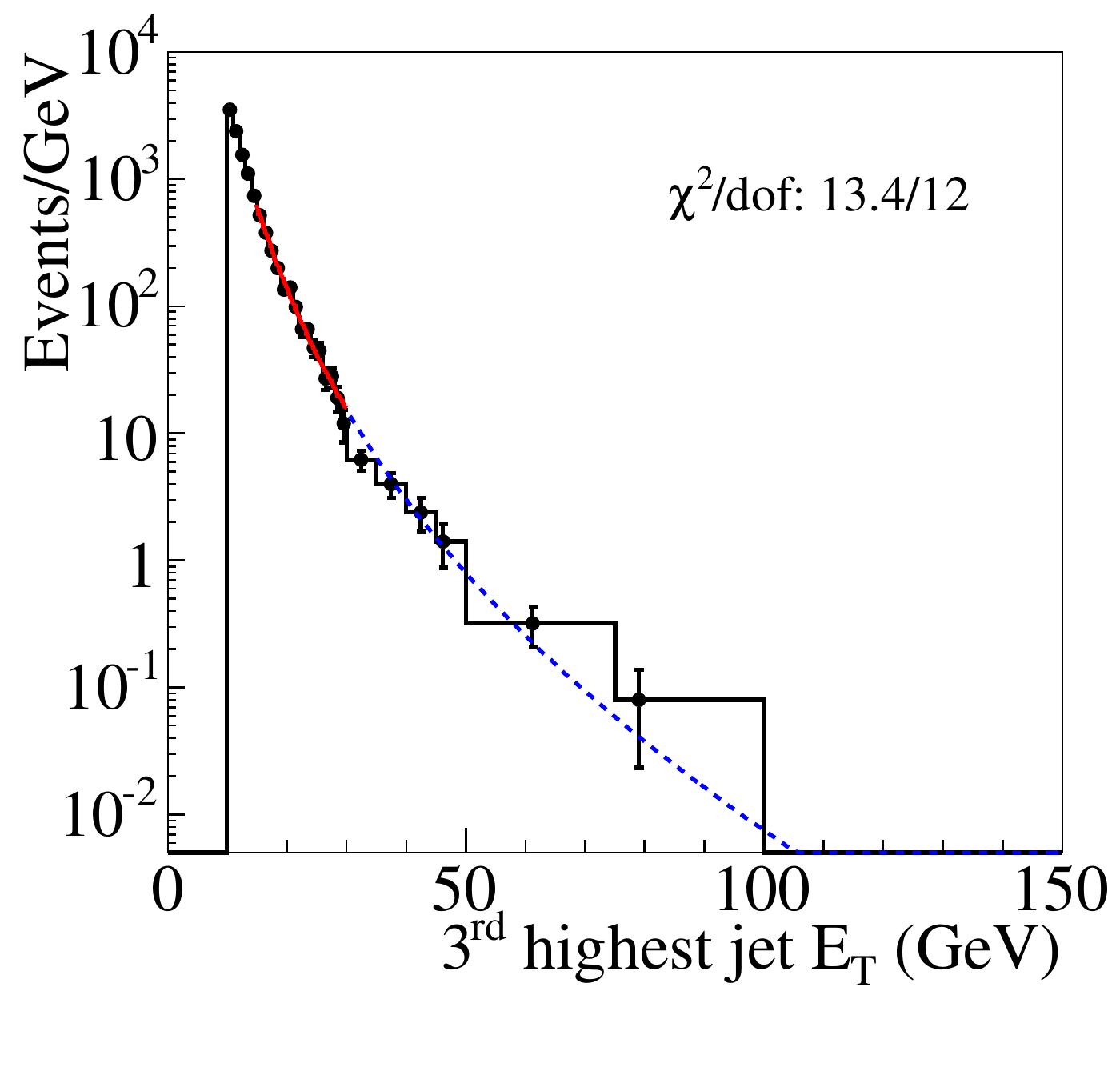}
\end{center}
\caption[]{
$E_T$ distribution of the third highest $E_T$ jet in ``$X$''+jet events selected with the jet triggers as described in the text.  The
distribution is fit to Eq.~(\ref{eqn:jetetparamfinal}) in the $15 < E_T < 30$~GeV region and extrapolated to the $E_T > 30$~GeV
region.
}
\label{fig:prd_jet20data_bkgfit_et2}
\end{figure}

We now predict the $\jtt$ shape.  Figures~\ref{fig:prd_jet20data_bkgfit_njet30_1} and \ref{fig:prd_jet20data_bkgfit_njet30_2} show
the fits to the jet $E_T$ spectra for events with $\njett = $ 1 and 2.  We extrapolate the parameter $p_1$ using the plot in
Fig.~\ref{fig:prd_jet20data_bkgfit_exp_vs_njet} to events with $\njett \geq 3$.  The $\njett$ shape is taken from the fit in
Fig.~\ref{fig:prd_jet20data_bkgfit_njet30_shape}.  Using these ingredients, the simple Monte Carlo program is used to obtain the
$\jtt$ shape, which is normalized to the prediction of $97$ events with $\njett \geq 3$.  The prediction and total uncertainty is
shown overlaid with the actual distribution in ``$X$''+jet data in Fig.~\ref{fig:prd_jet20data_bkgfit_jt_unblind}.  The distribution
clearly agrees well within the uncertainty envelope.  

Because the $\jtt$ uncertainties in each bin are correlated, an independent data/background comparison in each bin is not
straightforward.  Rather, we test the shape agreement once using the (arbitrarily chosen) region of $\jtt > 200$~GeV.  Above 200 GeV,
\jettrigjtcutbkgfitpred\ events are expected and \jettrigjtcutobserve\ events are observed.

\begin{figure}
\begin{center}
\includegraphics[width=2.5in]{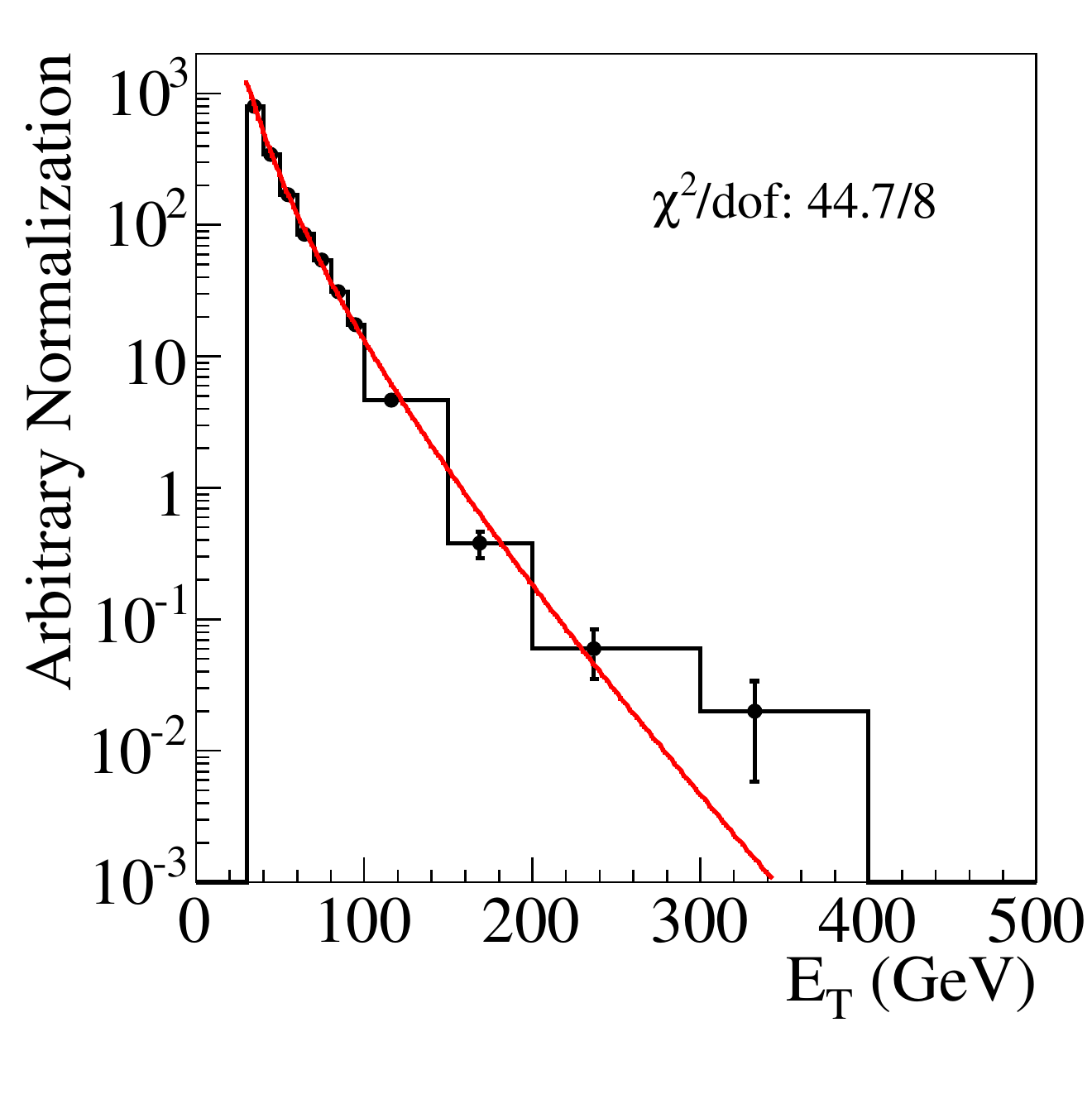}
\end{center}
\caption[]{
$E_T$ distribution of jets in $\njett = 1$ ``$X$''+jet events, selected with the jet triggers as described in the text.  The
distribution is fit to Eq.~(\ref{eqn:jetetparamfinal}) in the $E_T > 30$~GeV region.
}
\label{fig:prd_jet20data_bkgfit_njet30_1}
\end{figure}

\begin{figure}
\begin{center}
\includegraphics[width=2.5in]{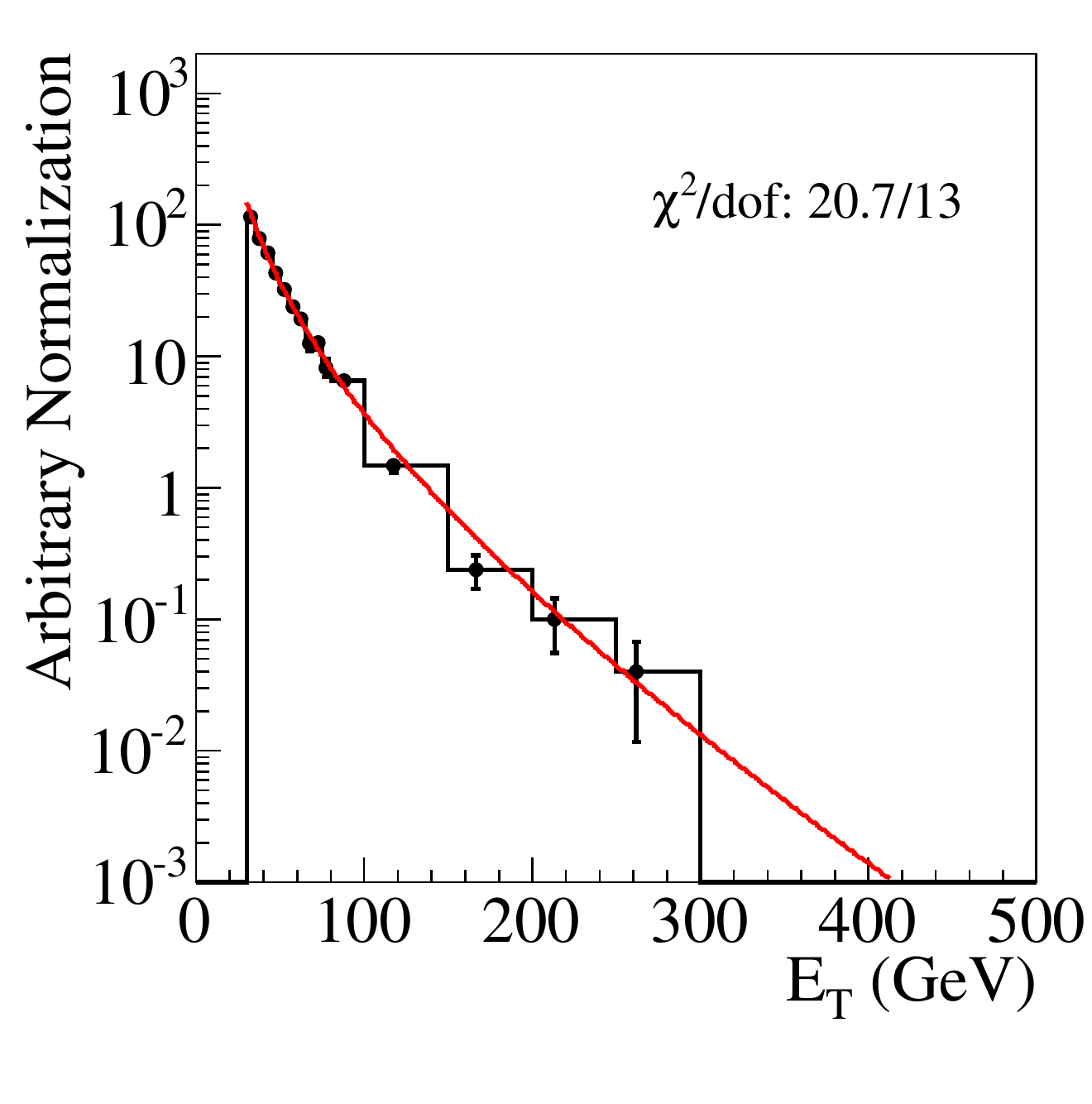}
\end{center}
\caption[]{
$E_T$ distribution of jets in $\njett = 2$ ``$X$''+jet events selected with the jet triggers as described in the text.  The
distribution is fit to Eq.~(\ref{eqn:jetetparamfinal}) in the $E_T > 30$~GeV region with the parameter $p_2$ fixed to that obtained
from the fit in Fig.~\ref{fig:prd_jet20data_bkgfit_njet30_1}.
}
\label{fig:prd_jet20data_bkgfit_njet30_2}
\end{figure}

\begin{figure}
\begin{center}
\includegraphics[width=2.5in]{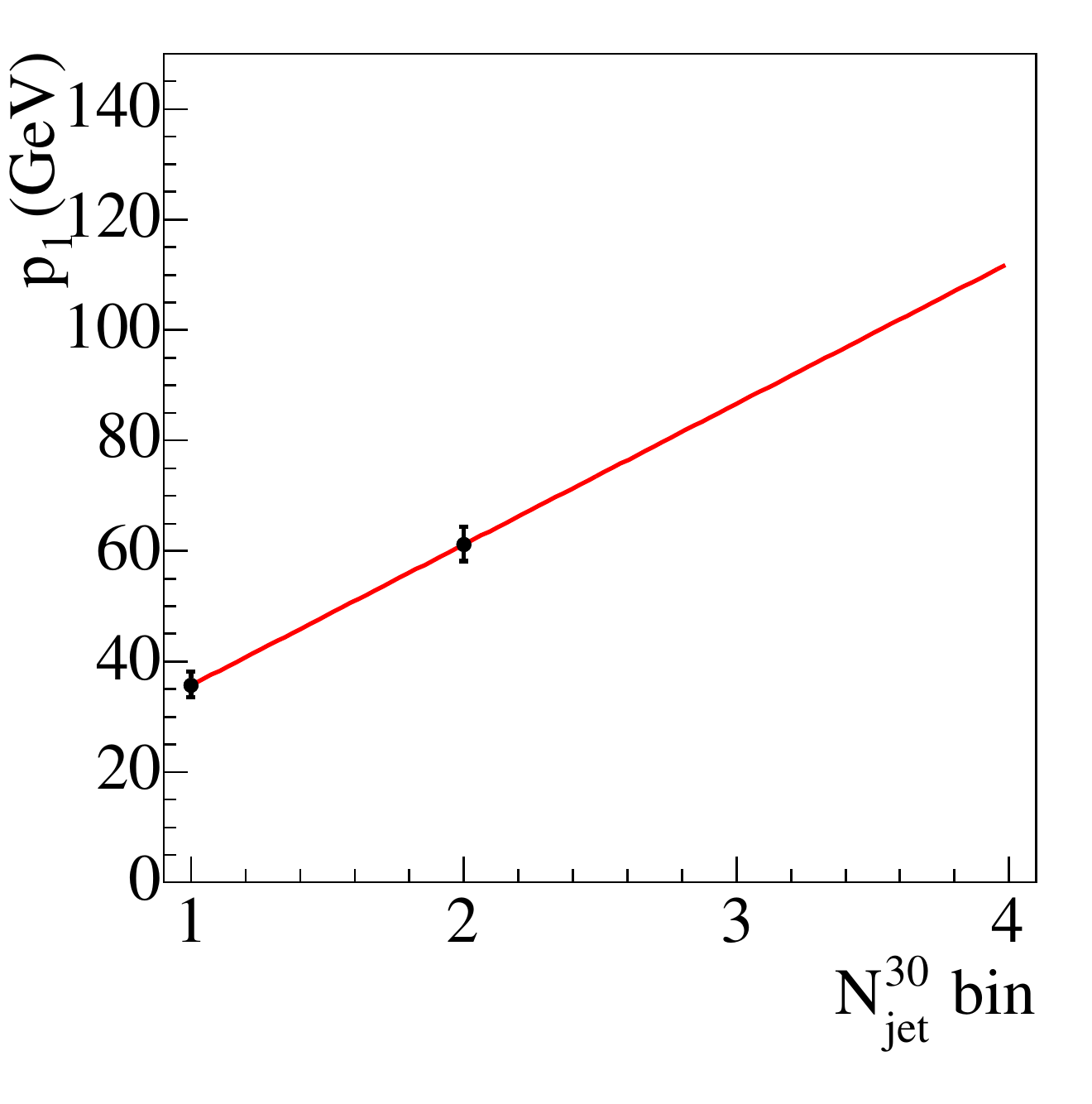}
\end{center}
\caption[]{
The extrapolation of the exponential parameter $p_1$ vs. $\njett$ in ``$X$''+jet events selected with the jet triggers as described
in the text.
}
\label{fig:prd_jet20data_bkgfit_exp_vs_njet}
\end{figure}

\begin{figure}
\begin{center}
\includegraphics[width=2.5in]{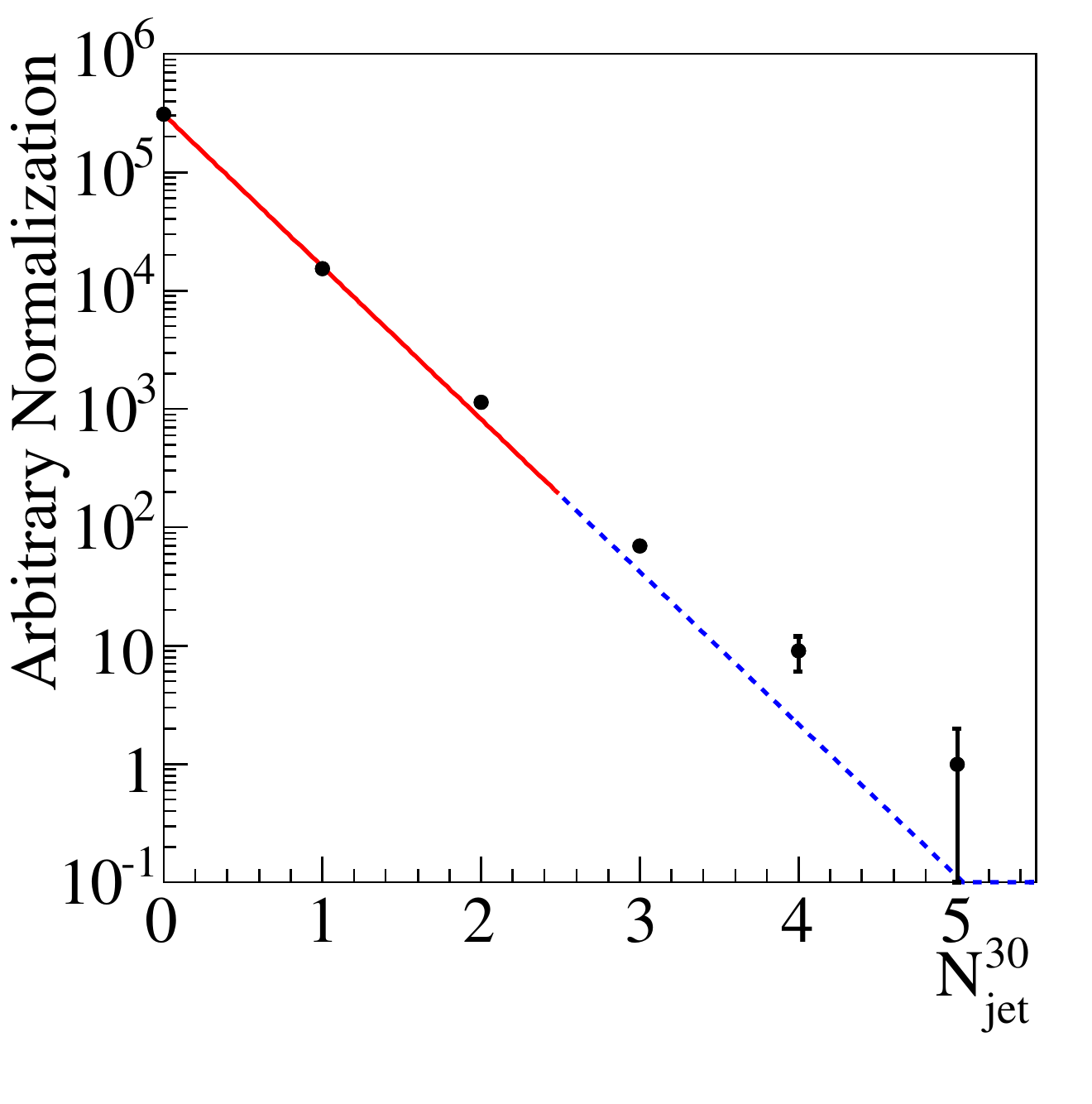}
\end{center}
\caption[]{
$\njett$ distribution in ``$X$''+jet events selected with the jet triggers as described in the text.  The distribution is fit to an
exponential in the range $\njett \leq 2$.
}
\label{fig:prd_jet20data_bkgfit_njet30_shape}
\end{figure}

\begin{figure}
\begin{center}
\includegraphics[width=2.5in]{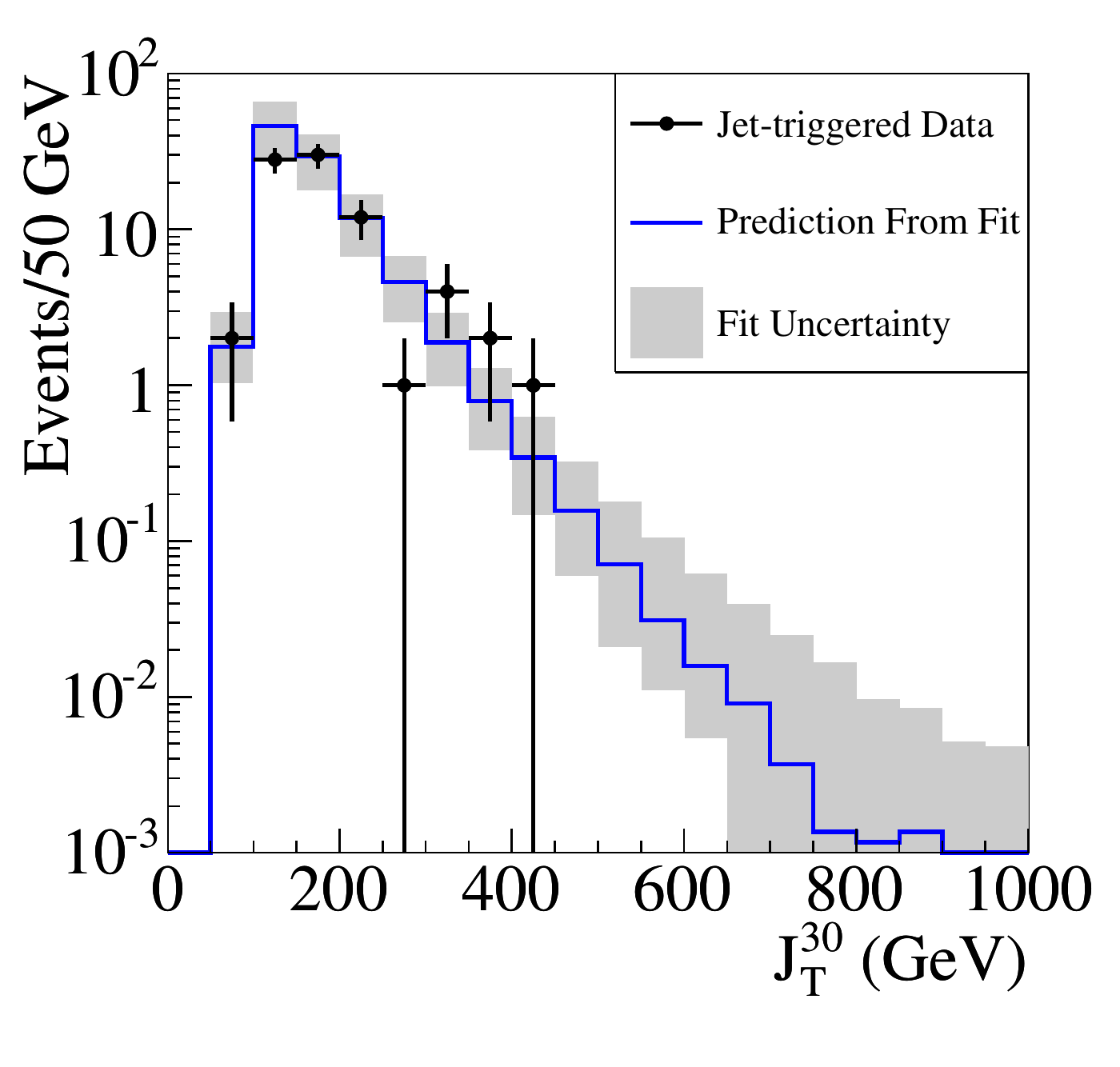}
\end{center}
\caption[]{
The prediction (blue line) and uncertainty (gray band) for the \jtt\ distribution of ``$X$''+jet events selected with the jet
triggers as described in the text.  The prediction is compared to the actual distribution (black points with errors).  The
observation agrees with the prediction.
}
\label{fig:prd_jet20data_bkgfit_jt_unblind}
\end{figure}

The background extrapolation method can accurately predict the normalization and shape of the $\jtt$ distribution in the jet
triggered sample.  However, because of the prescale, this sample has relatively low statistics despite the large cross section of QCD
multi-jet processes.  To obtain a higher statistics sample of multi-jet data, we can use the electron triggers, which are not
prescaled.  In this sample we construct an ``$X$'' by pairing the triggered electron with a ``fake'' electron, which is an EM
calorimeter cluster that is reconstructed as an electron but fails the low hadronic energy requirement.  ``$X$'' events selected in
this way are dominated by QCD dijet events.  Again, $M_{X} > 70\ \gev/c^2$ is required to remove the invariant mass turn-on.
Additionally the invariant mass region $81 < M_{X} < 101\ \gev/c^2$ is vetoed to remove real \zee\ events.
Figure~\ref{fig:prd_inverte_mee_nm1} shows the plot of the invariant mass before these requirements.

\begin{figure}
\begin{center}
\includegraphics[width=2.5in]{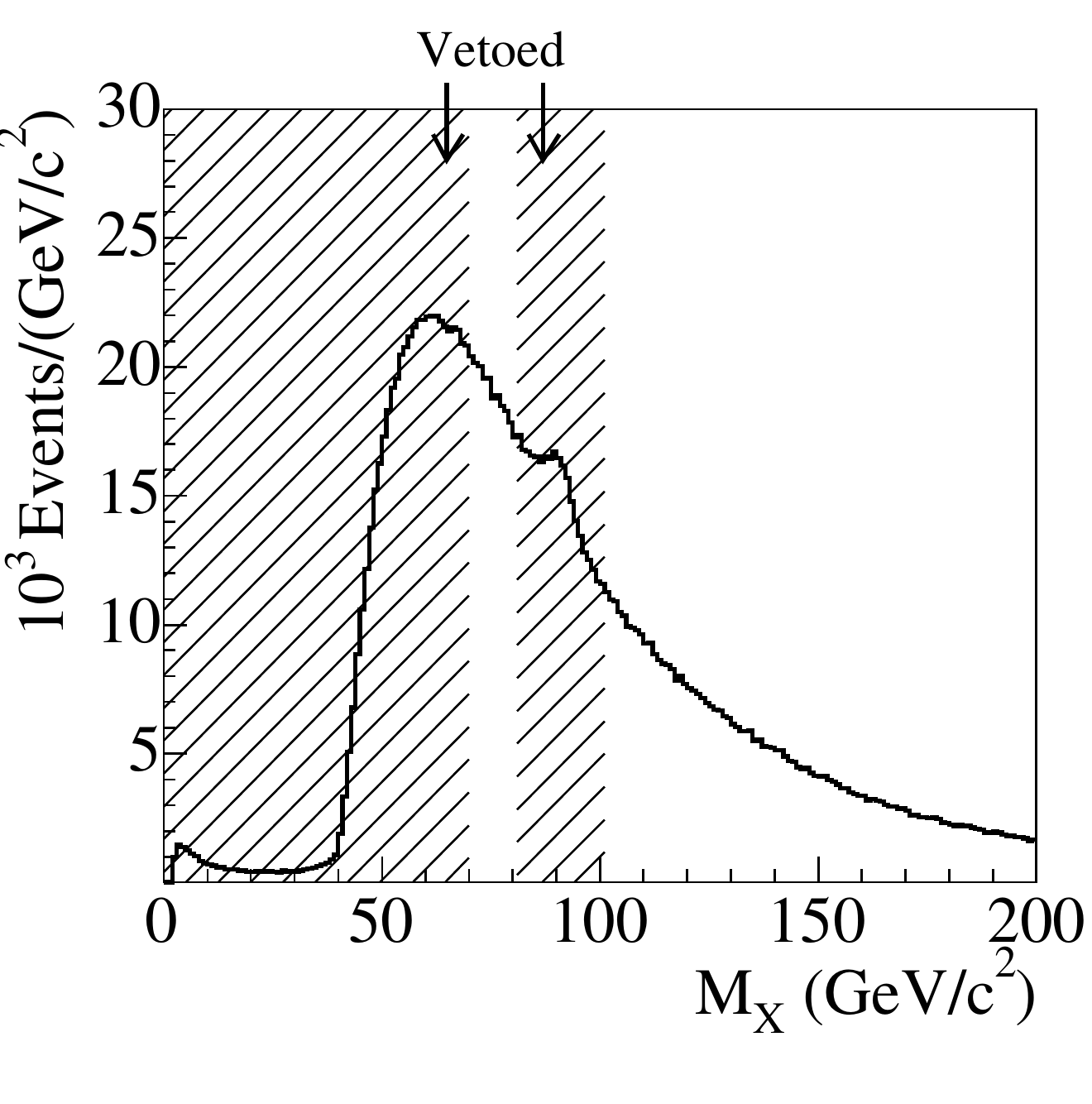}
\end{center}
\caption[]{
Distribution of $M_X$ in ``$X$''+jet events selected from the electron triggers as described in the text.  The shaded regions are
removed; that is, events with $M_X > 70\ \gev/c^2$ are selected, and the $81 < M_{X} < 101\ \gev/c^2$ region is vetoed.
}
\label{fig:prd_inverte_mee_nm1}
\end{figure}

\newcommand{\invertebkgfitpred}{$4427^{+354}_{-310}$}
\newcommand{\inverteobserve}{$4509$}
\newcommand{\invertesigma}{$0.23 \sigma$}
\newcommand{\invertejtcutbkgfitpred}{$1412^{+477}_{-212}$}
\newcommand{\invertejtcutobserve}{$1128$}
\newcommand{\invertejtcutnsigma}{$-1.3\sigma$}

Given this ``$X$'' selection, the remaining jets in the event are used to validate the procedure.
Figure~\ref{fig:prd_inverte_bkgfit_et2} shows the third highest $E_T$ jet distribution.  We extrapolate this distribution above 30
GeV using Eq.~(\ref{eqn:jetetparamfinal}).  A prediction of \invertebkgfitpred\ (statistical uncertainty only) events with $\njett
\geq 3$ is obtained.  \inverteobserve\ events are observed.  Approximating the Poisson distribution of the number of observed events
as a Gaussian, this is a \invertesigma\ level of consistency.

\begin{figure}
\begin{center}
\includegraphics[width=2.5in]{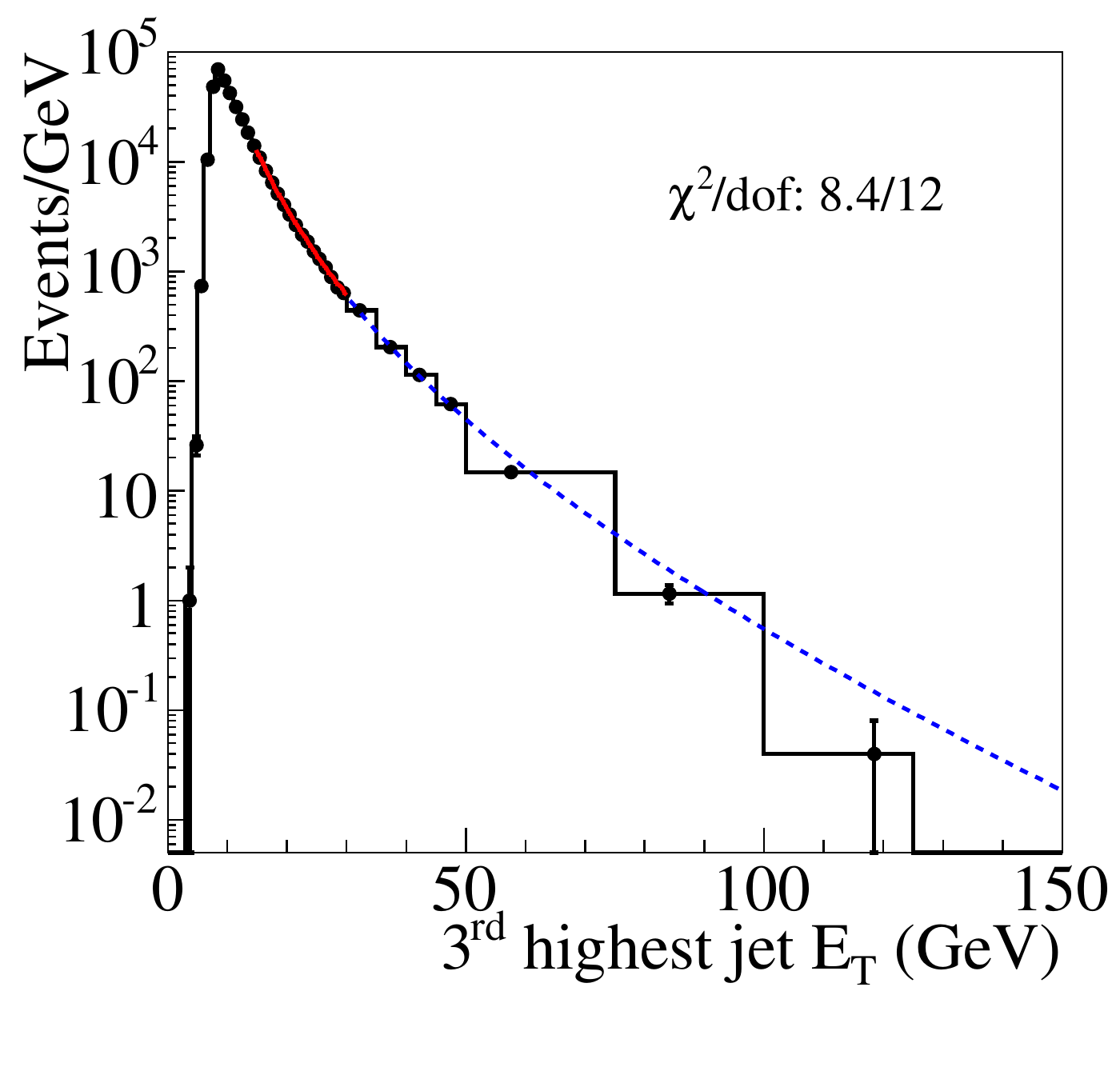}
\end{center}
\caption[]{
$E_T$ distribution of the third highest $E_T$ jet in ``$X$''+jet events selected with the electron triggers as described in the text.
The distribution is fit to Eq.~(\ref{eqn:jetetparamfinal}) in the $15 < E_T < 30$~GeV region and extrapolated to the $E_T > 30$
GeV region.
}
\label{fig:prd_inverte_bkgfit_et2}
\end{figure}

The $\jtt$ shape is predicted using the previously described procedure of extrapolating the jet $E_T$ distributions from events with
$\njett = $ 1 and 2 to $\njett \geq 3$.  The normalized prediction and its uncertainty are compared to the actual distribution in the
data in Fig.~\ref{fig:prd_inverte_bkgfit_jt_unblind}.  The distribution agrees well within the uncertainty envelope.  Above 200 GeV,
\invertejtcutbkgfitpred\ events are expected; \invertejtcutobserve\ events are observed, for a \invertejtcutnsigma\ level of
consistency.  The background prediction is compared to the number of observed events as a function of the \jtt\ cut in
Table~\ref{tab:inverte_databkg}.  The prediction agrees well over the entire \jtt\ distribution.

\begin{table}[htb]
\begin{center}
\begin{tabular}{ccc}
\hline \hline
Minimum $\jtt$ cut & Total Bkg. (events)                    & Data (events) \\ \hline
$50$               &  $ 4430    ^{ + 1270  }_{ - 600    } $ &  4509   \\
$100$              &  $ 4380    ^{ + 1250  }_{ - 590    } $ &  4463   \\
$150$              &  $ 2810    ^{ +  830  }_{ - 360    } $ &  2602   \\
$200$              &  $ 1410    ^{ +  480  }_{ - 210    } $ &  1128   \\
$250$              &  $  667    ^{ +  281  }_{ - 133    } $ &   436   \\
$300$              &  $  312    ^{ +  172  }_{ -  81.8  } $ &   170   \\
$350$              &  $  146    ^{ +  106  }_{ -  47.4  } $ &    62   \\
$400$              &  $   68.7  ^{ +  64.8 }_{ -  26.2  } $ &    27   \\
$450$              &  $   32.8  ^{ +  38.9 }_{ -  14.3  } $ &    15   \\
$500$              &  $   16.2  ^{ +  23.3 }_{ -   8.4  } $ &     6   \\
$550$              &  $    7.9  ^{ +  14.5 }_{ -   4.5  } $ &     3   \\
$600$              &  $    3.9  ^{ +   8.8 }_{ -   2.5  } $ &     0   \\
\hline \hline
\end{tabular}
\end{center}
\caption[]{
The ``X''+jet data (selected with the electron triggers as described in the text) vs. \jtt, compared with the background prediction.
}
\label{tab:inverte_databkg}
\end{table}

\begin{figure}
\begin{center}
\includegraphics[width=2.5in]{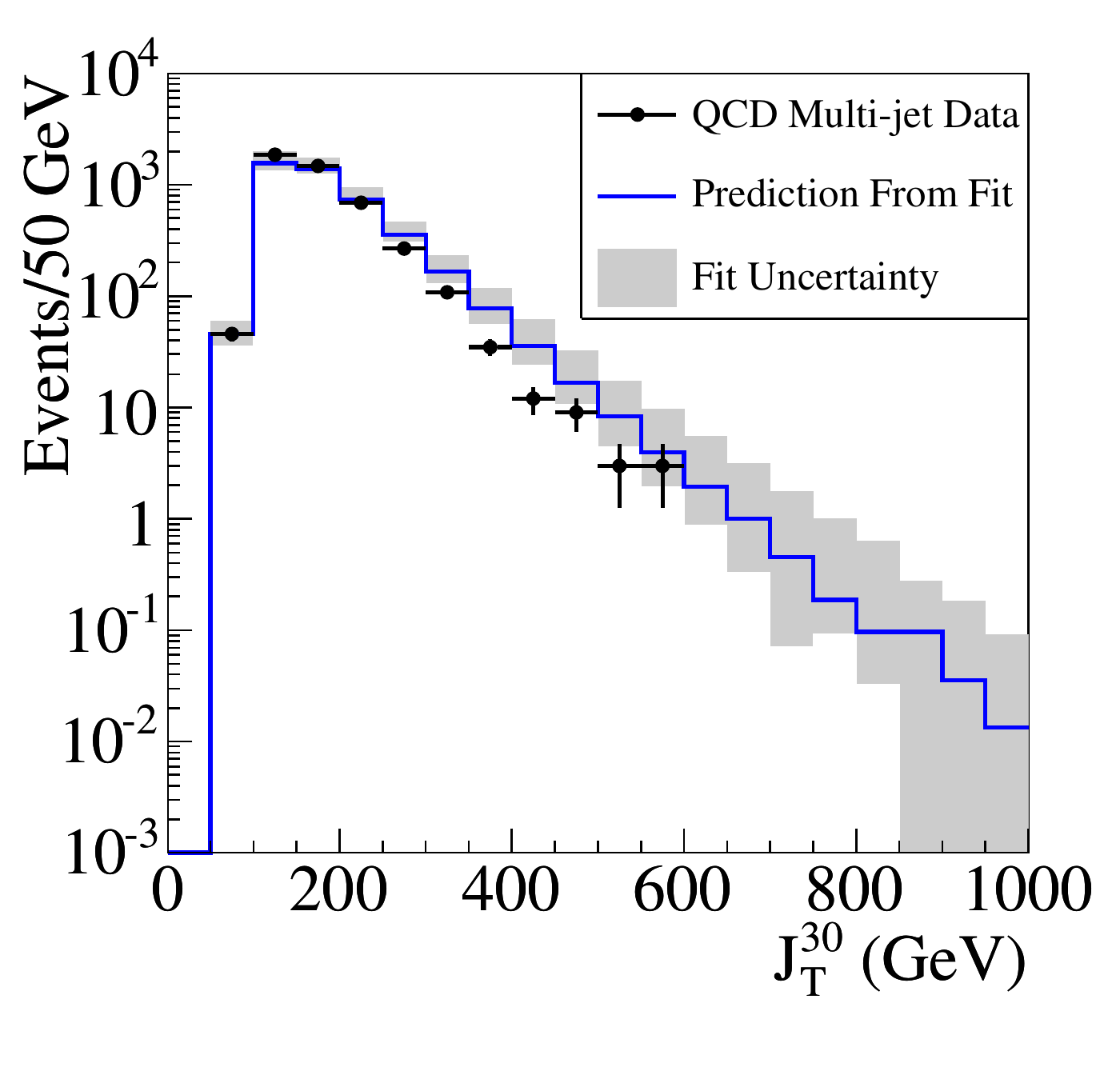}
\end{center}
\caption[]{
The prediction (blue line) and uncertainty (gray band) for the \jtt\ distribution of ``$X$''+jet events selected with the electron
triggers as described in the text.  The prediction is compared to the actual distribution (black points with errors).  The
observation agrees with the prediction, with a maximum fluctuation downward of $1.9 \sigma$.  The data are below the prediction
for several point because the shape uncertainty is correlated between bins.
}
\label{fig:prd_inverte_bkgfit_jt_unblind}
\end{figure}

We have seen that the background extrapolation performs well enough in this high-statistics validation sample.  Because of the
high-statistics, this sample can be divided into subsamples and test the prediction method many times over.  The electron-triggered
multi-jet data is divided into 50 subsamples to check the background estimation with a sample size similar to that expected in the
\Z+jet data.

To validate the third highest $E_T$ jet extrapolation, we evaluate the consistency between the fit prediction and the observation in
each subsample.  The pull distribution from these calculations is observed to be consistent with a Gaussian with mean 0 and width of
1, indicating that the mean prediction and the uncertainties are correctly calculated for the $\njett \geq 3$ prediction.  On
average, the background prediction is $3 \pm 5$\% low relative to the data.  That is, the background prediction underestimates the
background, but by an amount consistent with zero.  This is consistent with the fit done in standard model \Z\ Monte Carlo simulation
in Sec.~\ref{subsec:njetnorm}, in which the background prediction was $31 \pm 16$\% low relative to the data.

To validate the \jtt\ shape prediction, in each subsample we evaluate the consistency between the fit prediction and the observation
using a cut of $\jtt > 200$~GeV.  In this case, the resulting pull distribution was inconsistent with a Gaussian with mean 0 and
width 1.  We find that the background prediction overestimates the number of observed events, and that the uncertainty is overly
conservative, after correcting for this bias.  On average, the background prediction is $23 \pm 7$\% high relative to the data.
However, we find that this bias is covered by the uncertainties, with an average uncertainty on the background prediction of $47$\%.
To clarify, these biases are only present in the \jtt\ shape prediction, and not in the $\njett \geq 3$ prediction.

To compare the jet kinematics in each of the validation samples (both the ``X'' events selected from jet triggers and the ``X''
events selected from the electron triggers) to the \Z+jet data, the \jtt\ distribution of each is plotted, without the $\njett \geq
3$ requirement, in Fig.~\ref{fig:prd_zpt_comparison}.  The overall shape of each is the same, although they are slightly
different---for example, electron-triggered ``X''+jet data have a harder spectrum.  However, the background estimation takes these
differences into account in the fit procedure.

\begin{figure}
\begin{center}
\includegraphics[width=2.5in]{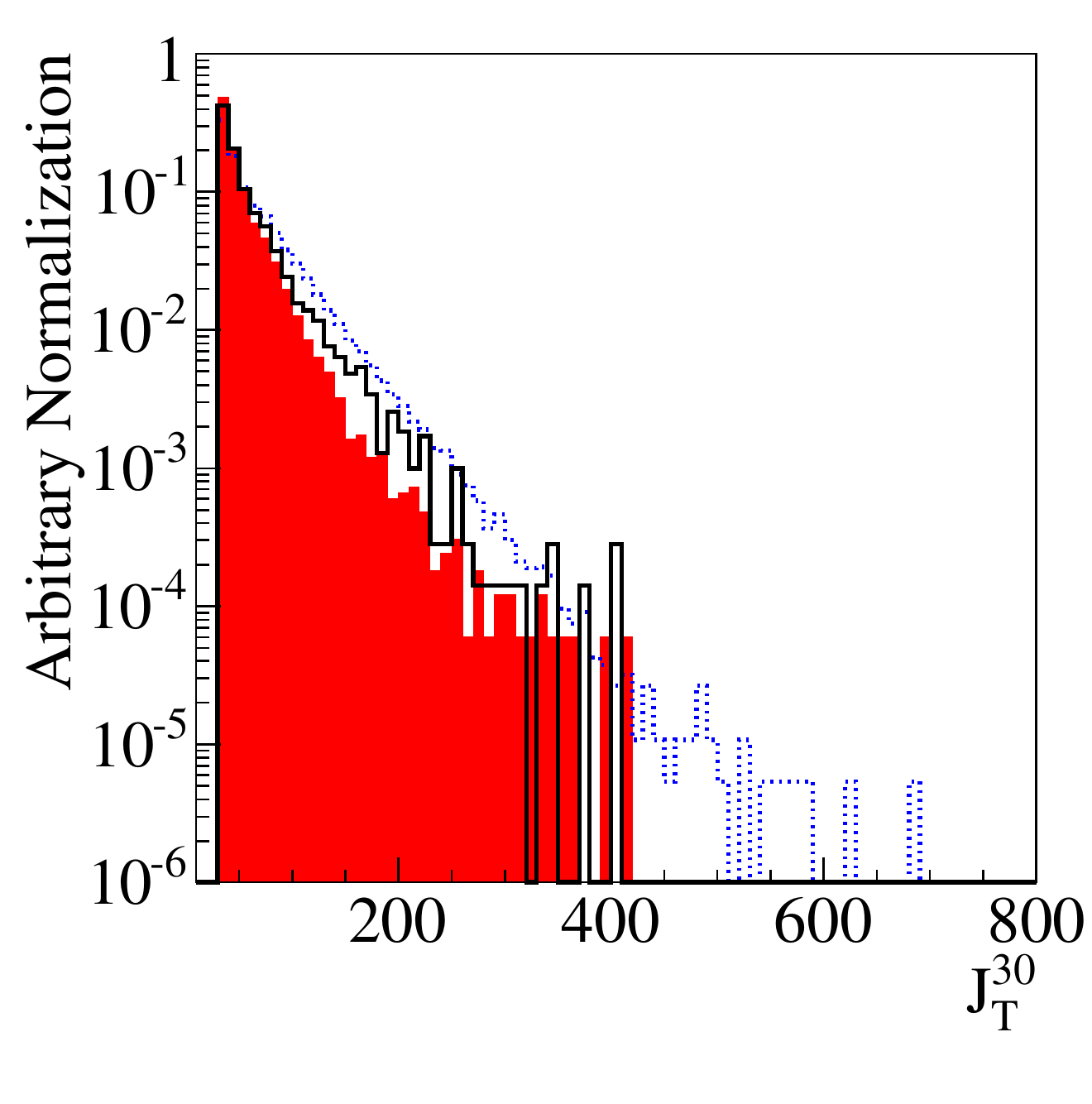}
\end{center}
\caption[]{
The \jtt\ distribution without the $\njett \geq 3$ requirement in the \Z+jet data (black line), compared to ``X''+jet data selected
with the jet triggers (red histogram) and to ``X''+jet data selected with the electron triggers (dotted blue line).
}
\label{fig:prd_zpt_comparison}
\end{figure}

These validations show that the fit prediction method correctly calculates the background when there is no signal present.  To verify
that it calculates the background correctly in the presence of signal, we use $W$+jet data.

\subsubsection{$W$+jet Data}
\label{subsubsec:fitwjetval}

The tree-level single $W$ diagrams and the physics that gives rise to additional jets is similar to $Z$+jet production, and so
similar behavior in the $W$+jet data is expected.  However, in the $W$+jet data, in addition to the single-$W$ production there is
also a heavy quark signal from the top quark, producing $W$ bosons via $t\bar{t} \rightarrow WW b\bar{b}$.  This sample provides a
useful and interesting validation of the method---it is a real data sample that can test whether or not the background fit procedure
performs properly in the presence of a signal similar to that of the search.

$W$ events in the $W \rightarrow \mu \nu$ channel are selected by requiring exactly one ``tight'' muon and missing transverse energy
(\met).  The \met\ is measured using the vector sum of the calorimeter tower transverse energies and the muon $p_T$.  $\met > 25$~GeV
is required.  Since only a single muon is required, this is the so-called ``lepton+jets'' channel of the top quark selected with only
kinematic information, and without tagging $b$-jets \cite{bib:toplepjets}.  

\newcommand{\datawjetbkgfit}{$439^{+20}_{-20}\ \mathrm{(stat.)}\ ^{+30}_{-24}\ \mathrm{(syst.)}$}
\newcommand{\datawjetobs}{762}
\newcommand{\nttbar}{$323^{+34}_{-34}\ \mathrm{(stat.)}\ ^{+30}_{-24}\ \mathrm{(syst.)}$}
\newcommand{\ttbaracceptance}{$3.41 \pm 0.02\%$}
\newcommand{\ttbarlumi}{$1.04\ \ifb$}
\newcommand{\ttbarxsec}{$9 \pm 1\ \mathrm{pb\ (stat.\ uncert.\ only)}$}

Using this $W$+jet selection, we test the extraction of the top signal for events with $\njett \geq 3$ using only data as a
validation of the method for predicting the \Z+jet background.  We expect standard model $W$+jet to be the dominant background for
\ttbar\ after the \njett\ requirement.  In single $W$+jet Monte Carlo simulation with no \ttbar\ component, the method does predict
the actual Monte Carlo distribution well.  We then apply the same method to the $W$+jet data, fitting the third highest $E_T$ jet
distribution to Eq.~(\ref{eqn:jetetparamfinal}) in Fig.~\ref{fig:prd_wjet_bkgfit_et2}.  In this case, the extrapolation does not
describe the data well.

The extrapolation predicts \datawjetbkgfit\ events; \datawjetobs\ events are observed.  We make the hypothesis that this excess is
due to the top quark, and test this by checking that the cross section is consistent with that expected for \ttbar.  The excess of
the data above the background gives the number of \ttbar\ candidates, \nttbar.  Using \ttbar\ Monte Carlo events gives an estimate
for the product of acceptance and efficiency of \ttbaracceptance.  The luminosity of the muon-triggered sample is \ttbarlumi.  A
cross section of \ttbarxsec\ \cite{bib:topxsecnote} is therefore obtained.  The proximity to the previous measured cross section in
this channel at CDF using 194 \ipb, $6.6 \pm±1.1\ \mathrm{(stat.)} \pm±1.5\ \mathrm{(syst.)\ pb}$ \cite{bib:toplepjets}, indicates
that the excess is consistent with the background+\ttbar\ hypothesis, and that the fit procedure is accurately predicting the
background from single $W$+jet production in the presence of signal.

\begin{figure}
\begin{center}
\includegraphics[width=2.5in]{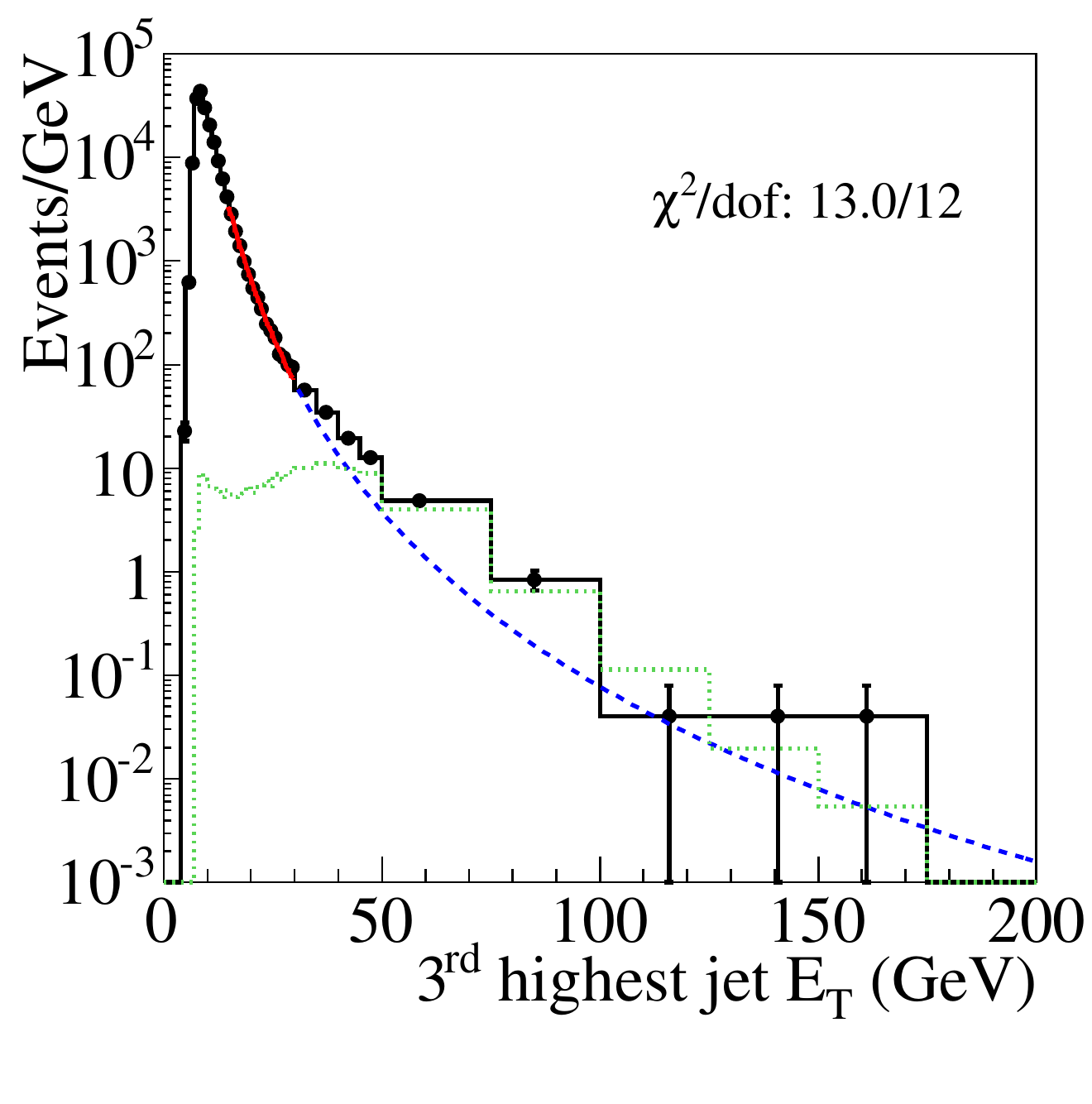}
\end{center}
\caption[]{
$E_T$ distribution of the third highest $E_T$ jet in $W$+jet events (black line and points).  The distribution is fit to
Eq.~(\ref{eqn:jetetparamfinal}) in the $15 < E_T < 30$~GeV region and extrapolated to the $E_T > 30$~GeV region.  The dotted green
line shows the contribution from \ttbar\ at the ``measured'' cross section of 9 pb.  There is very little contribution from \ttbar\
within the fit region.  The extrapolated distribution is inconsistent with the background-only hypothesis, but consistent with the
background plus \ttbar\ hypothesis.
}
\label{fig:prd_wjet_bkgfit_et2}
\end{figure}

A prediction is now made for the $\jtt$ shape of the $W$+jet background.  Figures~\ref{fig:prd_wjet_bkgfit_njet30_1}
and~\ref{fig:prd_wjet_bkgfit_njet30_2} show the fits to the jet $E_T$ spectra for events with $\njett = $ 1 and 2;
Fig.~\ref{fig:prd_wjet_bkgfit_exp_vs_njet} shows the parameter $p_1$ extrapolation; Fig.~\ref{fig:prd_wjet_bkgfit_njet30_shape} shows
the $\njett$ shape fit.  We use these shapes to obtain the $\jtt$ shape and errors, add the expected contribution from \ttbar\ using
Monte Carlo simulation (normalized to the ``measured'' cross section of 9 pb), and compare this to the actual distribution in data in
Fig.~\ref{fig:prd_wjet_bkgfit_jt_unblind}.  The observed data are well described by the total $\jtt$ prediction, verifying that the
fit procedure can predict the $\jtt$ shape of the background in the presence of signal.

\begin{figure}
\begin{center}
\includegraphics[width=2.5in]{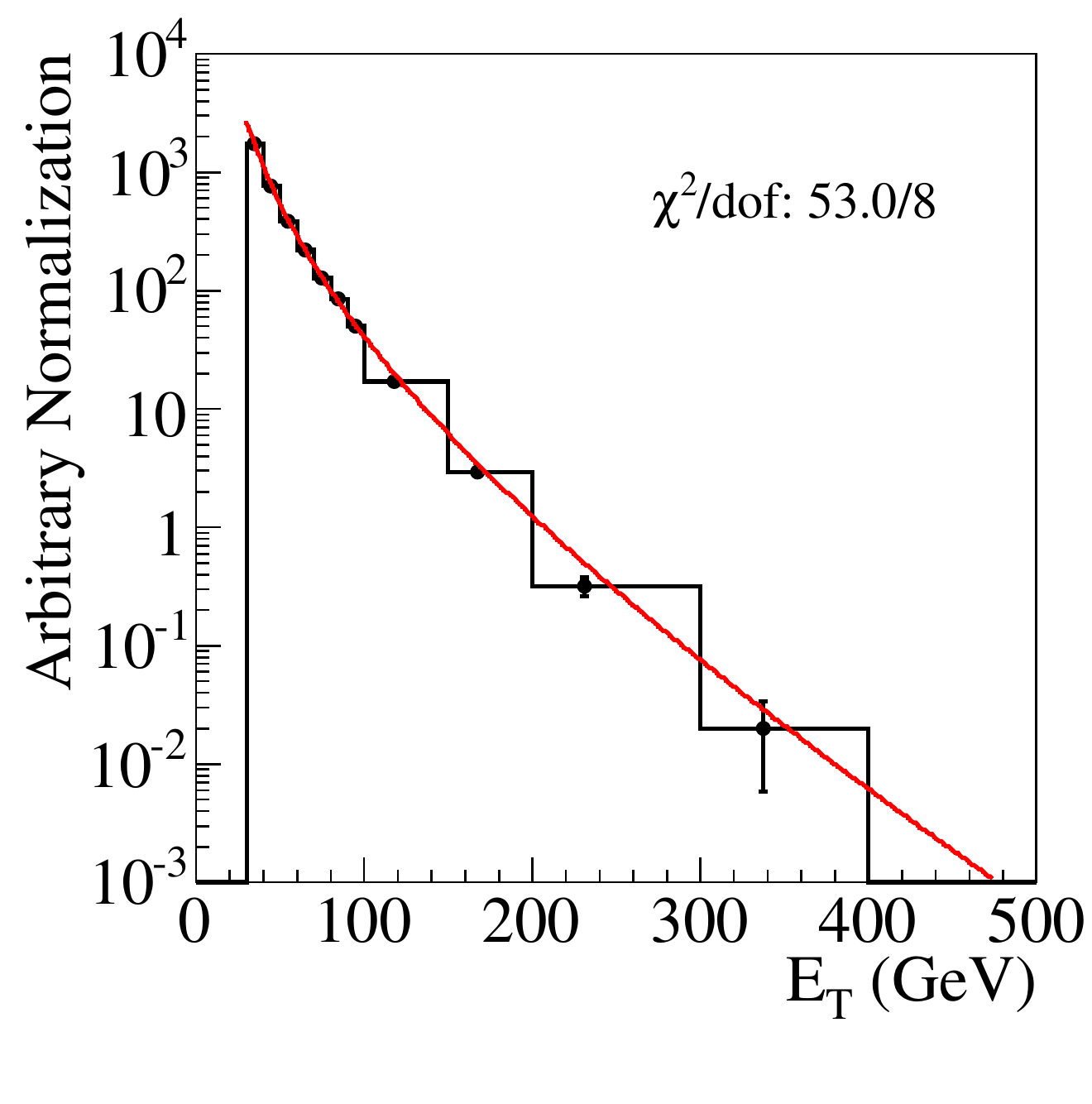}
\end{center}
\caption[]{
$E_T$ distribution of jets in $\njett = 1$ $W$+jet events.  The distribution is fit to Eq.~(\ref{eqn:jetetparamfinal}) in the $E_T >
30$~GeV region.
}
\label{fig:prd_wjet_bkgfit_njet30_1}
\end{figure}

\begin{figure}
\begin{center}
\includegraphics[width=2.5in]{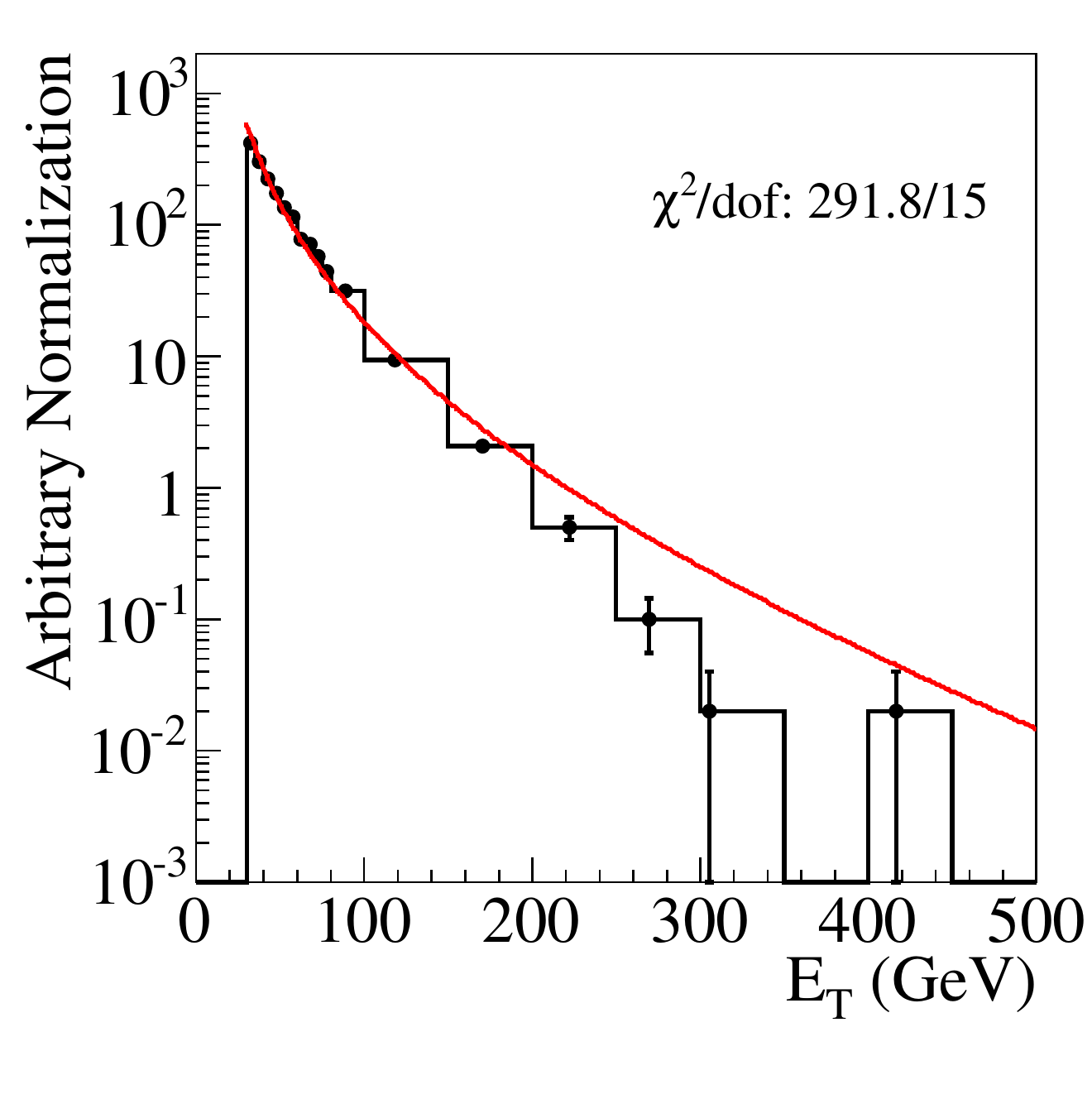}
\end{center}
\caption[]{
$E_T$ distribution of jets in $\njett = 2$ $W$+jet events.  The distribution is fit to Eq.~(\ref{eqn:jetetparamfinal}) in the $E_T >
30$~GeV region with the parameter $p_2$ fixed to that obtained from the fit in Fig.~\ref{fig:prd_wjet_bkgfit_njet30_1}.
}
\label{fig:prd_wjet_bkgfit_njet30_2}
\end{figure}

\begin{figure}
\begin{center}
\includegraphics[width=2.5in]{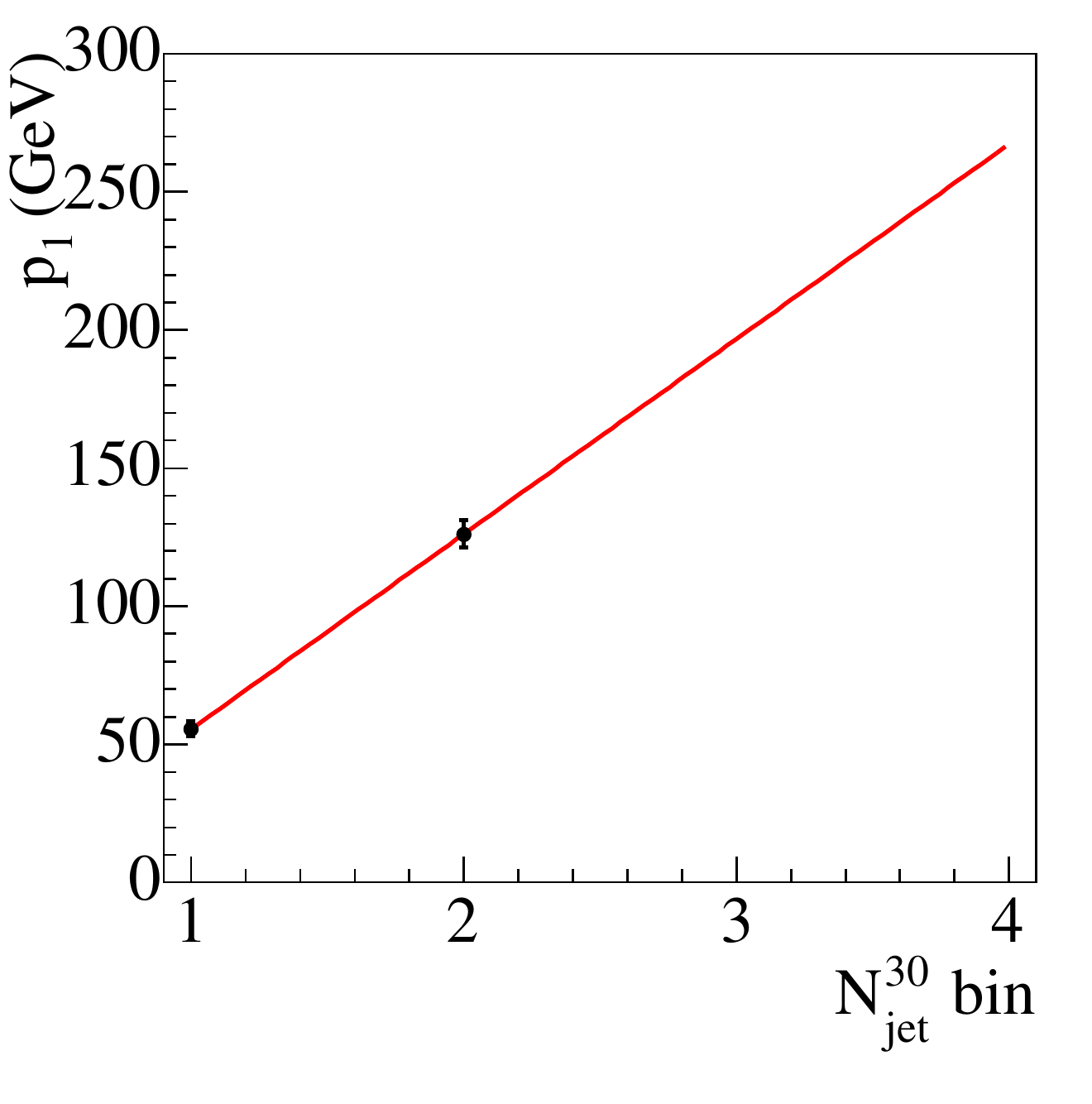}
\end{center}
\caption[]{
The extrapolation of the exponential parameter $p_1$ vs. $\njett$ in $W$+jet events.
}
\label{fig:prd_wjet_bkgfit_exp_vs_njet}
\end{figure}

\begin{figure}
\begin{center}
\includegraphics[width=2.5in]{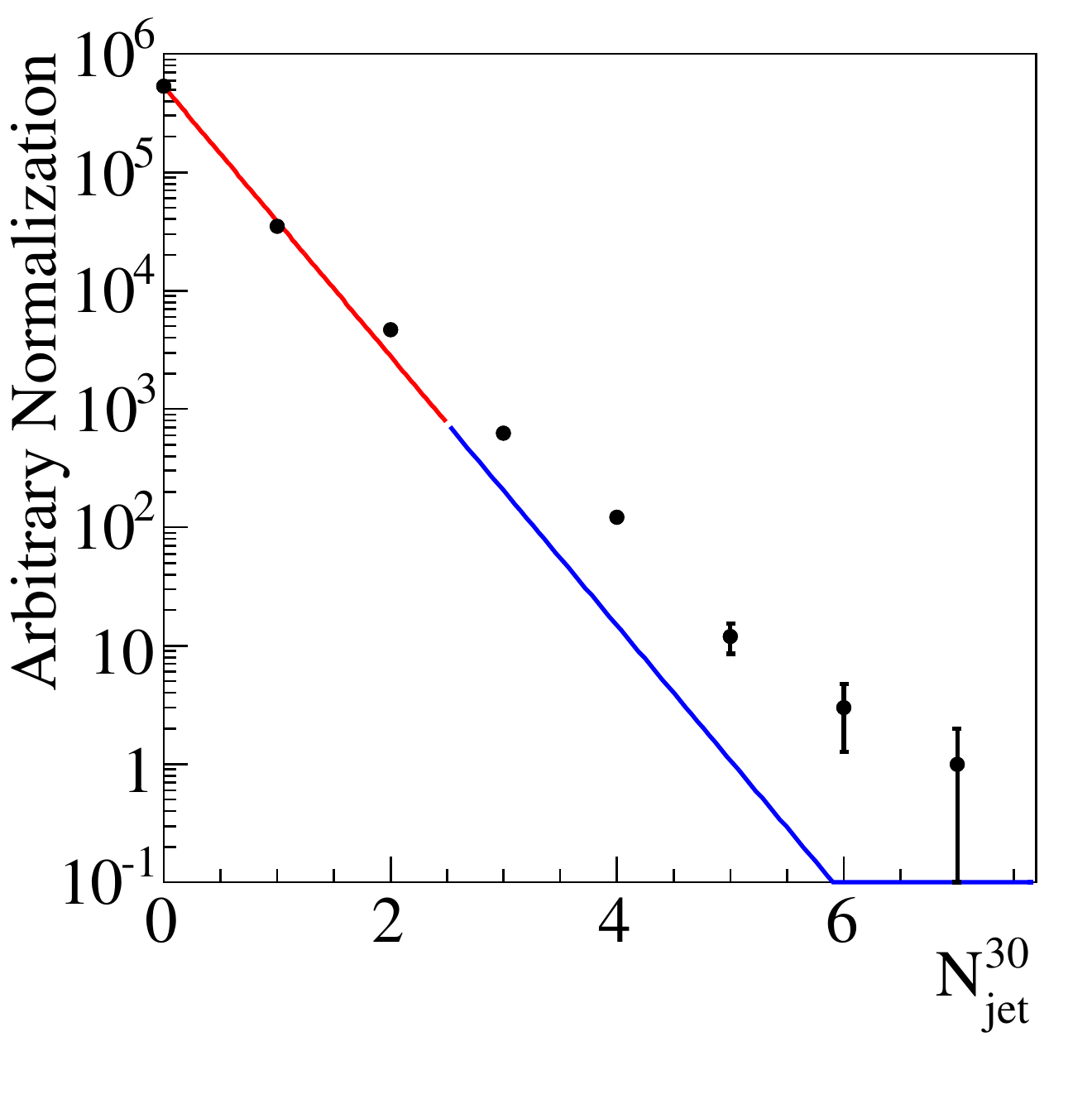}
\end{center}
\caption[]{
$\njett$ distribution in $W$+jet events.  The distribution is fit to an exponential in the range $\njett \leq 2$.
}
\label{fig:prd_wjet_bkgfit_njet30_shape}
\end{figure}

\begin{figure}
\begin{center}
\includegraphics[width=2.5in]{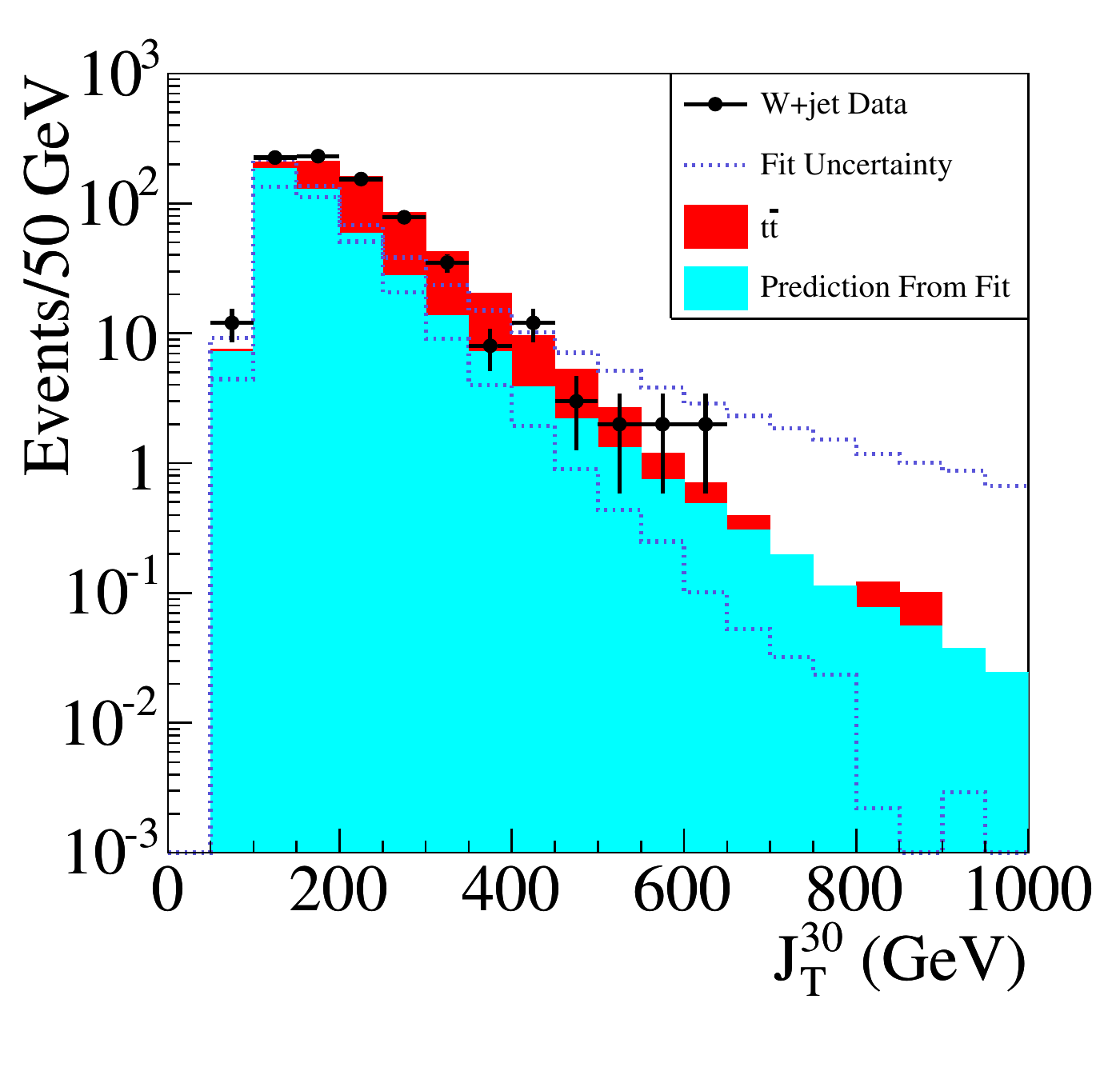}
\end{center}
\caption[]{
The prediction (cyan histogram) and uncertainty (dotted lines) for the \jtt\ distribution of $W$+jet events.  The expectation from
\ttbar\ is added to the prediction.  The data (points with errors) agree with the background plus \ttbar\ hypothesis.
}
\label{fig:prd_wjet_bkgfit_jt_unblind}
\end{figure}

While the predicted shape of the $\jtt$ distribution agrees with the data well (after adding the expected contribution from \ttbar),
the total uncertainty on the background prediction becomes extremely large at high $\jtt$.  The \jtt\ distribution for \ttbar\ peaks
near 200 GeV, where the uncertainty is small, but it is instructive to understand the reason for the increased uncertainty at very
large \jtt.  This large error is completely dominated by a poor parameterization of the $E_T$ distribution of jets in $\njett = 2$
events.  Since, in Fig.~\ref{fig:prd_wjet_bkgfit_njet30_2}, the fitted parameterization poorly describes the data, changing the range
from nominal (our method for determining the size of the mis-parameterization uncertainty) will make a large difference in the fit.
However, this is not a problem with the parameterization in Eq.~(\ref{eqn:jetetparamfinal}), because if the same spectrum is fit
without fixing the power law parameter to the value observed in events with $\njett = 1$, the quite reasonable fit, shown in
Fig.~\ref{fig:prd_wjet_bkgfit_njet30_2_nofixpow}, is obtained.  That is, the parameterization still describes the $\njett = 2$ $E_T$
spectrum well, but our method of fixing the power law parameter in this fit to that observed from the $\njett = 1$ $E_T$ spectrum
does not describe the behavior of the changing jet $E_T$ distributions as a function of $\njett$ well in this sample.  In the other
validation samples in data and Monte Carlo simulations, and particularly in the fits of the $Z$+jet data, we find no such large
systematic effect from a mis-parameterization in the $\njett = 2$ $E_T$ distribution.  This issue therefore does not affect this
analysis, but it suggests the background prediction procedure could be enhanced with a more sophisticated parameter extrapolation,
perhaps by extrapolating both parameters $p_1$ and $p_2$ simultaneously.

\begin{figure}
\begin{center}
\includegraphics[width=2.5in]{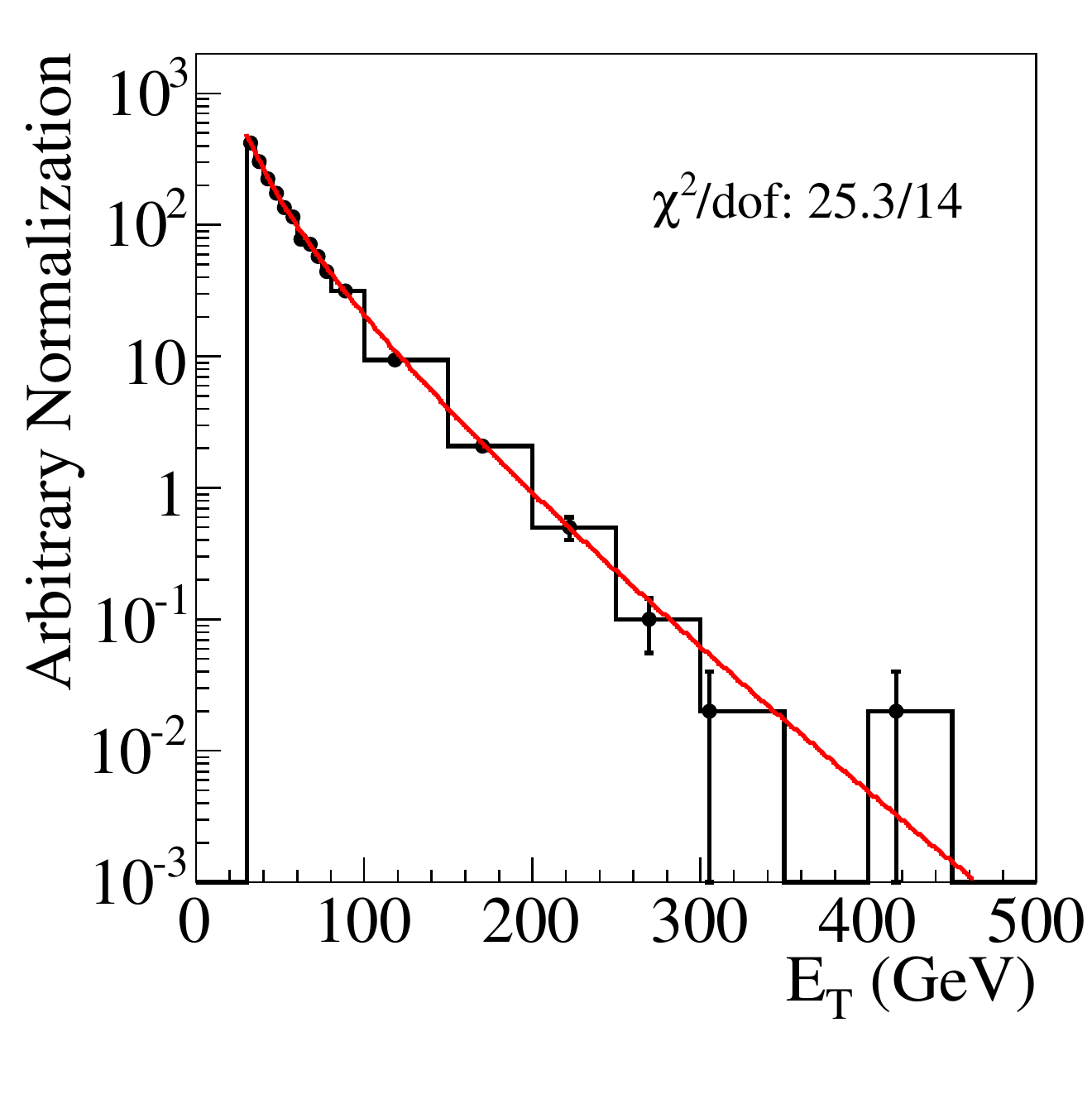}
\end{center}
\caption[]{
$E_T$ distribution of jets in $\njett = 2$ $W$+jet events.  The distribution is fit to Eq.~(\ref{eqn:jetetparamfinal})
in the $E_T > 30$~GeV region without fixing the parameter $p_2$.
}
\label{fig:prd_wjet_bkgfit_njet30_2_nofixpow}
\end{figure}

\subsubsection{Signal Injection Studies}
\label{subsubsec:fitsiginject}

The studies in data indicate the fit method adequately predicts the background, without and with the presence of signal.  We would
also like to understand at what point, if any, signal contamination causes an unacceptably large change to the background prediction.
That is, we need to verify that the background extrapolation does not ``fit-away'' the signal, as the jet $E_T$ distributions may be
substantially changed if there is a large amount of signal in the fitted regions.

To study this effect we use standard model Z Monte Carlo events with $b'\rightarrow bZ$ Monte Carlo events added at a variety of
signal masses.  An equivalent luminosity of $1\ \ifb$ of Monte Carlo events is used to understand the effect with the approximate
amount of statistics that is present in the data.  For this study ${BR}(b' \rightarrow bZ) = 100\%$ is assumed; reducing this
branching ratio will only reduce the effect of a signal bias.

For example, the predicted \jtt\ distributions, generated with and without $m_{b'}=200\ \gev/c^2$ Monte Carlo signal events added to
the \Z+jet background fit, are shown in Fig. \ref{fig:prd_siginject_m200_jt}.  The difference between the background predictions with
and without signal is small compared to the actual number of Monte Carlo events, indicating that signal does not bias the fit to a
large degree at this mass point.

As expected, as the $b'$ mass increases the fit becomes less biased from the presence of signal; as the $b'$ mass decreases, the fit
becomes more biased.  At a $b'$ mass of 150 $\gev/c^2$, we found an increase in signal bias, but sensitivity to this mass point is
still retained (at a significance of $4.8 \sigma$).  At a $b'$ mass of 100 $\gev/c^2$, however, we found that the signal was
completely fit away.  We therefore do not set limits below 150 $\gev/c^2$.  We note that this search is still sensitive to models
with masses near 100 $\gev/c^2$, as long as the cross sections are sufficiently small as to not bias the fit.  In general, though,
lower masses produce more signal contamination than higher masses, as both the cross sections are larger and the $E_T$ distributions
have larger fractions within the fit regions.  Sensitivity to these lower masses could be increased by lowering $E_T$ thresholds and
$\njet$ cuts, and applying similar fit procedures with the altered selection.

\begin{figure}
\begin{center}
\includegraphics[width=2.5in]{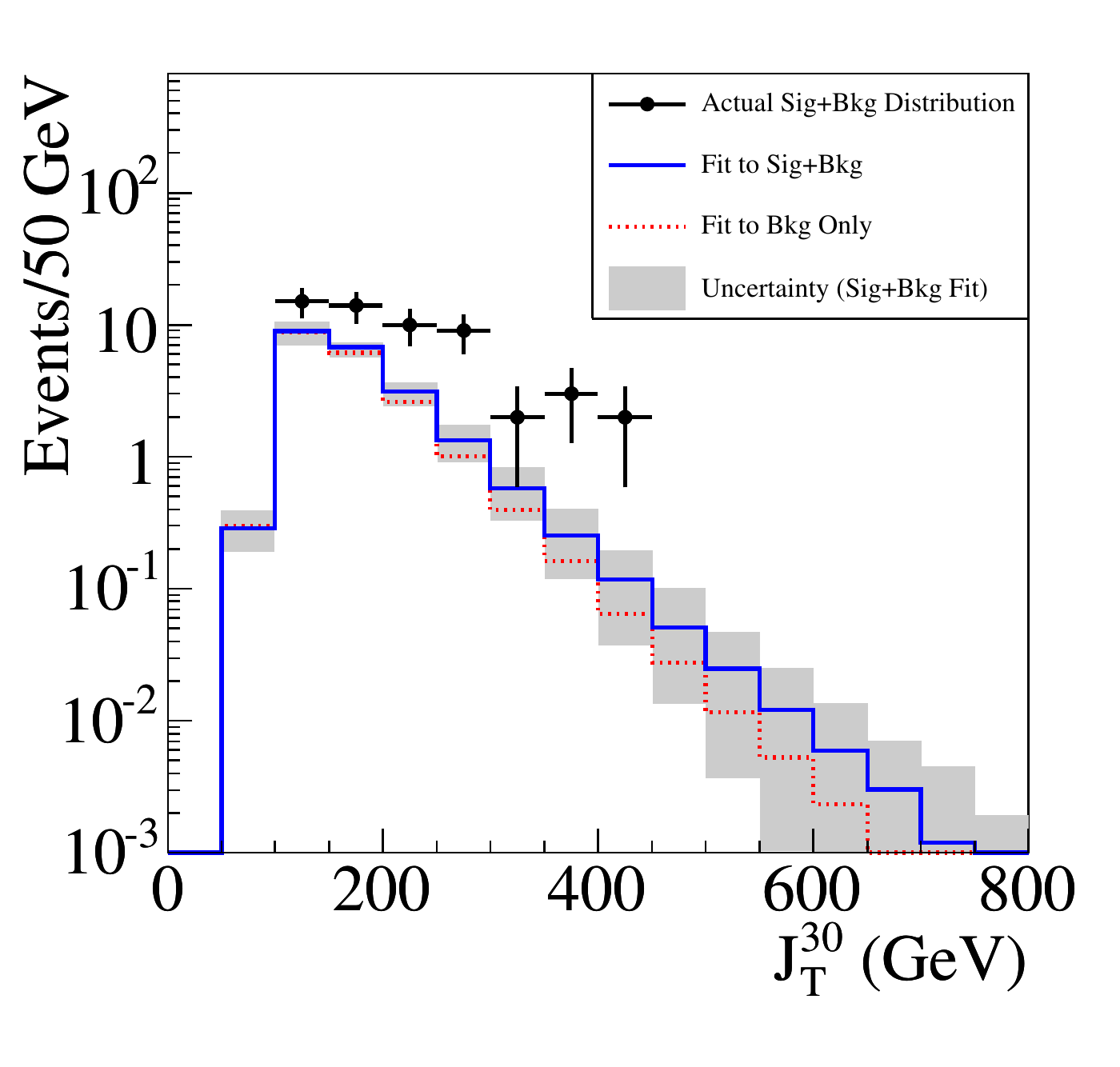}
\end{center}
\caption[]{
Prediction for the $\jtt$ distribution in standard model \zmm\ events, with and without the presence of a 200 $\gev/c^2$ $b'$ signal
introduced.  The difference between the two predictions is small compared to the excess of signal at large \jtt.
}
\label{fig:prd_siginject_m200_jt}
\end{figure}

\subsection{Application of Technique to the Signal Sample}
\label{subsec:applyzjet}

\newcommand{\zdatabkgfit}{$72.2^{+9.8}_{-11.1}$}
\newcommand{\zdataobs}{$80$}

We now apply the fit technique to the combined \mbox{\zee}\ and \zmm\ data to predict the background from \Z+jet final states.  The
third highest $E_T$ jet distribution is shown in Fig.~\ref{fig:prd_zdata_bkgfit_et2_blind}, with events that have $\njett \geq 3$
removed.  We fit in the region $15 < E_T < 30$~GeV, and extrapolate to the region $E_T > 30$~GeV.  We predict \zdatabkgfit\ events
with $\njett \geq 3$.

To obtain the \jtt\ shape of the \Z+jet background, we fit the jet $E_T$ distributions of events with $\njett = $ 1 and 2, and
linearly extrapolate the fit parameter $p_1$ to events with $\njett \geq 3$.  The fit to the $\njett = 1$ jet $E_T$ spectrum is shown
in Fig.~\ref{fig:prd_zdata_bkgfit_njet30_1}, the fit to the $\njet = 2$ jet $E_T$ spectrum in
Fig.~\ref{fig:prd_zdata_bkgfit_njet30_2}, and the extrapolation of the fit parameter in Fig.~\ref{fig:prd_zdata_bkgfit_exp_vs_njet}.
The fit to the $\njett$ distribution in the 0, 1, and 2 jet bins in Fig.~\ref{fig:prd_zdata_bkgfit_njet30_shape} is used as an
estimate of the shape of the $\njett$ distribution in the 3 and higher jet bins.  With these ingredients, the simple Monte Carlo
program is used to obtain the expected $\jtt$ shape, which is then normalized to the prediction for the total number of $\njett \geq
3$ background events, \zdatabkgfit.  The \jtt\ distribution prediction and its total statistical+systematic uncertainty is shown in
Fig.~\ref{fig:prd_zdata_bkgfit_jt_blind}.

\begin{figure}
\begin{center}
\includegraphics[width=2.5in]{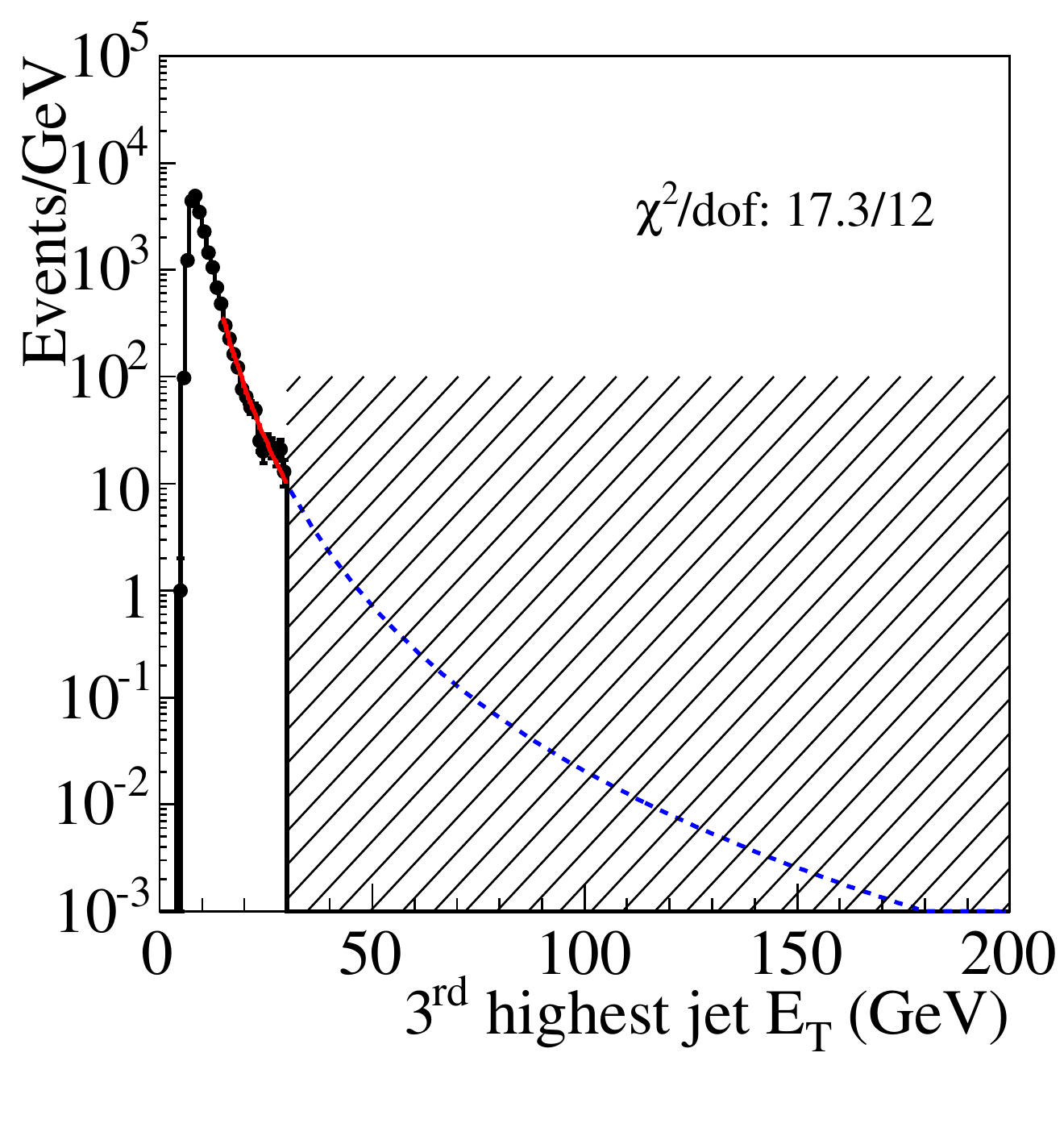}
\end{center}
\caption[]{
$E_T$ distribution of the third highest $E_T$ jet in \zee\ and \zmm\ events with $\njett \leq 2$.  The distribution is fit to
Eq.~(\ref{eqn:jetetparamfinal}) in the $15 < E_T < 30$~GeV region and extrapolated to the $E_T > 30$~GeV region.  Events with
$\njett \geq 3$ (equivalent to $E_T > 30$~GeV, the hatched region) are removed from the distribution.
}
\label{fig:prd_zdata_bkgfit_et2_blind}
\end{figure}

\begin{figure}
\begin{center}
\includegraphics[width=2.5in]{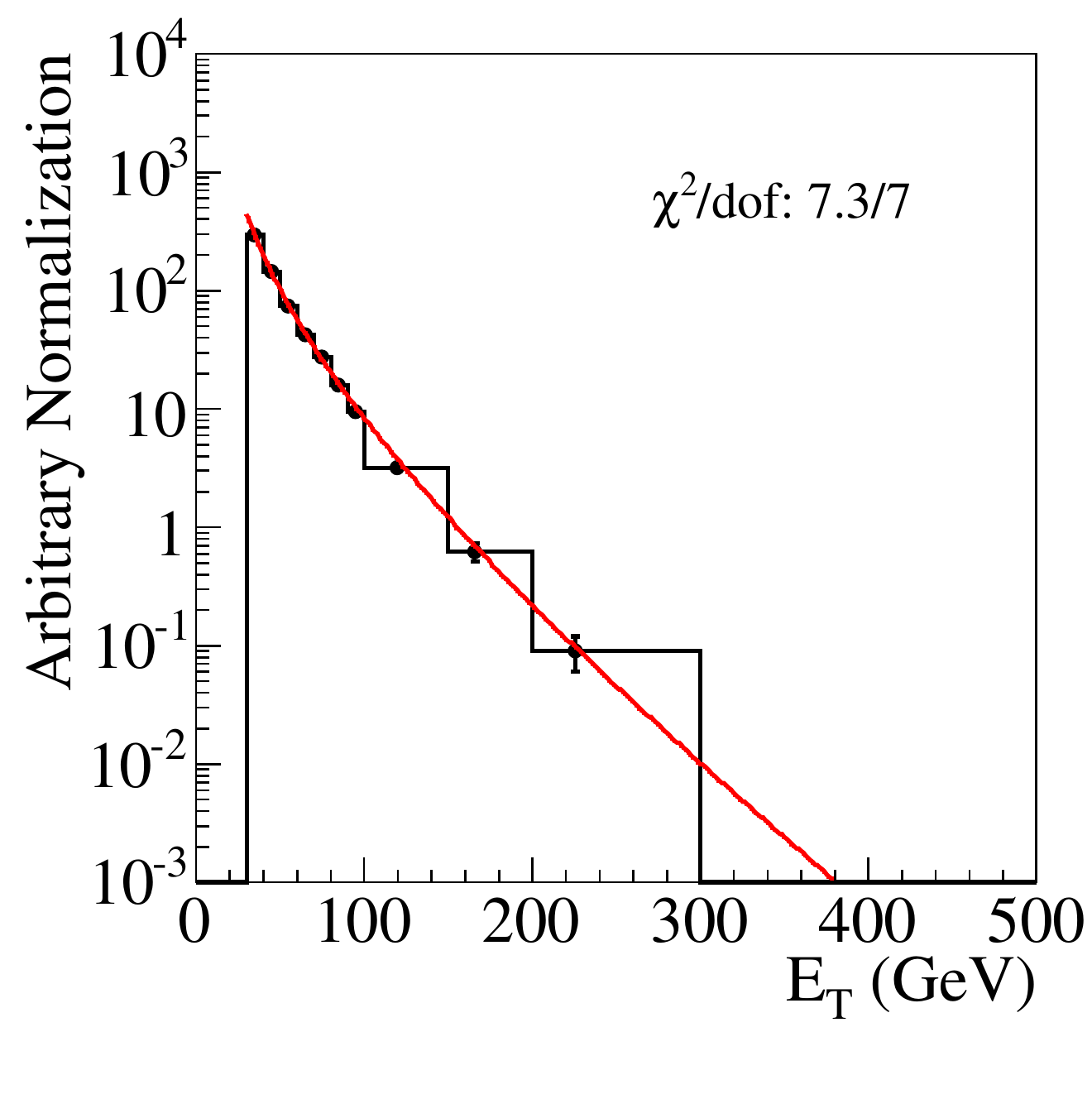}
\end{center}
\caption[]{
$E_T$ distribution of jets in $\njett = 1$ \zee\ and \zmm\ events.  The distribution is fit to Eq.~(\ref{eqn:jetetparamfinal}) in
the $E_T > 30$~GeV region.
}
\label{fig:prd_zdata_bkgfit_njet30_1}
\end{figure}

\begin{figure}
\begin{center}
\includegraphics[width=2.5in]{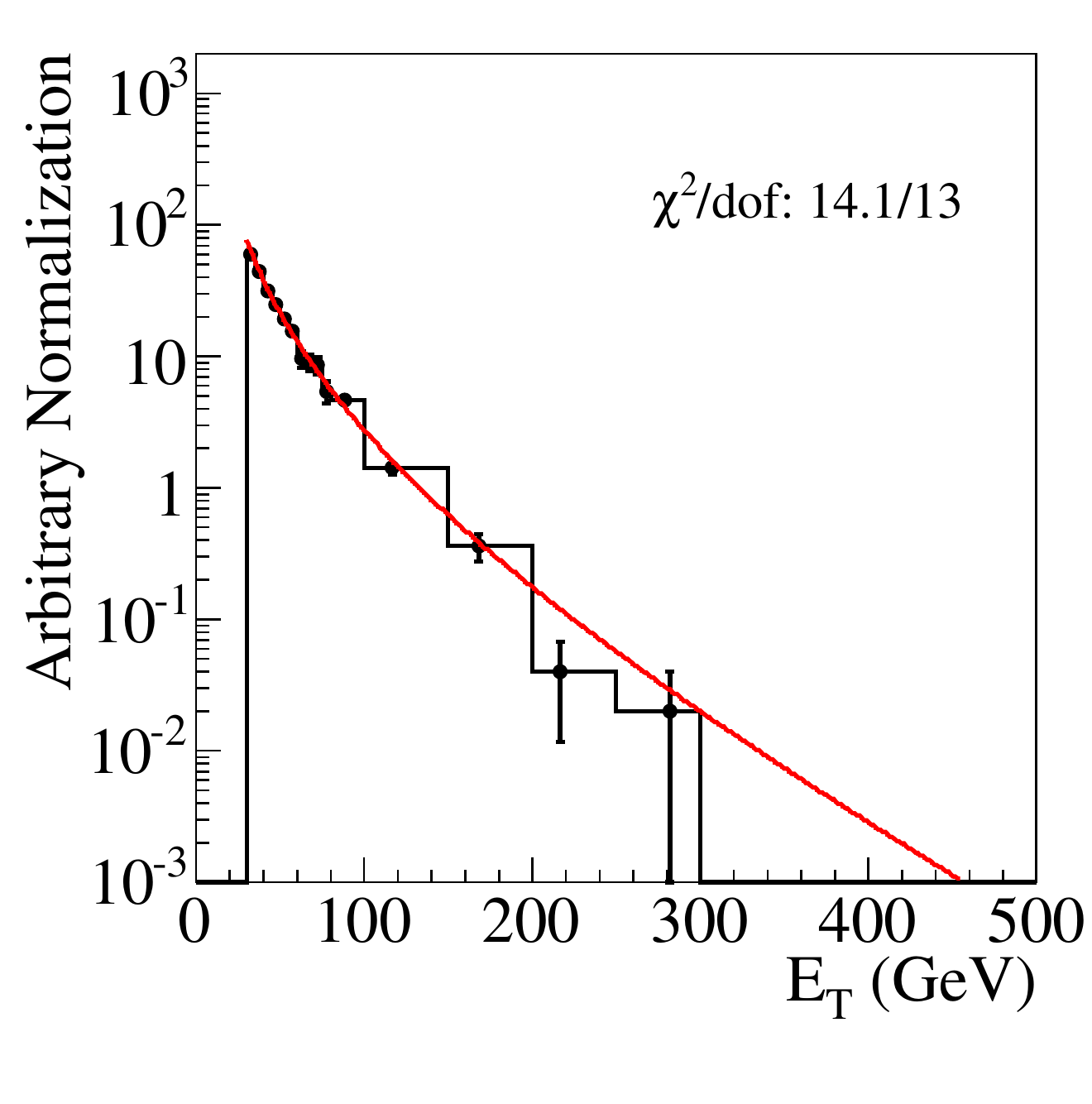}
\end{center}
\caption[]{
$E_T$ distribution of jets in $\njett = 2$ \zee\ and \zmm\ events.  The distribution is fit to Eq.~(\ref{eqn:jetetparamfinal}) in
the $E_T > 30$~GeV region with the parameter $p_2$ fixed to that obtained from the fit in Fig.~\ref{fig:prd_zdata_bkgfit_njet30_1}.
}
\label{fig:prd_zdata_bkgfit_njet30_2}
\end{figure}

\begin{figure}
\begin{center}
\includegraphics[width=2.5in]{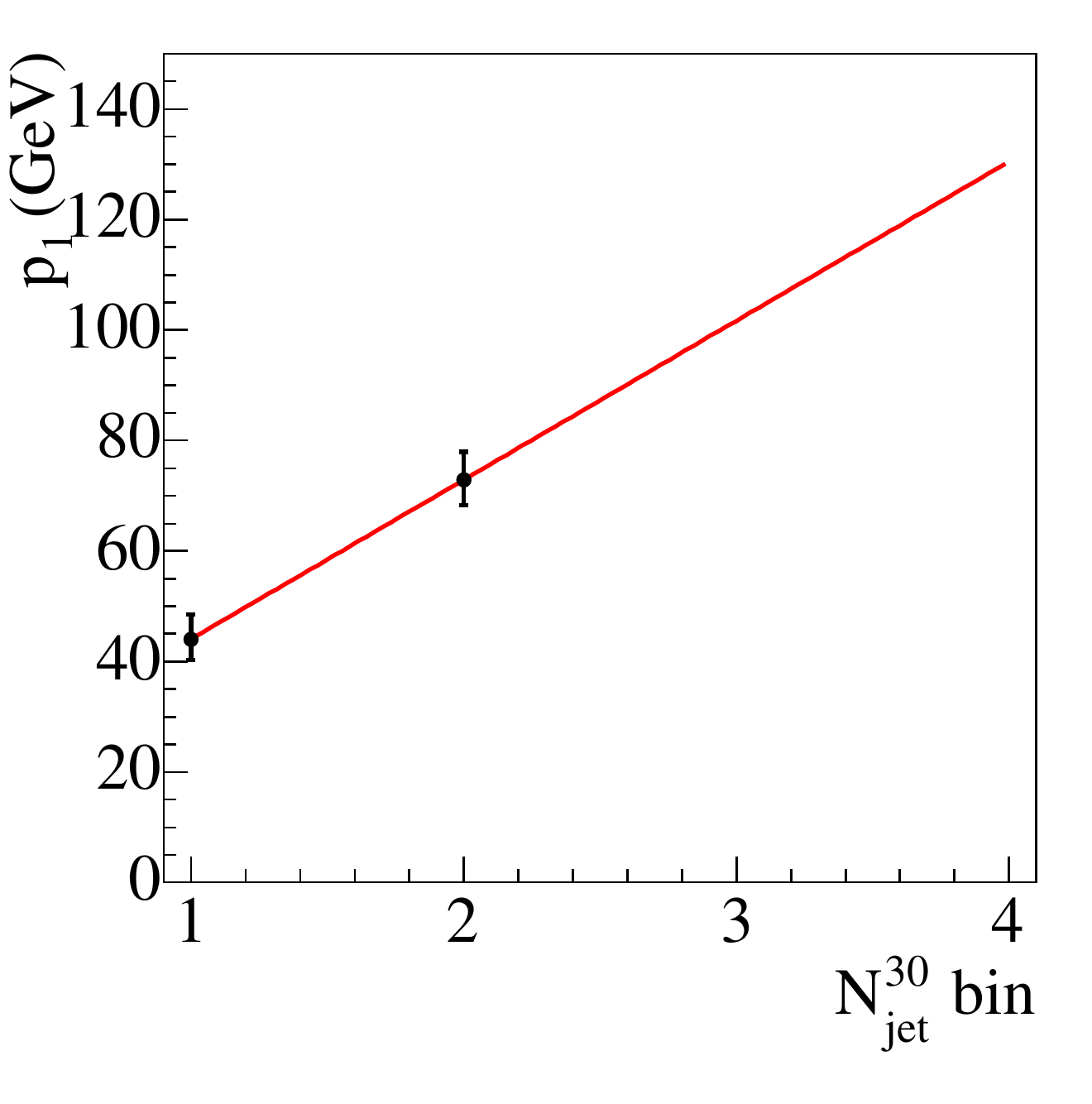}
\end{center}
\caption[]{
The extrapolation of the exponential parameter $p_1$ vs. $\njett$ in \zee\ and \zmm\ events.
}
\label{fig:prd_zdata_bkgfit_exp_vs_njet}
\end{figure}

\begin{figure}
\begin{center}
\includegraphics[width=2.5in]{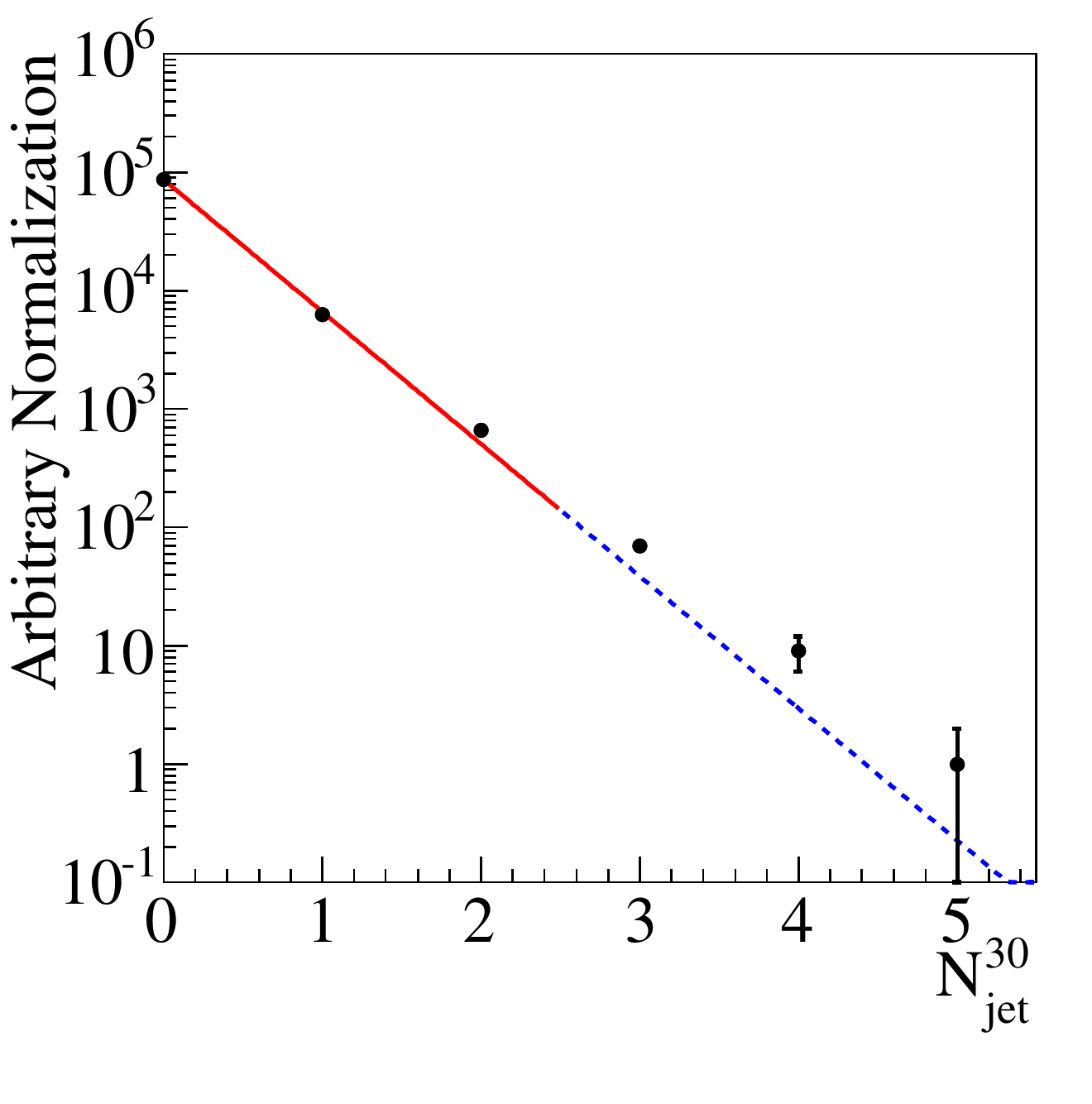}
\end{center}
\caption[]{
$\njett$ distribution in \zee\ and \zmm\ events.  The distribution is fit to an exponential in the range $\njett \leq 2$.
}
\label{fig:prd_zdata_bkgfit_njet30_shape}
\end{figure}

\begin{figure}
\begin{center}
\includegraphics[width=2.5in]{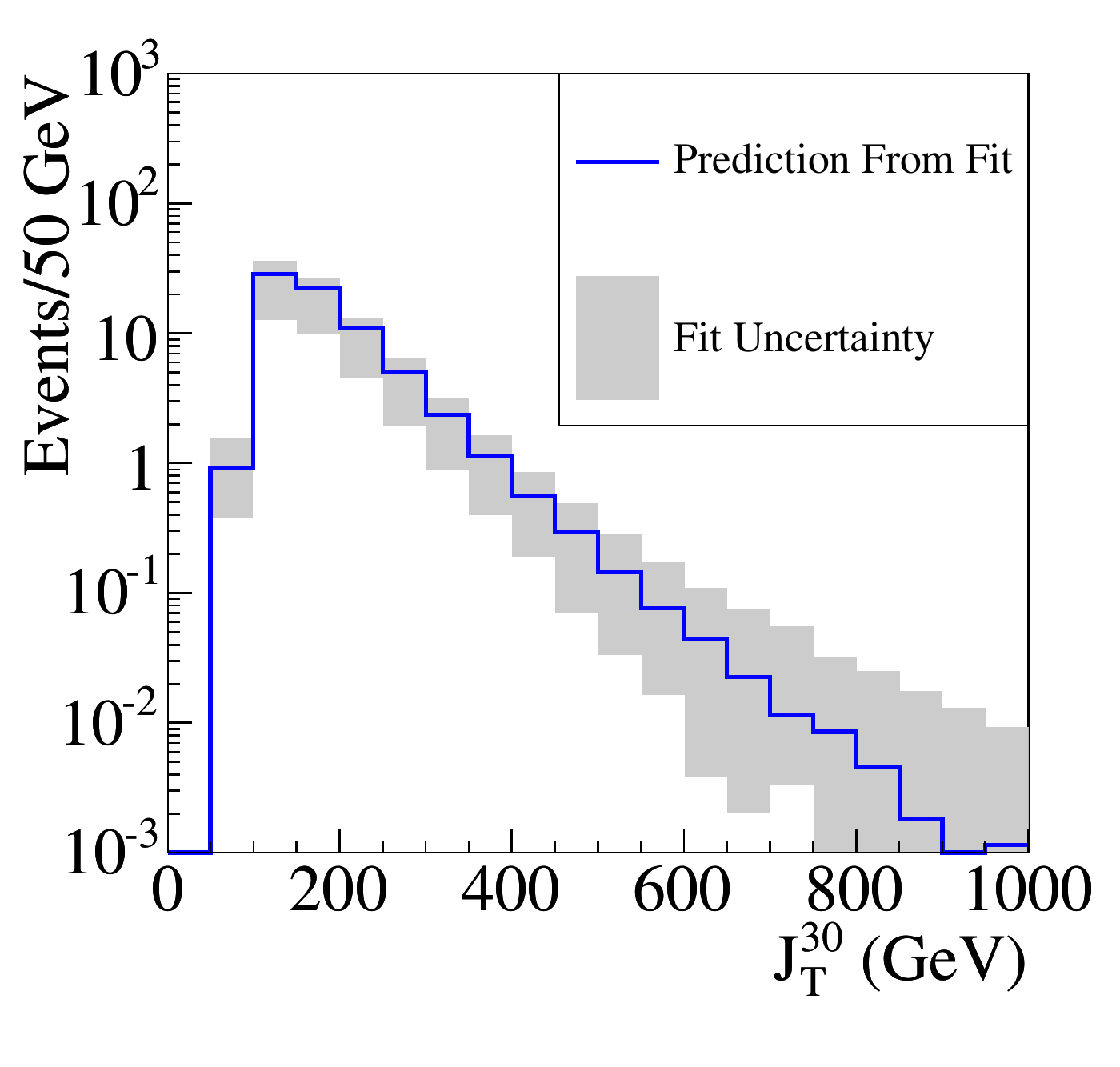}
\end{center}
\caption[]{
The prediction (blue line) and uncertainty (gray band) for the \jtt\ distribution of \zee\ and \zmm\ events.
}
\label{fig:prd_zdata_bkgfit_jt_blind}
\end{figure}

\section{Remaining Backgrounds}
\label{sec:bkgb}

After having estimated the contribution from \Z+jet with the above technique, the remaining backgrounds listed in Sec.~\ref{sec:bkga}
are now estimated.

The second background, multi-jet fakes, has approximately the same shape as the \Z+jet background, and is therefore included in the
fit procedure.  This shape similarity is demonstrated when validating the procedure using multi-jet data in
Sec.~\ref{subsubsec:fitqcdval} above.  Since this background is already included in the \Z+jet background estimate, no further
determination of it is needed.

Nonetheless, its size is independently measured to confirm that it is small relative to the \Z+jet background.  To obtain an upper
bound on the multi-jet background, the sidebands of the \mll\ distribution for events with $\njett \geq 3$ are used.  We attribute
all of the events in the sidebands to multi-jet fakes, and interpolate from the sidebands into the $81 < \mll < 101\ \gev/c^2$
region.  Using this method, less than $11 \pm 2$ events from multi-jet fakes are predicted.  The small size relative to the \Z+jet
background, \zdatabkgfit, indicates that this background is relatively unimportant.

While the third background, from multi-jet events occurring simultaneously with cosmic rays, is also included in the fit procedure as
the jet $E_T$ spectra are similar to the \Z+jet background, its size is again independently measured.  This background is rejected
using timing information from the COT.  That information is also used to estimate this background using the number of events rejected
with the timing cut, combined with a measurement of the rejection efficiency in a sample of cosmic rays with high-purity.  We find a
negligible background.

The remaining backgrounds are not included in the fit procedure since they contain jets from the decays of massive particles and so
the jet $E_T$ spectra do not follow the parameterization in Eq.~(\ref{eqn:jetetparamfinal}).  They can be estimated with Monte Carlo
simulations normalizing to the expected standard model cross sections.  All remaining backgrounds are negligible relative to the
\Z+jet background, the largest being from $WZ$, with an estimated contribution of $1.6 \pm 0.1$ events.  Each of the background
contributions to the $\njett \geq 3$ region is summarized in Table~\ref{tab:listbkg}.  As the backgrounds from $WZ$, $ZZ$, and
\ttbar\ are negligible compared to the \Z+jet background, they are excluded in the background estimation vs. \jtt.

\begin{table}
\begin{center}
\begin{tabular}{cc}
\hline \hline
Process     & Background            \\ \hline
\Z+jet      & \zdatabkgfit          \\ 
\multirow{2}{*}{Multi-jet fakes}                
& $<11 \pm 2$ (included                       \\
& in \Z+jet fit)             \\
Cosmics     & negligible     \\
$WZ$        & $1.6 \pm 0.1$         \\
$ZZ$        & $0.7 \pm 0.1$         \\
$t\bar{t}$  & $0.8 \pm 0.1$         \\ \hline
Total       & $75.3^{+9.8}_{-11.1}$ \\
\hline \hline
\end{tabular}
\caption[]{
Summary of all backgrounds after selecting events with $\njett \geq 3$, independent of \jtt.
}
\label{tab:listbkg}
\end{center}
\end{table}

\section{Results}
\label{sec:results}

\newcommand{\totalbkg}{$75.3^{+9.8}_{-11.1}$}

We now compare the background prediction to the observation in the \Z+jet data.  From the third highest $E_T$ jet extrapolation,
\totalbkg\ events with $\njett \geq 3$ are predicted, and \zdataobs\ events are observed.  In Fig.~\ref{fig:prd_zdata_bkgfit_et2},
the extrapolation is shown overlaid with the data.  The data agree with the extrapolation well.  The predicted \jtt\ distribution is
compared to that observed in data in Fig.~\ref{fig:prd_zdata_bkgfit_jt_unblind}.  Again, the data agree with the prediction quite
well.  The predicted and observed number of events integrated above various $\jtt$ cut values are listed in
Table~\ref{tab:jtdatasig}.  We search for an excess above the prediction at each \jtt\ cut value.  Even when ignoring the systematic
uncertainties, the maximum difference upward has a significance of $+0.9 \sigma$; the maximum difference downward has a significance
of $-1.4 \sigma$.

\begin{figure}
\begin{center}
\includegraphics[width=2.5in]{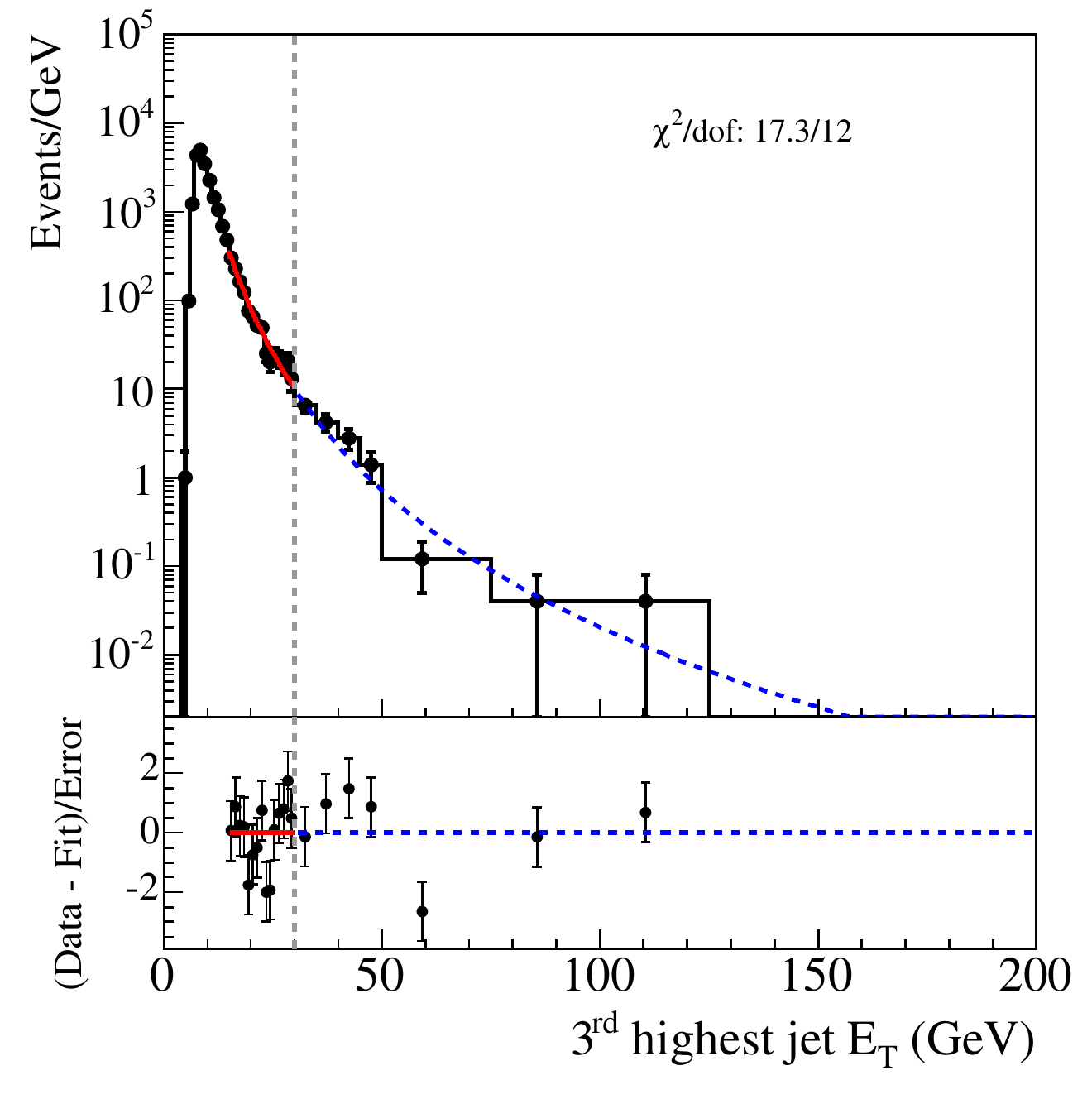}
\end{center}
\caption[]{
$E_T$ distribution of the third highest $E_T$ jet in \zee\ and \zmm\ events.  The fit from
Fig.~\ref{fig:prd_zdata_bkgfit_et2_blind} is overlaid.  The fit extrapolation matches the distribution above 30 GeV well.
}
\label{fig:prd_zdata_bkgfit_et2}
\end{figure}

\begin{figure}
\begin{center}
\includegraphics[width=2.5in]{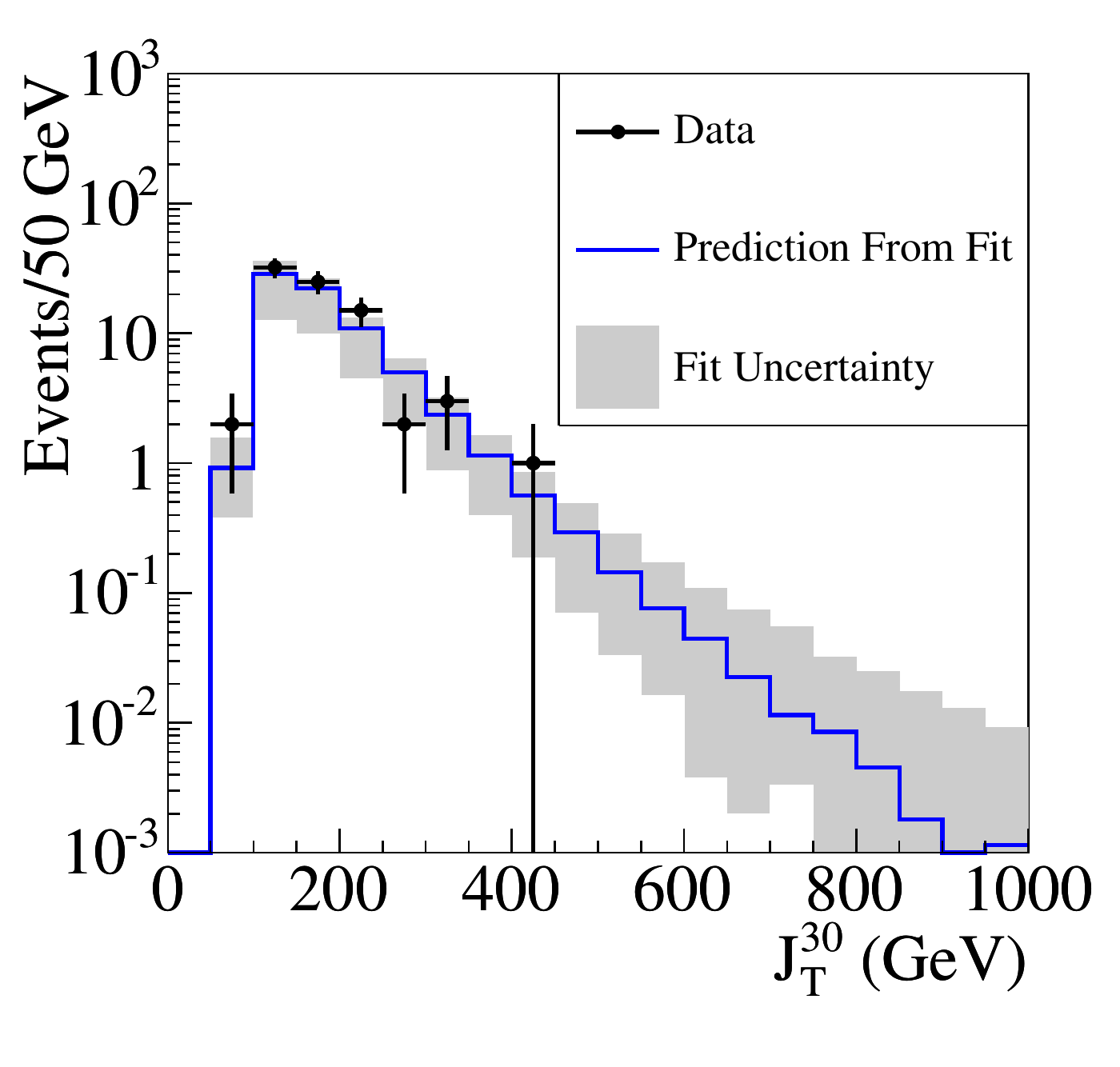}
\end{center}
\caption[]{
The \jtt\ prediction and uncertainty from Fig.~\ref{fig:prd_zdata_bkgfit_jt_blind} compared to the observed distribution (black
points and errors) in \zee\ and \zmm\ events with $\njett \geq 3$.  The prediction agrees well with the data.
}
\label{fig:prd_zdata_bkgfit_jt_unblind}
\end{figure}

\begin{table}[htb]
\begin{center}
\begin{tabular}{ccc}
\hline \hline
Minimum $\jtt$ cut & Total Bkg. (events) & Data (events) \\ \hline
$50$    & $72.2 ^{+17.9}_{-41.3}$ & 80 \\
$100$   & $71.3 ^{+17.3}_{-40.7}$ & 78 \\
$150$   & $42.8 ^{+9.6} _{-24.8}$ & 46 \\
$200$   & $20.6 ^{+5.6} _{-12.6}$ & 21 \\
$250$   & $9.7  ^{+3.6} _{-6.2}$  & 6  \\
$300$   & $4.7  ^{+2.3} _{-3.1}$  & 4  \\
$350$   & $2.3  ^{+1.5} _{-1.6}$  & 1  \\
$400$   & $1.2  ^{+1.0} _{-0.9}$  & 1  \\
$450$   & $0.6  ^{+0.7} _{-0.5}$  & 0  \\
$500$   & $0.3  ^{+0.5} _{-0.3}$  & 0  \\
\hline \hline
\end{tabular}
\end{center}
\caption[]{
The data compared to the \Z+jet background fit prediction vs. $\jtt$.
}
\label{tab:jtdatasig}
\end{table}

Given that there is no significant excess present in the data, a cross section limit is set using the fourth generation model.  At
each $b'$ mass, the counting experiment is evaluated with the requirement $\jtt > m_{b'} c^2$.  The limit is set at a 95\% confidence
level by integrating a likelihood obtained using a Bayesian technique that smears the Poisson-distributed background with Gaussian
acceptance and mean background uncertainties \cite{bib:conwaybayes}.  The background and its uncertainty are taken from the fit
prediction (listed in Table~\ref{tab:jtdatasig}); the product of acceptance and efficiency is taken from Monte Carlo simulation, with
correction factors applied to match the observed efficiency of leptons in $\zll$ data.  The uncertainty on the product of acceptance
and efficiency is 10\%, with the dominant source from a jet energy scale uncertainty of 6.7\% \cite{bib:jetcorr}, the second dominant
from a luminosity uncertainty of 5.9\%, and the remainder from Monte Carlo event statistics and imperfect knowledge of lepton
identification efficiencies \cite{bib:cdfxsecprd}, parton distribution functions \cite{bib:pdfparam}, and initial and final state
radiation.

The 95\% confidence level cross section limit as a function of mass is shown in Fig.~\ref{fig:talk_limit_waccsyst_nocomp}.  In models
with different acceptances, the acceptances of the fourth generation model (for these values, see Appendix \ref{apdx:bpacc}) simply
need to be factored out and the acceptances of those models should be included.

To set a mass limit on the fourth generation model, the $b'$ cross section is calculated at leading order using \textsc{pythia}, with
the assumption that $BR(b' \rightarrow b Z) = 100\%$.  With this assumption, the mass limit observed is $m_{b'} > 268\ \gev/c^2$.
The previous search on this model in the $bZ$ channel obtained a limit of $m_{b'} > 199\ \gev/c^2$ \cite{bib:runibp}, with a
selection catered to the specific $b'$ model by tagging $b$-jets using displaced vertices.

\begin{figure}
\begin{center}
\includegraphics[width=2.5in]{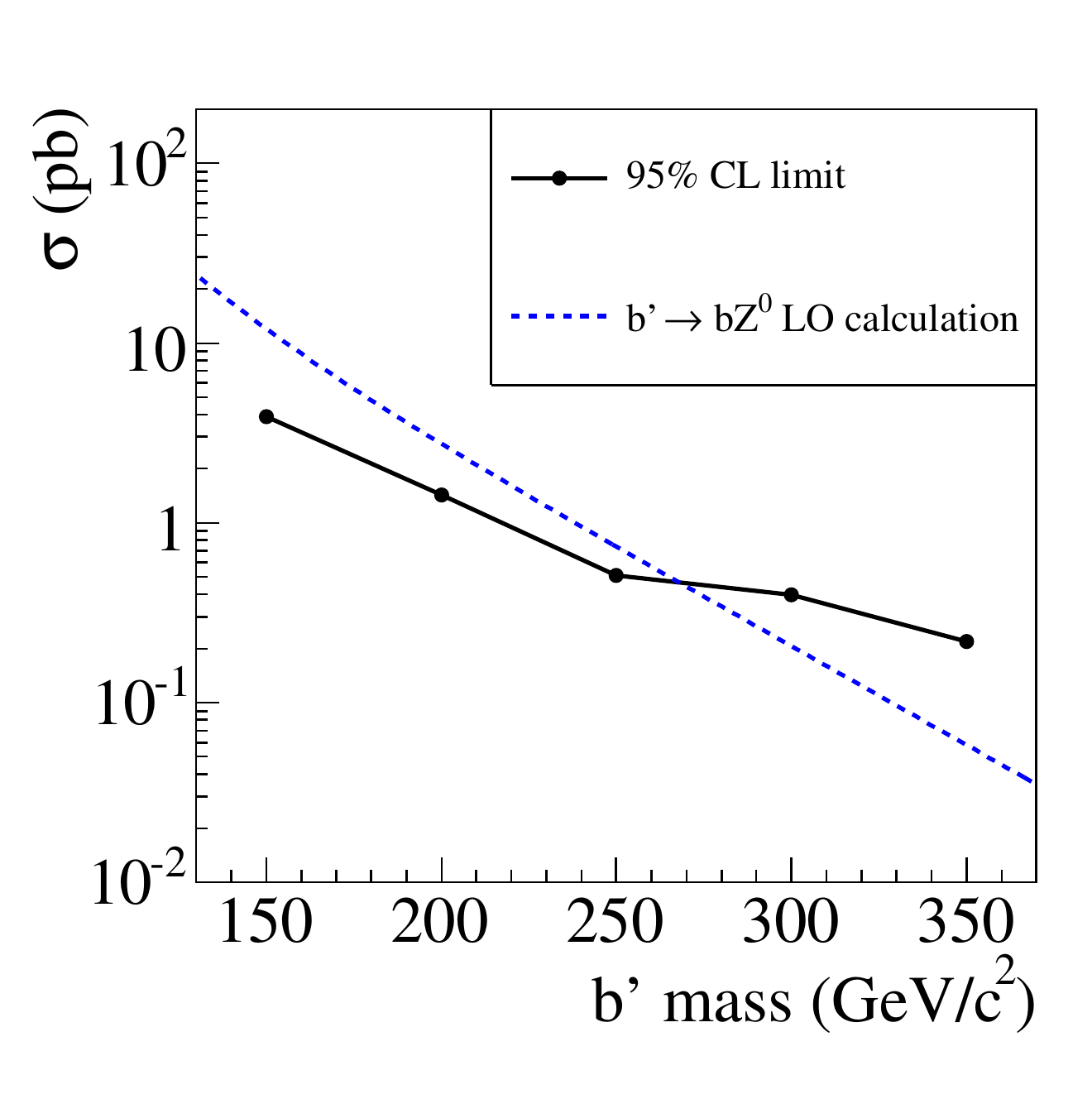}
\end{center}
\caption[]{
Cross section limit vs. $b'$ mass, set at a confidence level of 95\%.  In the acceptance calculation $BR(b' \rightarrow bZ) \equiv
\beta = 100$\% was assumed.  If $\beta < 100$\%, the acceptance would scale by the factor $1-(1-\beta)^2$, since the $b'$ is produced
in pairs and only one of them is required to decay to a \Z\ with our selection.  In addition, non-$Z$ decays could change the
acceptance of the $\njett \geq 3$ cut.
}
\label{fig:talk_limit_waccsyst_nocomp}
\end{figure}

\section{Conclusion}
\label{sec:conc}

We have presented the results of a search for new particles decaying to \Z\ bosons and jets.  We developed and validated a new
technique to predict the dominant background from the data alone.  This technique complements the phenomenological-based method of
predicting backgrounds via Monte Carlo calculations of higher-order matrix elements and non-perturbative soft parton showers.  The
technique presented here has advantages of not requiring careful tuning of phenomenological parameters when comparing to data and
not requiring the many resource-consuming iterations of Monte Carlo detector simulations.  The speed with which it can be applied
makes it an attractive tool for calculation of backgrounds in jet-rich environments at future experiments, including those at the
Large Hadron Collider.


In the application of the technique on CDF \Z+jet data, no significant excess above background was seen.  A cross section limit was
therefore set on a fourth generation model as a function of mass.  A mass limit of $m_{b'} > 268\ \gev/c^2$ using a leading-order
$b'$ cross section calculation with the assumption that $BR(b' \rightarrow b Z) = 100\%$ was set at a 95\% confidence level.

\begin{appendix}

\section{Acceptance of $b'$ Model}
\label{apdx:bpacc}

In Table~\ref{tab:bpacc} the acceptance times efficiency to select $b' \rightarrow bZ$ events (assuming $BR(b' \rightarrow
bZ)=100$\%) after the kinematic cuts is shown.  As these acceptances include a factor from $BR(\zll)$, they are maximally
$BR(\zee)+BR(\zmm) = 6.7$\%.

\begin{table}[h]
\begin{center}
\begin{tabular}{ cc }
\hline \hline
$b'$ mass (GeV) & Acceptance (\%) \\ \hline
150 & 1.05 \\
200 & 1.44 \\
250 & 1.61 \\
300 & 1.66 \\
350 & 1.77 \\
\hline \hline
\end{tabular}
\end{center}
\caption{Acceptances to select $b' \rightarrow bZ$ events versus mass, after applying the $\njett \geq 3$ and $\jtt > m_{b'} c^2$
requirements.  These include a factor from the branching ratio of \zee\ and \zmm.  If this factor is removed, the acceptances range
from 8--14\%.  $BR(b' \rightarrow bZ)=100$\% was assumed.}
\label{tab:bpacc}
\end{table}

\end{appendix}

\begin{acknowledgments}

We thank the Fermilab staff and the technical staffs of the participating institutions for their vital contributions. This work was
supported by the U.S. Department of Energy and National Science Foundation; the Italian Istituto Nazionale di Fisica Nucleare; the
Ministry of Education, Culture, Sports, Science and Technology of Japan; the Natural Sciences and Engineering Research Council of
Canada; the National Science Council of the Republic of China; the Swiss National Science Foundation; the A.P. Sloan Foundation; the
Bundesministerium f\"ur Bildung und Forschung, Germany; the Korean Science and Engineering Foundation and the Korean Research
Foundation; the Science and Technology Facilities Council and the Royal Society, UK; the Institut National de Physique Nucleaire et
Physique des Particules/CNRS; the Russian Foundation for Basic Research; the Comisi\'on Interministerial de Ciencia y
Tecnolog\'{\i}a, Spain; the European Community's Human Potential Programme; the Slovak R\&D Agency; and the Academy of Finland.

\end{acknowledgments}


\begin{thebibliography}{99}
\bibitem{bib:lumiave} The integrated luminosity of the electron-triggered sample is 1.08 $\ifb$; that of the muon-triggered sample is
1.04 $\ifb$; the uncertainty on each is 5.9\%, described in \cite{bib:clc}.  In the averaging, we weight each luminosity with the
acceptance of a 200 GeV $b'$ signal after applying the $\njett \geq 3$ cut described in Sec.~\ref{subsec:kinsel}.  After applying
this cut, the acceptances of the $b'\rightarrow\zee$ and $b'\rightarrow\zmm$ channels are approximately the same, so the weights are
approximately equal.
\bibitem{bib:thresh} In this section, it is useful to make a distinction between a ``threshold'' and a ``cut.''  We define a
``threshold'' as a requirement on a jet $E_T$; we define a cut as a requirement on a variable made from individual jets.  So, in the
equation $\njett \geq 3$, we say that we have a 30 GeV threshold on the jet $E_T$, and we have a cut of at least 3 jets.
\bibitem{bib:1p5explan} The full sensitivity figure of merit obtained from Eq.~(8) of \cite{bib:sigbkg1p5} is more complicated than
that used here.  The one used here, \srbfom, is a reduced formula for the full figure of merit, in the case where the significance of
discovery and the limit confidence level (both optimized for with this figure of merit) correspond to $3\sigma$ probabilities, and
where the number of signal and background events can be described by Gaussian (as opposed to Poisson) distributions.
\bibitem{bib:senslowmass} For the $m_{b'} = 100\ \gev/c^2$ mass point, the target selection has a signal of $450$ on a background of
$229$, giving a figure of merit $S/(1.5+\sqrt{B}) = 27$.  With the simple cuts, there is a signal of $64$ on a background of $20$,
giving a figure of merit $S/(1.5+\sqrt{B}) = 11$.  While the sensitivity figure of merit is smaller, a signal of $64$ on a background
of $20$ is still adequate for a discovery.
\bibitem{bib:fnlikemaxfit} We use unbinned likelihood-maximization fits.  In practice, rather than maximizing the likelihood $L$, the
quantity $-\log{L}$ is minimized.  When comparing unbinned fits with binned histograms, we place the x-value of each bin at the
average of the entries in that bin.
\bibitem{bib:fnnofitnjetfour} As we do not expect many events in data with $\njett \geq 4$, a detailed checking of the parameter
extrapolation to these jet multiplicities is not necessary.  We have verified that the extrapolation is consistent with this
distribution with the statistics present in Monte Carlo.  In addition, this extrapolation is implicitly validated when the validation
of the method as a whole is done in Sec.~\ref{subsec:fitval}.
\bibitem{bib:fnprescale} Since the cross section for jet events with $E_T > 20$ GeV is extremely large, not all events of this type
are able to be kept by the data acquisition system.  Only a fraction of these events are kept; the inverse of the fraction of events
kept is known as the prescale.  For the single jet trigger with $E_T > 20$ GeV, the total prescale used is approximately $500$.
\bibitem{bib:fnprobcalc} 
This probability calculation is done by integrating the distribution of the expected number of events above the observed value.  For
the distribution of the expected number of events, we use a Poisson distribution with a mean equal to the fit prediction ($97$ in
this case) smeared with a Gaussian centered at zero with a width equal to the background prediction's uncertainty ($\pm 27$ in
this case).  If the number of data events is higher than the background prediction, we use the upper uncertainty in the Gaussian
smearing; we use the lower uncertainty if the data is below the background prediction.  Additionally we convert this probability to
units of standard deviation by inverting:

\[
\int_n^{\infty} \frac{1}{\sqrt{2\pi}} e^{-x^2/2} dx = p 
\].
\bibitem{bib:topxsecnote}
The statistical uncertainty is the statistical uncertainty on the background prediction added in quadrature with the
statistical uncertainty on the number of observed events, $\sqrt{ \datawjetobs }$.  We do not give a full systematic uncertainty, as
we do not evaluate the uncertainty on the acceptance of \ttbar.  The systematic uncertainty on the cross section from the background
prediction alone is 1.5 pb.
\end{thebibliography}
\end{document}